\input harvmac
\input epsf

\noblackbox

\newcount\figno

\figno=0
\def\fig#1#2#3{
\par\begingroup\parindent=0pt\leftskip=1cm\rightskip=1cm\parindent=0pt
\baselineskip=11pt
\global\advance\figno by 1
\midinsert
\epsfxsize=#3
\centerline{\epsfbox{#2}}
\vskip 12pt
\centerline{{\bf Figure \the\figno} #1}\par
\endinsert\endgroup\par}
\def\figlabel#1{\xdef#1{\the\figno}}
\def\pano{\par\noindent}

\def\bigno{\bigskip\noindent}

\font\cmss=cmss10
\font\cmsss=cmss10 at 7pt

\def\rlx{\relax\leavevmode}
\def\inbar{\vrule height1.5ex width.4pt depth0pt}
\def\IC{\relax\,\hbox{$\inbar\kern-.3em{\rm C}$}}
\def\IR{\relax{\rm I\kern-.18em R}}
\def\IN{\relax{\rm I\kern-.18em N}}
\def\IP{\relax{\rm I\kern-.18em P}}
\def\ZZ{\rlx\leavevmode\ifmmode\mathchoice{\hbox{\cmss Z\kern-.4em Z}}
 {\hbox{\cmss Z\kern-.4em Z}}{\lower.9pt\hbox{\cmsss Z\kern-.36em Z}}
 {\lower1.2pt\hbox{\cmsss Z\kern-.36em Z}}\else{\cmss Z\kern-.4em Z}\fi}

\def\narrowplus{\kern -.04truein + \kern -.03truein}
\def\narrowminus{- \kern -.04truein}
\def\narrowminussub{\kern -.02truein - \kern -.01truein}

\def\o#1{\overline{#1}}
\def\la{\langle}
\def\ra{\rangle}



\lref\AngelantonjCT{
C.~Angelantonj and A.~Sagnotti,
{\it Open Strings},
hep-th/0204089.
}

\lref\GorbatovPW{
E.~Gorbatov, V.~S.~Kaplunovsky, J.~Sonnenschein, S.~Theisen and S.~Yankielowicz,
{\it On Heterotic Orbifolds, M-theory and Type I$\, '$ Brane Engineering},
JHEP {\bf 0205}, 015 (2002)
hep-th/0108135.
}

\lref\KaplunovskyIA{
V.~Kaplunovsky, J.~Sonnenschein, S.~Theisen and S.~Yankielowicz,
{\it On the Duality between Perturbative Heterotic
Orbifolds and M-theory on  $T^4/\ZZ_N$},
Nucl.\ Phys.\ B {\bf 590} (2000) 123, 
hep-th/9912144.
}

\lref\DoranVE{
C.~F.~Doran, M.~Faux and B.~A.~Ovrut,
{\it Four-dimensional N = 1 Super Yang-Mills Theory from an M
theory Orbifold},
hep-th/0108078.
}

\lref\FauxSP{
M.~Faux, D.~L\"ust and B.~A.~Ovrut,
{\it An M-theory Perspective on Heterotic K3 Orbifold Compactifications},
hep-th/0010087.
}

\lref\FauxDV{
M.~Faux, D.~L\"ust and B.~A.~Ovrut,
{\it Local Anomaly Cancellation, M-theory Orbifolds and Phase-transitions},
Nucl.\ Phys.\ B {\bf 589} (2000) 269, 
hep-th/0005251.
}

\lref\FauxHM{
M.~Faux, D.~L\"ust and B.~A.~Ovrut,
{\it Intersecting Orbifold Planes and Local
Anomaly Cancellation in  M-theory},
Nucl.\ Phys.\ B {\bf 554} (1999) 437, 
hep-th/9903028.
}


\lref\LercheJB{
W.~Lerche, C.~A.~Lutken and C.~Schweigert,
{\it D-branes on ALE Spaces and the ADE
Classification of Conformal Field Theories},
Nucl.\ Phys.\ B {\bf 622} (2002) 269,
hep-th/0006247.
}

\lref\KasteID{
P.~Kaste, W.~Lerche, C.~A.~Lutken and J.~Walcher,
{\it D-branes on K3-fibrations},
Nucl.\ Phys.\ B {\bf 582} (2000) 203,
hep-th/9912147.
}

\lref\rangles{M.~Berkooz, M.~R.~Douglas and R.~G.~Leigh, {\it Branes Intersecting
at Angles}, Nucl. Phys. B {\bf 480} (1996) 265, hep-th/9606139.
}

\lref\KleinVU{
M.~Klein,
{\it Couplings in Pseudo-Supersymmetry}, hep-th/0205300.
}

\lref\LeighJQ{
R.~G.~Leigh,
{\it Dirac-Born-Infeld Action From Dirichlet Sigma Model},
Mod.\ Phys.\ Lett.\ A {\bf 4} (1989) 2767.
}

\lref\FradkinQD{
E.~S.~Fradkin and A.~A.~Tseytlin,
{\it Nonlinear Electrodynamics From Quantized Strings},
Phys.\ Lett.\ B {\bf 163} (1985) 123.
}

\lref\berlin{R.~Blumenhagen, B.~K\"ors and D.~L\"ust,
{\it Moduli Stabilization for Intersecting Brane Worlds in Type 0'
String Theory}, hep-th/0202024.
}

\lref\rrab{R.~Rabadan, {\it Branes at Angles, Torons, Stability and
Supersymmetry}, Nucl.\ Phys.\ B {\bf 620} (2002) 152, hep-th/0107036.
}

\lref\rbgkb{R.~Blumenhagen, L.~G\"orlich and B.~K\"ors,
{\it A New Class of Supersymmetric Orientifolds with D-Branes at
Angles}, hep-th/0002146.
}

\lref\rbgkc{R.~Blumenhagen, L.~G\"orlich and B.~K\"ors, {\it
Supersymmetric 4D Orientifolds of Type IIA with D6-branes at Angles},
JHEP {\bf 0001} (2000) 040, hep-th/9912204.
}

\lref\rfhs{S.~F\"orste, G.~Honecker and R.~Schreyer, {\it Supersymmetric
$\ZZ_N \times \ZZ_M$ Orientifolds in 4-D with D-branes at Angles},
Nucl. Phys. B {\bf 593} (2001) 127, hep-th/0008250.
}

\lref\rfhstwo{S.~F\"orste, G.~Honecker and R.~Schreyer, {\it Orientifolds
with Branes at Angles}, JHEP {\bf 0106} (2001) 004, hep-th/0105208.
}

\lref\rbgklnon{R.~Blumenhagen, L.~G\"orlich, B.~K\"ors and D.~L\"ust,
{\it Noncommutative Compactifications of Type I Strings on Tori with Magnetic
Background Flux}, JHEP {\bf 0010} (2000) 006, hep-th/0007024.
}

\lref\IbanezDJ{
L.~E.~Ibanez,
{\it Standard Model Engineering with Intersecting Branes}, 
hep-ph/0109082.
}

\lref\rbgklmag{R.~Blumenhagen, L.~G\"orlich, B.~K\"ors and D.~L\"ust,
{\it Magnetic Flux in Toroidal Type I Compactification}, Fortsch. Phys. 49
(2001) 591, hep-th/0010198.
}

\lref\rba{C.~Angelantonj, R.~Blumenhagen, {\it Discrete Deformations in
Type I Vacua}, Phys. Lett. B {\bf 473} (2000) 86,
hep-th/9911190.
}

\lref\ras{C.~Angelantonj, A.~Sagnotti, {\it Type I
Vacua and Brane Transmutation}, hep-th/0010279.
}

\lref\raads{C.~Angelantonj, I.~Antoniadis, E.~Dudas, A.~Sagnotti, {\it Type I
Strings on Magnetized Orbifolds and Brane Transmutation},
Phys. Lett. B {\bf 489} (2000) 223, hep-th/0007090.
}

\lref\rbkl{R.~Blumenhagen, B.~K\"ors and D.~L\"ust,
{\it Type I Strings with $F$ and $B$-Flux}, JHEP {\bf 0102} (2001) 030,
hep-th/0012156.
}

\lref\rbgkl{R.~Blumenhagen, L.~G\"orlich, B.~K\"ors and D.~L\"ust,
{\it Asymmetric Orbifolds, Noncommutative Geometry and Type I
Vacua}, Nucl.\ Phys.\ B {\bf 582} (2000) 44, hep-th/0003024.
}

\lref\rbgka{R.~Blumenhagen, L.~G\"orlich and B.~K\"ors,
{\it Supersymmetric Orientifolds in 6D with D-Branes at Angles},
Nucl. Phys. B {\bf 569} (2000) 209, hep-th/9908130.
}

\lref\rcvetica{M.~Cvetic, G.~Shiu and  A.~M.~Uranga,  {\it Three-Family
Supersymmetric Standard-like Models from Intersecting Brane Worlds}
Phys. Rev. Lett. {\bf 87} (2001) 201801,  hep-th/0107143.
}

\lref\rcveticb{M.~Cvetic, G.~Shiu and  A.~M.~Uranga,  {\it
Chiral Four-Dimensional N=1 Supersymmetric Type IIA Orientifolds from
Intersecting D6-Branes}, Nucl. Phys. B {\bf 615} (2001) 3, hep-th/0107166.
}

\lref\rott{R.~Blumenhagen, B.~K\"ors, D.~L\"ust and T.~Ott, {\it
The Standard Model from Stable Intersecting Brane World Orbifolds},
Nucl. Phys. B {\bf 616} (2001) 3, hep-th/0107138.
}

\lref\rottb{R.~Blumenhagen, B.~K\"ors, D.~L\"ust and T.~Ott, {\it
Intersecting Brane Worlds on Tori and Orbifolds}, hep-th/0112015.
}

\lref\KorsKU{
B.~K\"ors,
{\it Open Strings In Magnetic Background Fields},
Fortsch.\ Phys.\  {\bf 49} (2001) 759.
}

\lref\HaackJZ{
M.~Haack and J.~Louis,
{\it M-theory Compactified on Calabi-Yau Fourfolds with Background Flux},
Phys.\ Lett.\ B {\bf 507} (2001) 296, 
hep-th/0103068.
}

\lref\PolchinskiSM{
J.~Polchinski and A.~Strominger,
{\it New Vacua for Type II String Theory},
Phys.\ Lett.\ B {\bf 388} (1996) 736, 
hep-th/9510227.
}

\lref\MichelsonPN{
J.~Michelson,
{\it Compactifications of Type IIB Strings to Four Dimensions with  
non-trivial Classical Potential},
Nucl.\ Phys.\ B {\bf 495}, 127 (1997)
hep-th/9610151.
}

\lref\LouisUY{
J.~Louis and A.~Micu,
{\it Heterotic String Theory with Background Fluxes},
Nucl.\ Phys.\ B {\bf 626}, 26 (2002)
hep-th/0110187.
}

\lref\rbonna{S.~F\"orste, G.~Honecker and R.~Schreyer, {\it
Orientifolds with Branes at Angles}, JHEP {\bf 0106} (2001) 004,
hep-th/0105208.
}

\lref\rbonnb{G.~Honecker, {\it Intersecting Brane World Models from
D8-branes on $(T^2 \times T^4/\ZZ_3)/\Omega R_1$ Type IIA Orientifolds},
JHEP {\bf 0201} (2002) 025, hep-th/0201037.
}

\lref\rqsusy{D.~Cremades, L.~E.~Ibanez and F.~Marchesano, {\it
SUSY Quivers, Intersecting Branes and the Modest Hierarchy Problem},
hep-th/0201205.
}

\lref\rqsusyb{D.~Cremades, L.~E.~Ibanez and F.~Marchesano, {\it
     Intersecting Brane Models of Particle Physics and the Higgs Mechanism},
      hep-th/0203160.
}

\lref\AldazabalPY{
G.~Aldazabal, L.~E.~Ibanez and A.~M.~Uranga,
{\it Gauging Away the Strong CP Problem},
hep-ph/0205250.
}

\lref\BeckerNN{
K.~Becker, M.~Becker, M.~Haack and J.~Louis,
{\it
Supersymmetry Breaking and alpha' Corrections to Flux induced  Potentials},
hep-th/0204254.
}

\lref\LouisNY{
J.~Louis and A.~Micu,
{\it
Type II Theories Compactified on Calabi-Yau Threefolds in
the Presence  of Background Fluxes}, hep-th/0202168.
}

\lref\SchomerusWZW{
S.~Fredenhagen and V.~Schomerus,
{\it Branes on Group Manifolds, Gluon Condensates, and twisted K-theory},
JHEP\ {\bf 0104} (2001) 007, hep-th/0012164.
}

\lref\MooreKinst{
J.~Maldacena, G.~Moore, and N.~Seiberg, 
{\it D-Brane Instantons and K-Theory Charges},
JHEP\ {\bf 0111} (2001) 062, hep-th/0108100.
}

\lref\deWitXG{
B.~de Wit, D.~J.~Smit and N.~D.~Hari Dass,
{\it Residual Supersymmetry Of Compactified D = 10 Supergravity}, 
Nucl.\ Phys.\ B {\bf 283} (1987) 165.
}

\lref\AgataZH{
G.~Dall'Agata,
{\it Type IIB Supergravity Compactified on a Calabi-Yau Manifold with H-fluxes},
JHEP {\bf 0111} (2001) 005, hep-th/0107264.
}

\lref\rbachas{C.~Bachas, {\it A Way to Break Supersymmetry}, hep-th/9503030.
}

\lref\rafiruph{G.~Aldazabal, S.~Franco, L.~E.~Ibanez, R.~Rabadan, A.~M.~Uranga,
{\it Intersecting Brane Worlds}, JHEP {\bf 0102} (2001) 047, hep-ph/0011132.
}

\lref\rafiru{G.~Aldazabal, S.~Franco, L.~E.~Ibanez, R.~Rabadan, A.~M.~Uranga,
{\it $D=4$ Chiral String Compactifications from Intersecting Branes}, 
J.\ Math.\ Phys.\  {\bf 42} (2001) 3103, hep-th/0011073.
}

\lref\rimr{L.~E.~Ibanez, F.~Marchesano, R.~Rabadan, {\it Getting just the
Standard Model at Intersecting Branes}, 
JHEP {\bf 0111} (2001) 002, hep-th/0105155.
}

\lref\belrab{J.~Garcia-Bellido and R.~Rabadan, {\it Complex Structure Moduli
Stability in Toroidal Compactifications}, 
JHEP {\bf 0205} (2002) 042, hep-th/0203247.
}

\lref\rcim{D.~Cremades, L.~E.~Ibanez and F.~Marchesano, {\it
    Standard Model at Intersecting D5-branes: Lowering the String Scale},
     hep-th/0205074.
}

\lref\rkokoa{C.~Kokorelis, {\it GUT Model Hierarchies from Intersecting Branes},
hep-th/0203187.
}

\lref\rkokob{C.~Kokorelis, {\it New Standard Model Vacua from Intersecting Branes},
hep-th/0205147.
}

\lref\HoneckerDJ{
G.~Honecker,
{\it Non-supersymmetric Orientifolds with D-branes at Angles}, hep-th/0112174.
}

\lref\rcls{M.~Cvetic, P.~Langacker, and G.~Shiu, {\it
 Phenomenology of A Three-Family Standard-like String Model},
hep-ph/0205252.
}

\lref\rgkp{S.~B.~Giddings, S.~Kachru and J.~Polchinski, {\it
Hierarchies from Fluxes in String Compactifications},
hep-th/0105097.
}

\lref\rkst{S.~Kachru, M.~Schulz and S.~Trivedi, {\it
             Moduli Stabilization from Fluxes in a Simple IIB Orientifold},
              hep-th/0201028. 
}

\lref\radd{N.~Arkani-Hamed, S.~Dimopoulos, and G.~Dvali, {\it The Hierarchy
Problem and New Dimensions at a Millimeter}, Phys. Lett. B {\bf 429} (1998)
263, hep-ph/9803315.
}

\lref\raadd{I.~Antoniadis, N.~Arkani-Hamed, S.~Dimopoulos, and G.~Dvali, {\it
New Dimensions at a Millimeter to a Fermi and Superstrings at a TeV},
Phys. Lett. B {\bf 436} (1998) 257, hep-ph/9804398.
}

\lref\roemel{C.~R\"omelsberger,
{\it (Fractional) Intersection Numbers, Tadpoles and Anomalies},
hep-th/0111086.
}

\lref\rv{C.~Voisin, Journ\'ees de G\'eom\'etrie Alg\'ebrique d'Orsay,
eds. A.~Beauville et al.
(Orsay, 1992), Ast\'erisque No. 218 (1993) 273.
}

\lref\rb{C.~Borcea, {\it K3 surfaces with involution and mirror pairs of
Calabi-Yau manifolds}, in: Essays on Mirror Manifolds II,
eds. B.~Greene and S.-T.~Yau,
(International Press and AMS 1997).
}

\lref\ruranga{A.~M.~Uranga, {\it D-brane, Fluxes and Chirality},
hep-th/0201221.
}

\lref\rurangab{A.~M.~Uranga, {\it Localized Instabilities at Conifolds},
hep-th/0204079.
}

\lref\rscruccab{C.~A.~Scrucca and M.~Serone,
   {\it Anomalies and Inflow on D-branes and O-planes},
   Nucl. Phys. B {\bf 556} (1999) 197, hep-th/9903145.
}

\lref\rscrucca{J.~F.~Morales, C.~A.~Scrucca and M.~Serone,
        {\it Anomalous Couplings for D-branes and O-planes},
      Nucl. Phys. B {\bf 552} (1999) 291, hep-th/9812071.
}


\lref\AldazabalMR{
G.~Aldazabal, A.~Font, L.~E.~Ibanez and G.~Violero,
{\it D = 4, N = 1, Type IIB Orientifolds},
Nucl.\ Phys.\ B {\bf 536} (1998) 29, hep-th/9804026.
}

\lref\DixonJC{
L.~J.~Dixon, J.~A.~Harvey, C.~Vafa and E.~Witten,
{\it Strings On Orbifolds 2},
Nucl.\ Phys.\ B {\bf 274} (1986) 285.
}

\lref\DixonJW{
L.~J.~Dixon, J.~A.~Harvey, C.~Vafa and E.~Witten,
{\it Strings On Orbifolds},
Nucl.\ Phys.\ B {\bf 261} (1985) 678.
}

\lref\BailinIE{
D.~Bailin, G.~V.~Kraniotis and A.~Love,
{\it Standard-like Models from Intersecting D4-branes},
Phys.\ Lett.\ B {\bf 530} (2002) 202,
hep-th/0108131.
}

\lref\VafaXN{
C.~Vafa,
{\it Evidence for F-Theory},
Nucl.\ Phys.\ B {\bf 469} (1996) 403, hep-th/9602022.
}

\lref\MorrisonPP{
D.~R.~Morrison and C.~Vafa,
{\it Compactifications of F-Theory on Calabi--Yau Threefolds -- II},
Nucl.\ Phys.\ B {\bf 476} (1996) 437,
hep-th/9603161. 
}

\lref\MorrisonNA{
D.~R.~Morrison and C.~Vafa,
{\it Compactifications of F-Theory on Calabi--Yau Threefolds -- I},
Nucl.\ Phys.\ B {\bf 473} (1996) 74,
hep-th/9602114.
}

\lref\BerkoozDW{
M.~Berkooz and R.~G.~Leigh,
{\it A D = 4 N = 1 Orbifold of Type I Strings},
Nucl.\ Phys.\ B {\bf 483} (1997) 187,
hep-th/9605049.
}

\lref\rnik{V.~Nikulin, in: Proceedings of the International Congress of
Mathematicians (Berkeley 1986) 654.
}

\lref\MarinoAF{
M.~Marino, R.~Minasian, G.~W.~Moore and A.~Strominger,
{\it Nonlinear Instantons from Supersymmetric $p$-branes},
JHEP {\bf 0001} (2000) 005,
hep-th/9911206.
}


\lref\KakushadzeEG{
Z.~Kakushadze,
{\it On Four-dimensional N = 1 Type I Compactifications},
Nucl.\ Phys.\ B {\bf 535} (1998) 311, hep-th/9806008.
}

\lref\sagn{M.~Bianchi and A.~Sagnotti, {\it On the Systematics of Open String
Theories}, Phys. Lett. B {\bf 247} (1990) 517.
}

\lref\rgimpol{
E.~G.~Gimon and J.~Polchinski, {\it Consistency Conditions
for Orientifolds and D-Manifolds}, Phys.\ Rev.\ {\bf D54} (1996) 1667,
hep-th/9601038.
}

\lref\rgimjo{ E.~G.~Gimon and C.~V.~Johnson, {\it K3 Orientifolds},
Nucl.\ Phys.\ B {\bf 477} (1996) 715, hep-th/9604129.
}

\lref\rbluma{J.~D.~Blum and A.~Zaffaroni, {\it An Orientifold from F Theory},
Phys.Lett. B {\bf 387} (1996) 71, hep-th/9607019.
}

\lref\rblumb{J.~D.~Blum, {\it F Theory Orientifolds, M Theory Orientifolds and
Twisted Strings}, Nucl.Phys. B {\bf 486} (1997) 34, hep-th/9608053.
}

\lref\DabholkarKA{
A.~Dabholkar and J.~Park,
{\it A Note on Orientifolds and F-theory},
Phys.\ Lett.\ B {\bf 394} (1997) 302, hep-th/9607041.
}

\lref\rdabol{ A.~Dabholkar and J.~Park, {\it An Orientifold of Type
        IIB theory on K3}, Nucl. Phys. B {\bf 472} (1996) 207,
        hep-th/9602030;
        {\it Strings on Orientifolds}, Nucl. Phys. B {\bf 477}
        (1996) 701, hep-th/9604178.
}

\lref\rstanev{C.~Angelantonj, M.~Bianchi, G.~Pradisi, A.~Sagnotti and
      Y.~Stanev, {\it Chiral Asymmetry in Four-Dimensional Open-String Vacua},
  Phys.Lett. B {\bf 385} (1996) 96, hep-th/9606169.
}


\lref\rbdlr{I.~Brunner, M.~R.~Douglas, A.~Lawrence and C.~R\"omelsberger, {\it
  D-branes on the Quintic}, JHEP 0008 (2000) 015, hep-th/9906200.
}

\lref\raahv{B.~Acharya, M.~Aganagic, K.~Hori and  C.~Vafa, {\it
      Orientifolds, Mirror Symmetry and Superpotentials}, hep-th/0202208.
}

\lref\rakv{M.~Aganagic, A.~Klemm and C.~Vafa,
    {\it Disk Instantons, Mirror Symmetry and the Duality Web},
           Z.Naturforsch. {\bf A57} (2002) 1-28, hep-th/0105045.
}

\lref\rav{M.~Aganagic and C.~Vafa, {\it
         Mirror Symmetry, D-Branes and Counting Holomorphic Discs},
          hep-th/0012041.
}

\lref\rhiv{K.~Hori, A.~Iqbal and C.~Vafa, {\it
            D-Branes And Mirror Symmetry}, hep-th/0005247.
}

\lref\rvafa{C.~Vafa, {\it Extending Mirror Conjecture to Calabi-Yau with
            Bundles}, hep-th/9804131.
}

\lref\rkklma{S.~Kachru, S.~Katz, A.~Lawrence and J.~McGreevy, {\it
      Open String Instantons and Superpotentials}, Phys.Rev. {\bf D62} (2000)
      026001, hep-th/9912151.
}

\lref\rkklmb{S.~Kachru, S.~Katz, A.~Lawrence and J.~McGreevy, {\it
             Mirror Symmetry for Open Strings}, hep-th/0006047.
}

\lref\rhklm{S.~Hellermann, S.~Kachru, A.~Lawrence and J.~McGreevy, {\it
               Linear Sigma Models for Open Strings}, hep-th/0109069 .
}

\lref\rkachmca{S.~Kachru and  J.~McGreevy, {\it
            Supersymmetric Three-cycles and (Super)symmetry Breaking},
               Phys.Rev. {\bf D61} (2000) 026001,  hep-th/9908135.
}

\lref\rmayr{P.~Mayr, {\it N=1 Mirror Symmetry and Open/Closed String Duality},
 hep-th/0108229.
}

\lref\rmayrb{P.~Mayr, {\it Summing up Open String Instantons and N=1 String
Amplitudes}, hep-th/0203237.
}

\lref\rgovin{S.~Govindarajan, T.~Jayaraman and T.~Sarkar, {\it
Disc Instantons in Linear Sigma Models},  hep-th/0108234.
}

\lref\riqbal{A.~Iqbal, A.~K.~Kashani-Poor, {\it
    Discrete Symmetries of the Superpotential and Calculation of Disk Invariants},
         hep-th/0109214.
}

\lref\rblum{J.~D.~Blum, {\it Calculation of Nonperturbative Terms in Open
String Models}, hep-th/0112039.
}

\lref\rdouglas{M.~Douglas, {\it D-branes, Categories and N=1 Supersymmetry}, hep-th/0011017.
}

\lref\rrecknagel{A.~Recknagel and V.~Schomerus, {\it D-branes in Gepner
models}, Nucl. Phys. B {\bf 531} (1998) 185, hep-th/9712186.
}

\lref\ralda{G.~Aldazabal and A.M.~Uranga, {\it Tachyon-free Non-supersymmetric 
 Type IIB Orientifolds via Brane-Antibrane Systems}, JHEP {\bf 9910} (1999) 024,
 hep-th/9908072.
}


\lref\rghm{M.~Green, J.~A.~Harvey, G.~Moore, 
{\it I-Brane Inflow and Anomalous Couplings on D-Branes}
Class. Quant. Grav. {\bf 14} (1997) 47, hep-th/9605033.
}

\lref\DouglasBN{
M.~R.~Douglas,
{\it Branes within Branes}, hep-th/9512077.
}


\lref\Jbook{
D.~Joyce,
{\it Compact Manifolds of Special Holonomy},
Oxford University Press, 2000
}

\lref\AtiyahZZ{
M.~Atiyah, J.~M.~Maldacena and C.~Vafa,
{\it An M-theory Flop as a Large N Duality},
J.\ Math.\ Phys.\  {\bf 42} (2001) 3209, hep-th/0011256.
}

\lref\BrandhuberYI{
A.~Brandhuber, J.~Gomis, S.~S.~Gubser and S.~Gukov,
{\it Gauge Theory at Large N and New $G_2$ Holonomy Metrics},
Nucl.\ Phys.\ B {\bf 611} (2001) 179, hep-th/0106034.
}

\lref\GibbonsER{
G.~W.~Gibbons, D.~N.~Page and C.~N.~Pope,
{\it Einstein Metrics On S**3 R**3 And R**4 Bundles},
Commun.\ Math.\ Phys.\  {\bf 127} (1990) 529.
}

\lref\GubserMZ{
S.~S.~Gubser,
{\it TASI lectures: Special Holonomy in String Theory and M-theory},
hep-th/0201114.
}

\lref\CveticZX{
M.~Cvetic, G.~W.~Gibbons, H.~Lu and C.~N.~Pope,
{\it Cohomogeneity One Manifolds of Spin(7) and $G_2$ Holonomy},
Phys.\ Rev.\ D {\bf 65} (2002) 106004, hep-th/0108245.
}

\lref\rbs{
R.~L.~Bryant and S.~Salamon, {\it On the Construction of some Complete
Metrics with Exceptional Holonomy},
Duke Math. J. {\bf 58} (1989) 829.
}

\lref\rberbra{
P.~Berglund and A.~Brandhuber, {\it Matter from $G_2$ Manifolds},
hep-th/0205184.
}

\lref\PapadopoulosDA{
G.~Papadopoulos and P.~K.~Townsend,
{\it Compactification of D = 11 Supergravity on Spaces of Exceptional Holonomy},
Phys.\ Lett.\ B {\bf 357} (1995) 300, hep-th/9506150.
}

\lref\AcharyaGB{
B.~S.~Acharya,
{\it On Realising N = 1 Super Yang-Mills in M theory},
hep-th/0011089.
}

\lref\BrandhuberKQ{
A.~Brandhuber,
{\it $G_2$ Holonomy Spaces from Invariant Three-forms},
Nucl.\ Phys.\ B {\bf 629} (2002) 393, hep-th/0112113.
}

\lref\AtiyahQF{
M.~Atiyah and E.~Witten,
{\it M-theory Dynamics on a Manifold of $G_2$ Holonomy},
hep-th/0107177.
}

\lref\KachruJE{
S.~Kachru and J.~McGreevy,
{\it M-theory on Manifolds of $G_2$ Holonomy and Type IIA Orientifolds},
JHEP {\bf 0106} (2001) 027, hep-th/0103223.
}

\lref\NoyvertMC{
B.~Noyvert,
{\it Unitary Minimal Models of
SW(3/2,3/2,2) Superconformal Algebra and  Manifolds of $G_2$ Holonomy},
JHEP {\bf 0203} (2002) 030, hep-th/0201198.
}

\lref\EguchiIP{
T.~Eguchi and Y.~Sugawara,
{\it String Theory on $G_2$ Manifolds based on Gepner Construction},
Nucl.\ Phys.\ B {\bf 630} (2002) 132, hep-th/0111012.
}

\lref\RoibanCP{
R.~Roiban and J.~Walcher,
{\it Rational Conformal Field Theories with $G_2$ Holonomy},
JHEP {\bf 0112} (2001) 008, hep-th/0110302.
}

\lref\ShatashviliZW{
S.~L.~Shatashvili and C.~Vafa,
{\it Superstrings and Manifolds of Exceptional Holonomy},
hep-th/9407025.
}

\lref\BlumenhagenJB{
R.~Blumenhagen and V.~Braun,
{\it Superconformal Field Theories for Compact $G_2$ Manifolds},
JHEP {\bf 0112} (2001) 006, hep-th/0110232.
}

\lref\rkachmcb{S. Kachru and  J. McGreevy, {\it
             M-theory on Manifolds of $G_2$ Holonomy and Type IIA
             Orientifolds}, JHEP {\bf 0106} (2001) 027, hep-th/0103223.
}

\lref\rwitten{E.~Witten, {\it Anomaly Cancellation On Manifolds Of $G_2$
Holonomy}, hep-th/0108165.
}

\lref\rwa{B.~Acharya and E.~Witten, {\it Chiral Fermions from Manifolds of
$G_2$ Holonomy}, hep-th/0109152.
}

\lref\rgukova{S.~Gukov, C.~Vafa and E.~Witten, {\it CFT's From Calabi-Yau
Four-folds}, Nucl. Phys. B {\bf 584} (2000) 69 [Erratum-ibid. B608 (2001) 477],
hep-th/9906070.
}

\lref\rgukovb{S.~Gukov,  {\it Solitons, Superpotentials and Calibrations},
        Nucl. Phys. B {\bf 574} (2000) 169, hep-th/9911011.
}


\lref\rdka{J.~Distler, S.~Kachru, {\it $(0,2)$ Landau-Ginzburg Theory},
         Nucl. Phys. B {\bf 413} (1994) 213, hep-th/9309110.
}

\lref\rdkb{J.~Distler, S.~Kachru, {\it Singlet Couplings and $(0,2)$ Models},
              Nucl. Phys. B {\bf 430} (1994) 13, hep-th/9406090.
}

\lref\rsw{E.~Silverstein and E.~Witten, {\it Criteria for Conformal Invariance
of $(0,2)$ Models}, Nucl. Phys. B {\bf 444} (1995) 161, hep-th/9503212.
}

\lref\rbw{R.~Blumenhagen and A.~Wisskirchen, {\it
               Exactly Solvable $(0,2)$ Supersymmetric String Vacua With GUT
               Gauge Groups}, Nucl. Phys. B {\bf 454} (1995) 561, hep-th/9506104.
}

\lref\rbsw{R.~Blumenhagen, R.~Schimmrigk, and A.~Wisskirchen, {\it
           The $(0,2)$ Exactly Solvable Structure of Chiral Rings,
             Landau-Ginzburg Theories and Calabi-Yau Manifolds},
                 Nucl. Phys. B {\bf 461} (1996) 460, hep-th/9510055.
}


\lref\rsena{A.~Sen, {\it A Non-perturbative Description of the
Gimon-Polchinski Orientifold}, Nucl. Phys. B {\bf 489} (1997) 139,
hep-th/9611186.
}

\lref\rsenb{A.~Sen, {\it F-theory and the Gimon-Polchinski Orientifold},
   Nucl. Phys. B498 (1997) 135, hep-th/9702061.
}

\lref\BryantTori{
R.~L.~Bryant,
{\it Some Examples of Special Lagrangian Tori},
math.DG/9902076
}

\lref\BanksNJ{
T.~Banks, M.~R.~Douglas and N.~Seiberg,
{\it Probing F-theory with Branes},
Phys.\ Lett.\ B {\bf 387} (1996) 278,
hep-th/9605199.
}

\lref\SenVD{
A.~Sen,
{\it F-theory and Orientifolds},
Nucl.\ Phys.\ B {\bf 475} (1996) 562,
hep-th/9605150.
}

\lref\BlumenhagenUA{
R.~Blumenhagen, B.~K\"ors, D.~L\"ust and T.~Ott,
{\it Hybrid Inflation in Intersecting Brane Worlds},
hep-th/0202124.
}

\lref\VafaWI{
C.~Vafa,
{\it Superstrings and Topological Strings at Large $N$},
J.\ Math.\ Phys.\  {\bf 42} (2001) 2798,
hep-th/0008142.
}

\lref\TaylorII{
T.~R.~Taylor and C.~Vafa,
{\it RR flux on Calabi-Yau and Partial Supersymmetry Breaking},
Phys.\ Lett.\ B {\bf 474} (2000) 130, 
hep-th/9912152.
}

\lref\AganagicGS{
M.~Aganagic and C.~Vafa,
{\it Mirror Symmetry, D-branes and Counting Holomorphic Discs}
hep-th/0012041.
}

\lref\AganagicNX{
M.~Aganagic, A.~Klemm and C.~Vafa,
{\it Disk Instantons, Mirror Symmetry and the Duality Web},
Z.\ Naturforsch.\ A {\bf 57} (2002) 1,
hep-th/0105045.
}

\lref\AcharyaAG{
B.~Acharya, M.~Aganagic, K.~Hori and C.~Vafa,
{\it Orientifolds, Mirror Symmetry and Superpotentials},
hep-th/0202208.
}

\lref\LercheCW{
W.~Lerche and P.~Mayr,
{\it On N = 1 Mirror Symmetry for Open Type II Strings},
hep-th/0111113.
}

\lref\OoguriBV{
H.~Ooguri and C.~Vafa,
{\it Knot Invariants and Topological Strings},
Nucl.\ Phys.\ B {\bf 577} (2000) 419, 
hep-th/9912123.
}

\lref\FerraraUZ{
S.~Ferrara, C.~Kounnas, D.~L\"ust, and F.~Zwirner,
{\it Duality Invariant Partition Functions and
Automorphic Superpotentials for $(2,2)$ String Compactifications},
Nucl.\ Phys.\ B {\bf 365} (1991) 431.
}

\lref\MayrHH{
P.~Mayr,
{\it On Supersymmetry Breaking in String Theory and
its Realization in Brane Worlds},
Nucl.\ Phys.\ B {\bf 593} (2001) 99, 
hep-th/0003198.
}

\lref\CurioAE{
G.~Curio, A.~Klemm, B.~K\"ors and D.~L\"ust,
{\it Fluxes in Heterotic and Type II String Compactifications},
Nucl.\ Phys.\ B {\bf 620} (2002) 237, 
hep-th/0106155.
}

\lref\CurioSC{
G.~Curio, A.~Klemm, D.~L\"ust and S.~Theisen,
{\it On the Vacuum Structure of Type II String
Compactifications on  Calabi-Yau Spaces with H-fluxes},
Nucl.\ Phys.\ B {\bf 609} (2001) 3,
hep-th/0012213.
}

\lref\MoorePN{
G.~W.~Moore,
{\it Arithmetic and Attractors},
hep-th/9807087.
}

\lref\MooreZU{
G.~W.~Moore,
{\it Attractors and Arithmetic},
hep-th/9807056.
}

\lref\BehrndtJN{
K.~Behrndt, G.~Lopes Cardoso, B.~de Wit, R.~Kallosh, D.~L\"ust and T.~Mohaupt,
{\it Classical and Quantum N = 2 Supersymmetric Black Holes},
Nucl.\ Phys.\ B {\bf 488} (1997) 236,
hep-th/9610105.
}

\lref\DenefSV{
F.~Denef,
{\it Attractors at Weak Gravity},
Nucl.\ Phys.\ B {\bf 547} (1999) 201, 
hep-th/9812049.
}

\lref\CurioDZ{
G.~Curio, B.~K\"ors and D.~L\"ust,
{\it Fluxes and Branes in Type II Vacua and
M-theory Geometry with $G_2$ and  $Spin(7)$ Holonomy},
hep-th/0111165.
}


\lref\TatarFlux{
K.~Dasgupta, K.~Oh, J.~Park and R.~Tatar,
{\it Geometric Transition versus Cascading Solution}, 
JHEP\ {\bf 0201} (2002) 031,
hep-th/0110050.
}

\lref\Stefanski{
B.~Stefa\'nski,~jr,
{\it Gravitational Couplings of D-branes and O-planes}, 
Nucl.\ Phys.\ B {\bf 548} (1999) 275,
hep-th/9812088.
}

\lref\refPradisi{
G.~Pradisi,
{\it Type I Vacua from Diagonal $\ZZ_3$-Orbifolds}, 
Nucl.\ Phys.\ B {\bf 575} (2000) 134,
hep-th/9912218.
}

\Title{\vbox{
 \hbox{HU--EP-02/23}
 \hbox{SPIN-02/17}
 \hbox{ITP-UU-02/26}
 \hbox{hep-th/0206038}}}
{\vbox{\vskip-1cm\centerline{Orientifolds of K3 and Calabi-Yau Manifolds}
\vskip 0.3cm
          \centerline{ with Intersecting D-branes}
}}
\centerline{Ralph Blumenhagen{$^1$}, Volker Braun{$^1$},
Boris K\"ors{$^2$}, and Dieter L\"ust{$^1$} }
\bigskip
\centerline{$^1$ {\it Humboldt-Universit\"at zu Berlin, Institut f\"ur
Physik,}}
\centerline{\it Invalidenstrasse 110, 10115 Berlin, Germany}
\centerline{\tt e-mail:
blumenha, braun, luest@physik.hu-berlin.de}
\medskip
\centerline{$^2$ {\it Spinoza Institute, Utrecht University,}}
\centerline{\it Utrecht, The Netherlands}
\centerline{\tt email: kors@phys.uu.nl}
\bigskip

\centerline{\bf Abstract}
\noindent
We investigate orientifolds of type II string
theory on K3 and Calabi-Yau 3-folds with intersecting D-branes
wrapping special Lagrangian cycles. We determine quite
generically the chiral massless spectrum in terms of
topological invariants and discuss both
orbifold examples  and algebraic realizations in detail.
Intriguingly, the developed techniques provide an elegant way
to figure out the chiral sector of orientifold models
without computing any explicit string partition function.
As a new example we derive a non-supersymmetric Standard-like Model
from an orientifold of type IIA on the quintic Calabi-Yau 3-fold
with wrapped D6-branes.
In the case of supersymmetric intersecting brane models
 on Calabi-Yau manifolds we discuss
the D-term and F-term potentials, the effective gauge couplings
and the Green-Schwarz mechanism.
The mirror symmetric formulation of this construction is provided
within type IIB theory.
We finally include a short discussion about the lift of these
models from type IIB on K3 to F-theory and from type IIA on
Calabi-Yau 3-folds to M-theory on $G_2$ manifolds.


\Date{06/2002}
\newsec{Introduction}

One of the ultimate goals of string theory is to
provide a model from which one might derive the real low energy
physics as its effective theory.
Of course, we are still far away from a solution
to this problem, but recently new models
have been devised that qualitatively come
fairly close to the Standard Model in many respects.
The main new ingredient in these models is that
they contain intersecting D-branes and open strings
in a consistent manner, which
provide simple mechanisms to generate chiral fermions
and  to break supersymmetry
\refs{\rbgklnon\raads\rbgklmag\ras\rafiru\rafiruph\rbkl\rimr\rbonna\rrab
\rott\rcvetica\rcveticb\BailinIE\IbanezDJ\rottb\HoneckerDJ\rbonnb\rqsusy\berlin
\BlumenhagenUA\rqsusyb\rkokoa\belrab\rcim\rkokob\AldazabalPY\rcls-\KleinVU}.
They were therefore called intersecting brane worlds.
The main focus has so far been put on orientifolds
for which the
background geometry is realized as a torus or a toroidal
orbifold.
This restriction we want to remove in the present paper,
and generalize
the concept of intersecting brane worlds to generic
background geometries of K3 or Calabi-Yau 3-folds.
We construct six- and four-dimensional orientifold vacua of type II
string theories with D-branes wrapping cycles of middle dimension in the
internal compact space. The patterns of their intersections govern the
properties of the effective theory, such as its chiral fermion spectrum,
the breaking of supersymmetry and the scalar potential generated for
the moduli fields.

In brane world models one usually considers D-branes wrapping part
of the internal compactification space and filling out the
non-compact space-time. It is clear that a single D-brane wrapping
a cycle in some compact Calabi-Yau manifold is not a consistent
string background of such a type, as the total Ramond-Ramond
(RR)-charge has to vanish. One way to remedy the situation is to
consider non-compact Calabi-Yau manifolds allowing the RR-flux to
escape to infinity. This option is only applicable to local models
of gauge theories, as the gravitational excitations on the
internal space will in general not decouple from the effective
four-dimensional physics.
Another option within type II string theory is to use D-branes
together with their anti-branes, which will allow for
configurations with vanishing RR-charge, however at the price of
breaking supersymmetry \ralda. Since for a supersymmetric model not only
the RR-charges  must add up to zero but also the overall tension
of the branes must vanish, one must introduce objects with negative
tension to compensate for the branes. Such objects are provided by
orientifold planes, so that the natural arena for compact models
are orientifolds of type II string theory on Calabi-Yau manifolds.

Apparently, in such models we have to deal with the description
of D-branes wrapping internal cycles of the K3  respectively the
Calabi-Yau manifold. Independent from more phenomenological
considerations,
the physics of D-branes wrapped on supersymmetric cycles in some
Calabi-Yau manifold was studied in detail during
the last few years
\refs{\rbdlr\KasteID\rkklma\rhiv\rkklmb\LercheJB\rav\rakv\rmayr\rgovin
\rhklm\riqbal\LercheCW\rblum-\raahv}.
As a very striking result the notion of mirror symmetry
has been shown to generalize to such open string models,
in simple cases allowing to compute the corrections to
the ${\cal N}=1$ superpotential due to world-sheet disc
instantons by utilizing the
perturbative tree level result of the mirror configuration.
In principle these techniques should find a natural
application in the far more complicated
models we are going to discuss in this paper.
Thus, in generalizing the orientifold construction from the singular
orbifold backgrounds to generic Calabi-Yau spaces, we hope to
provide opportunities for contact between the two fields of
study in the future.

The rather simple toroidal and orbifold
intersecting brane worlds models have been of interest during the
last two years for their ability to produce phenomenologically
appealing string vacua in a simple way.
They allow for a bottom-up approach to search for both
supersymmetric and non-supersymmetric stringy realizations of the
Standard Model or grand unified extensions thereof
\refs{\rimr,\rott,\rcvetica,\rcveticb,\rcim}. One starts
with a conventional orientifold model, defined by taking the
quotient of a type II string theory by a group of symmetries of
the background and the world sheet parity $\Omega$. For simplicity, one can
restrict to the case that the orientifold planes just wrap middle dimensional
homological cycles of the underlying manifold. It is then possible
to introduce generic intersecting D-branes wrapping middle class
homological cycles, as well, in a way that the total RR-charge
cancels among the two. At the intersection of any such
D-branes chiral fermions are localized \rangles. It was shown that one can
actually maintain supersymmetry in particular models of
this kind \refs{\raads,\rbkl,\rcvetica,\rcveticb},
but generically it will be broken. If the
supersymmetric models are perturbed, a scalar potential will be
generated in the form of an F- or D-term and tachyons can arise. The latter
is not automatic and can also be avoided. The scalar potential then freezes
some of the K\"ahler and complex structure moduli. This phenomenon of moduli
stabilization can, of course, also be observed in
non-supersymmetric compactifications \refs{\rott,\berlin,\belrab}.

It will turn out that a major fraction of the usual
orientifold models studied extensively in six and four uncompactified
space-time dimensions can be considered as a small
subclass of the orientifolds with intersecting branes.
The standard compactifications
considered in the literature are recovered from the
more general class of intersecting brane models
by placing the D-branes parallel to the orientifold planes.

To appreciate some
of the results we will derive in this paper let us shortly review,
how the CFT construction of an orientifold model
on a toroidal orbifold background
proceeds technically.\footnote{$^1$}{For a recent and exhaustive review
on open strings and orientifold constructions see \AngelantonjCT.}
One considers a quotient of a type II string theory on a four- or
six-dimensional torus by a finite group which not only contains
discrete symmetries but also the world-sheet
parity transformation $\Omega$. In order to derive the RR-charge
cancellation conditions one needs to compute the Klein bottle
amplitude explicitly in order to extract the infra-red divergences in the
tree-channel. They are due to the exchange of
massless closed string modes, which couple to the RR charge of
the orientifold planes. Next one has to add appropriate open
string sectors, D-branes, into the theory to cancel the divergences
via the interference of the annulus and the M\"obius strip amplitudes
with the Klein bottle. This
means that the RR-charges of the orientifold planes and
the D-branes add up to zero. Due to the fact that the coupling of the
massless RR closed string modes to the orientifold planes and D-branes is of a
topological nature, the conditions that derive from the cancellation
requirement only restrict some discrete parameters of the model,
like the number of D-branes. 
But usually, some freedom remains in the construction, for instance in the
action of the discrete symmetries on the Chan-Paton factors.
Utilizing this information and fixing the arbitrariness by some choice
one is finally able to compute the massless spectrum.
It is not obvious that the resulting spectrum is given in terms of topological
invariants of the D-brane
geometry, as it involves a projection in the Chan-Paton gauge bundle 
on their world volume and the latter  seems to depend on the details of the model.

However, from other string models we are used to the fact that the
spectrum of chiral fermions is determined entirely  by the
topology of the compact manifold. For instance, the number of
generations in compactifications of the heterotic string on a
Calabi-Yau manifold is determined by its Euler characteristic.
In the same way, when toroidal intersecting brane models of this type
were first considered in \rbgklnon, it was shown by explicit world
sheet CFT computations that the chiral massless spectrum is determined
by the topological intersection numbers of the homological cycles the D-branes
are wrapping on. Since this relation between chiral fermions
and topology is manifest in the anomalies, we expect a similar
result to hold in any case. More concretely, one
expects that for orientifold models on smooth K3 and Calabi-Yau
3-folds the chiral fermion spectrum should be given by
topological invariants of the compactification space, of the
submanifolds wrapped by the
orientifold planes and the D-branes, and of the gauge bundles on the
respective D-branes. By restricting the branes and planes to preserve
supersymmetry in the weak sense, that is any single D-brane preserves some
supersymmetry, but not all branes and planes necessarily the same,  one can
simplify the situation considerably
(see also \refs{\rqsusy,\rqsusyb}). The branes then have to be wrapped
on certain calibrated cycles and the gauge bundles on their world
volume need to be flat. This restriction is natural in the orientifold
models which we shall be considering, and will only be
relaxed in the non-supersymmetric Standard Model on the quintic in
section (7.6). In particular,
it allows to write explicit formulas for the chiral fermion spectra.
We will see that the reason, why this has
not been apparent in most of the orientifold models studied so far,
is, that the actual computation has been performed at a singular
point in moduli space, where certain cycles have shrunken to zero
size.

Concretely, the spectrum of
chiral fermions is always given by the topological intersection
numbers of the cycles the D-branes are wrapped on.
We consider it the main result of this paper to make this
explicit and test the resulting spectra for their consistency
with a large number of examples. In order to
make a comparison to singular orbifold models one first has to
perform a blow-up of their singularities and only then
compute the intersection numbers on the resulting smooth manifold.
We will show that the computations for the
orientifold on the blown-up K3 or on Calabi-Yau 3-folds agree
completely with the chiral spectrum determined at the orbifold
point by the lengthy procedure outlined above.
For K3 these examples include T-dual versions of the model
commonly known as the Gimon-Polchinski model \rgimpol\ (even
though it was first constructed by Bianchi and Sagnotti
\sagn) and those studied by Gimon-Johnson \rgimjo.
Thus, the developments of this paper can be considered technically
as an alternative route towards the determination of the phenomenologically
important chiral massless spectrum avoiding the
computation of explicit string partition functions.

This paper is organized as follows. In section 2 we introduce the
general set-up of intersecting brane worlds on orientifolds of
smooth K3 and Calabi-Yau 3-folds. In section 3 we discuss K3
compactifications and present the generic result
for the chiral massless spectrum expressed in terms of topological
intersection numbers of 2-cycles in the K3. Anomaly cancellation
in six dimensions yields a non-trivial constraint on the
self-intersection number of the orientifold O7-planes for which we
also give a mathematical proof. In section 4 we revisit a number
of six-dimensional orientifold models studied so far in the literature and show
that the chiral spectra computed by utilizing our novel prescription
for the blown-up orbifolds agree with
the spectra obtained at the orbifold point. The discussed examples
include the original orientifold models of Gimon-Johnson, the
orientifolds on orbifolds of K3 studied in \refs{\rbgka,\rba,\refPradisi,\rbgkb} and a model on
the quartic. We study both non-supersymmetric and supersymmetric
models. For specific supersymmetric orbifold models we also discuss
the lift to F-theory on an elliptically fibered Calabi-Yau threefold.
Finally, we show that the compactification of F-theory on Voisin-Borcea
Calabi-Yau 3-folds is consistent with the proposed chiral spectrum,
as well. In section 5 we
apply our methods to orientifolds on smooth
Calabi-Yau 3-folds. Again the chiral matter content is given by
the homological intersection numbers of the D6-branes respectively
O6-planes. Section 6 is devoted to a detailed discussion of the
generalized Green-Schwarz mechanism in these models. In section 7
we present some blown-up orbifold examples and construct
on the Quintic Calabi-Yau manifold a non-supersymmetric Standard like Model.
We continue in section 8 with the
application of known results on the physics of
D-branes wrapping special Lagrangian cycles in Calabi-Yau
manifolds to supersymmetric intersecting brane worlds. This
includes the generation of F-term and D-term potentials as well as
a discussion of the gauge couplings. An equivalent mirror
symmetric formulation for these set-ups is discussed in section 9
and finally in section 10 we comment on the lift of these
configurations to M-theory and the geometric interpretation of
their phase transitions. Finally, section 11 contains our conclusions.

\vskip1cm

\newsec{Intersecting brane worlds}
\seclab\sIBWintro

The models we are going to study all start from
a supersymmetric type II compactification, either of type IIB on a
K3 surface or of type IIA on a Calabi-Yau 3-fold, which admits an
involution $\o\sigma$ of order 2. For the most part of the paper, we
will want it to be the complex conjugation,
\eqn\sigmadef{
\o\sigma : z_i \mapsto \bar{z}_i,\ i=1,\, ...\, ,d,
}
in local coordinates. It is anti-holomorphic in the sense that it
takes the holomorphic $d$-form to its conjugate. This geometric
involution is then combined with the world sheet parity $\Omega$.
Dividing out by $\Omega\o\sigma$ leads to the orientifold
backgrounds
\eqn\spacetime{
{\cal X} = \IR^{9-2d,1} \times {{\cal M}^{2d} \over \Omega \o\sigma}
}
we are going to study. Of course, these models have some relations
to type I string compactifications, which we are also going to use on
occasion. Our examples for ${\cal M}^{2d}$ will involve K3 and
Calabi-Yau orbifolds as well as algebraic models such as the
quartic in $\IC\IP^3$ and the quintic in $\IC\IP^4$.

As is well known, the fixed locus Fix$(\o\sigma)$ defines the location of
orientifold O$q$-planes of dimension $q+1=10-d$, which takes values
$q+1=8,7$ in the cases at hand.
They extend into $d$ internal directions and fill out the space-time.
Just by the definition of $\o\sigma$, Fix$(\o\sigma)$ is a special
Lagrangian (sLag) submanifold of the internal space. Define locally the
holomorphic $d$-form $\Omega_d$ and the K\"ahler form $J$ by
\eqn\nform{
\Omega_d = dz_1 \wedge \, ... \, \wedge dz_d , \quad
J = i \sum_{i=1}^d{ dz_i \wedge d\bar{z}_i} .
} From $\o\sigma( \Omega_d ) = \o{\Omega}_d$ and $\o\sigma( J ) = -J$ it
follows that
\eqn\slagone{
\Im(\Omega_d)\vert_{{\rm Fix}(\o\sigma)} =0  ,\quad
J\vert_{{\rm Fix}(\o\sigma)} =0  .
}
Furthermore the holomorphic $d$-form $\Omega_d$ satisfies
\eqn\holos{
-i^d \int_{\cal M} \Omega_d\wedge\o\Omega_d={\rm Vol}({\cal M}^{2d})
.
}
and this then determines
\eqn\slagtwo{
\Re(\Omega_d)\vert_{{\rm Fix}(\o\sigma)} = d{\rm vol}\vert_{{\rm Fix}(\o\sigma)} ,
}
which means, Fix$(\o\sigma)$ is calibrated with respect to $\Re(\Omega_d)$.
\noindent
For future reference we also introduce the normalized holomorphic
$d$-form
\eqn\normom{
\widehat \Omega_d={1\over \sqrt{{\rm Vol}({\cal M}^{2d})}}\Omega_d .
}

\subsec{Tadpole conditions and the scalar potential}

In the Klein bottle amplitude, which is, of course, only
computable at very special points in the moduli space where a CFT
description is available, the orientifold planes introduce IR
divergences which have to be canceled by similar contributions
from D$q$-branes. The  RR tadpole cancellation condition can be
deduced on very general grounds from the Chern-Simons action for
the D$q$-branes \refs{\DouglasBN\rghm\rscrucca\rscruccab-\Stefanski}
\eqn\bornina{
{\cal S}^{({\rm D}q)}_{\rm CS} =
\mu_q \int_{{\rm D}q}
{\rm ch}({\cal F})\wedge  \sqrt{{\hat {\cal A}({\cal R}_T) \over \hat {\cal A}({\cal R}_N)
   }} \wedge \sum_p{C_p} ,
}
and the orientifold planes
\eqn\bornori{
{\cal S}^{({\rm O}q)}_{\rm CS} =
Q_q \mu_q \int_{{\rm O}q}
    \sqrt{{\hat {\cal L}({\cal R}_T/4) \over \hat{\cal L}({\cal R}_N/4)
   }} \wedge \sum_p{C_p} .
}
Here the RR-charge quantum is given by
\eqn\chargeq{ \mu_q={1\over (2\pi )^q} (\alpha')^{-{q+1\over 2}} } and
the sum is over all RR-forms $C_p$ in the respective
theory.
The relative charge of the orientifold planes is given by
$Q_q=-2^{q-4}$. The Chern character, Dirac genus and the
Hirzebruch polynomial are defined via
\eqn\chern{\eqalign{
{\rm ch}({\cal F})&={\rm Tr}\, e^{i{\cal F}/2\pi}, \cr
\hat{\cal A}({\cal R})&=1-{p_1({\cal R})\over 24}
+\, \cdots =
1+\hat{\cal A}_4({\cal R})
+\, \cdots \, , \cr
\hat{\cal L}({\cal R}/4)&=1+{p_1({\cal R})\over 48} +\cdots = 1+\hat{\cal L}_4({\cal R}/4)+
 \, \cdots \, . \cr} }
Moreover, ${\cal R}_T$ and ${\cal R}_N$ denote the pullbacks of
the curvature two-forms of the tangent and normal bundle on the
brane. The physical gauge fields and curvatures are related to the
skew-hermitian ones in \bornina\ by ${\cal F}=-4i\pi^2\alpha' F$
and ${\cal R}=-4i\pi^2\alpha' R$. In all the models we shall be
considering, the D-branes will be wrapping on calibrated sLag
cycles of real  dimension 2 or 3. Since the supersymmetry
condition on such a sLag cycles states that the gauge bundle has
to be flat, we get the simplification ${\rm ch}({\cal
F})\vert_{{\rm D}q} ={\rm ch}_0({\cal F})\vert_{{\rm D}q}={\rm
rk}({\cal F})$. Moreover, the pull-back of the Pontryagin class
$p_1({\cal R})$ onto these lower dimensional branes reduces to
$p_1({\cal R})\vert_{{\rm D}q}=0$. The only contribution in the
CS-term \bornina\ now comes from $C_{q+1}$.

Let us denote the homology class of the fixed point set of the
orientifold projection by $\pi_{{\rm O}q} = [{\rm Fix}(\o\sigma)]
\in H_d({\cal M}^{2d})$. Similarly we denote the homology class a D$q_a$-brane
is wrapping  around by $\pi_a$. In
general, such a $\pi_a$ is not invariant under the $\Omega\o\sigma$
projection but is mapped to a different cycle $\pi'_a$. Therefore,
one is forced to introduce a D-brane wrapped on that cycle, too.
The part of the supergravity Lagrangian where the RR-field $C_{q+1}$
appears reads
\eqn\sugra{\eqalign{
{\cal S}=
-{1\over 4\kappa^2} \int_{\cal X} & { dC_{q+1} \wedge \star dC_{q+1} } +\mu_q
\sum_a{ N_a \int_{\IR^{9-2d,1}\times\pi_a} C_{q+1} } \cr
&+\mu_q \sum_a{ N_a \int_{\IR^{9-2d,1}\times\pi'_a} C_{q+1} }
+\mu_q Q_{q} \int_{ \IR^{9-2d,1}\times\pi_{{\rm O}q}  } C_{q+1},
}}
where the ten-dimensional gravitational coupling is
$\kappa^2={1\over 2} (2\pi)^7(\alpha')^4$. The resulting equation
of motion for the RR field strength $G_{q+2}=dC_{q+1}$ is
\eqn\equaasa{ {1\over \kappa^2}\,
     d\star G_{q+2}=\mu_q\sum_a N_a\, \delta(\pi_a)+
                    \mu_q\sum_a N_a\, \delta(\pi'_a)
        + \mu_q Q_q\,  \delta(\pi_{{\rm O}q}),}
where $\delta(\pi_a)$ denotes the Poincar\'e dual form of $\pi_a$.
Since the left hand side in
\equaasa\ is exact, the RR-tadpole cancellation
condition becomes in homology
\eqn\tadhom{
\sum_a  N_a\, (\pi_a + \pi'_a)
          +Q_q\,   \pi_{Oq}=0.
}
For D-branes not lying directly on top of the orientifold plane,
any stack of $N_a$ branes will then support one factor of the total gauge
group
\eqn\gauge{
G=\prod_{a} U(N_a) .
}
Only if a stack of D-branes is located within the fixed locus of
$\o\sigma$, also $SO(N_a)$ or $Sp(N_a)$ gauge groups can occur.
Since this case is less generic, we will restrict our attention to
unitary gauge groups for the most part of this paper. Compared to
type I string theory with a gauge groups $SO(32)$ in ten
dimensions \gauge\ can be interpreted as a particular pattern of
gauge symmetry breaking which involves a reduction of the rank of
the gauge group, as well. In principle, the tadpole condition
\tadhom\ involves as many linear relations as there are
independent cycles in $H_d({\cal M}^{2d},\IR)$. But, of course,
the action of $\o\sigma$ on ${\cal M}^{2d}$ also induces an action
$[\o\sigma]$ on the homology and cohomology. In particular,
$[\o\sigma]$ swaps $H^{r,s}$ and $H^{s,r}$, and therefore for
3-folds the number of conditions is halved in any case.

Similarly one can generally determine the disc level NSNS tadpoles.
These result from integrating the Dirac-Born-Infeld effective
action \refs{\FradkinQD,\LeighJQ}
\eqn\bornin{
{\cal S}_{\rm DBI}=-T_q\int_{{\cal M}_{q+1}}d^{q+1}x~e^{-\phi_{10}}
\sqrt{g_{q+1}}}
of the D$q$-branes with the tension given by $T_q=\mu_q$. The action
\bornin\  is proportional to the
volume of the D$q$-branes respectively the O$q$-plane, so that the
disc level scalar potential reads
\eqn\susy{
{\cal V}=T_q\, {e^{-\phi_{10-2d}}
\over \sqrt{{\rm Vol({\cal M}^{2d})}}}
               \left( \sum_a  N_a \left( {\rm Vol}({\rm D}q_a) + {\rm Vol}({\rm D}q'_a) \right) +
          Q_q\,   {\rm Vol}({\rm O}q)\right). }
Generically, this scalar potential is non-vanishing reflecting the
fact that intersecting branes in general break supersymmetry.
Non-supersymmetric compactifications can still be phenomenologically
relevant in the context of large transverse volume
compactifications \refs{\radd,\raadd}.
Note, that obstructions to these scenarios met with toroidal models \rbgklnon\
are no longer present in the more general backgrounds
we are going to consider. But since we
are eventually interested in supersymmetric intersecting brane
worlds, let us specialize to D-branes wrapping sLag cycles of K3
respectively Calabi-Yau 3-folds for the remainder of this section.
These can in principle be calibrated with respect to different
$d$-forms $\Re(e^{i\theta}\Omega_d)$, differing by a constant
phase factor, which means that each of them preserves
supersymmetry, but not necessarily all of them the same.
Effectively, supersymmetry can be broken completely.
For D-branes wrapping sLag cycles the scalar potential
\susy\ only depends on the complex structure moduli, as
in this case the scalar potential gets simplified to
\eqn\dbi{
{\cal V}=T_q\, e^{-\phi_{10-2d}} \left(
\sum_a{N_a \left| \int_{\pi_a} \widehat\Omega_d \right|} +
 \sum_a{N_a \left| \int_{\pi'_a} \widehat\Omega_d \right|}+
Q_q \left| \int_{\pi_{{\rm O}q}} \widehat\Omega_d \right |\right) .
}
Since $\widehat\Omega_d$ is closed, the integrals only depend on
the homology class of the world volume of the branes and planes.
Supersymmetric models in this context are then defined by
requiring that any single D$q_a$-brane conserves the same
supersymmetries as the orientifold plane. In other words, the
cycles the D$q$-branes wrap on must be calibrated with respect to
the same calibration form $\Re(\widehat\Omega_d)$ as the
O$q$-planes. In this particular case we can write \dbi\ as
\eqn\dbic{
{\cal V}=T_q\,  e^{-\phi_{10-2d}} \left(
\sum_a{N_a \int_{\pi_a+\pi_a'} \Re(\widehat\Omega_d )} +
Q_q  \int_{\pi_{{\rm O}q}} \Re(\widehat\Omega_d)\right)  .
}
Due to the RR-tadpole cancellation condition
\tadhom\ this vanishes, so that the scalar potential is zero,
as expected in the supersymmetric situation .

\subsec{Spectra of massless bulk modes}

It is instructive to study the action of $\Omega\o\sigma$ on the cohomology
in a bit more detail. While the fixed locus Fix$(\o\sigma) \subset
{\cal M}^{2d}$ may well be
empty, the set of $\o\sigma$ invariant d-cycles ${\rm Fix}([\o\sigma])
\subset H^d({\cal M}^{2d},\IR)$, never is, just as $\Re(\Omega_d)$
defines an invariant class in $H^{d}$. Of course, we always have
$[{\rm Fix}(\o\sigma)] \in {\rm Fix}([\o\sigma])$. The importance of
$[\o\sigma]$ lies in the fact that it already determines completely
the low energy spectrum of gravity fields. They arise from the
dimensional reduction of the respective type II string theory upon
expanding ten-dimensional fields in terms of harmonic forms on
${\cal M}^{2d}$.

For compactifications of type IIB string theory on K3 the
cancellation of the six-dimensional gravitational anomalies links
the closed and open string spectra. The rules are fairly simple:
Any harmonic $(1,1)$-form $\omega_i,\ i=1,\, ...\, , h^{(1,1)}$,
gives rise to a six-dimensional ${\cal N}=(0,1)$ hypermultiplet
with four scalars from the reduction of the metric and the RR
2-form in addition to  a tensormultiplet  containing  the
self-dual six-dimensional tensors that descend from the self-dual
RR 4-form and a scalar from the reduction of the NSNS 2-form.
Under the world-sheet parity transformation only the $(1,1)$ forms
are anti-invariant $\omega_i\mapsto -\omega_i$ implying the
following projections for massless closed string modes in the
$\Omega\o\sigma$ orientifold. If under $\o\sigma$ the 2-form
$\omega_i$ is anti-invariant we get a massless hypermultiplet and,
if the 2-form $\omega_i$ is invariant, a massless tensor-multiplet
survives the projection.
Since one tensor is part of the supergravity multiplet,
in summary we can write
\eqn\kthreespec{
n_{\rm H} + n_{\rm T} = 21 , \quad n_{\rm T} =
{\rm dim}\, \left( {\rm Fix}([\o\sigma]) \cap H^{(1,1)}({\cal M}^{2d}) \right) .
}
The extra tensormultiplets enter the anomaly
cancellation condition for the irreducible $R^4$ term and need to
be balanced by chiral matter in the open string spectrum. We
postpone the concrete computation of anomalies  until we
specialize to particular dimensions in the following sections.
Very generally, it is determined by the number of intersections of
the D-branes on the internal space. Roughly speaking, any single
intersection point of two branes supports a single chiral fermion
in the effective theory, which transforms as a bifundamental under
the two respective gauge groups on the branes. For flat branes
this is confirmed by the quantization of open strings with the
appropriate boundary conditions. In a non-trivial background
one can employ geometrical methods at large volume, where the
classical geometry is valid. Then, chiral fermions are identified
as zero-modes of the Dirac operator. By the Atiyah-Singer index
theorem, these are related to topological invariants of the
background manifold and the gauge bundle on top.

For Calabi-Yau threefolds the story is very similar. The bulk
supersymmetry is reduced by the $\Omega\o\sigma$ projection from
${\cal N}=2$ to ${\cal N}=1$. This implies that all bulk ${\cal
N}=2$ superfields are truncated to ${\cal N}=1$ superfields by the
$\Omega\o\sigma$ projection. Before the projection there were
$h^{(1,1)}$ abelian vector multiplets and $h^{(2,1)}$
hypermultiplets. On the one hand the $h^{(1,1)}$ vector multiplets
consist of 1 scalar coming from the  dimensional reduction of the
gravity field (the K\"ahler modulus), another scalar resulting
from the reduction of the NSNS 2-form and a four-dimensional
vector from the reduction of the RR 3-form along the 2-cycle.
Similar to the K3 case, if the $(1,1)$ form is invariant under
$\Omega\o\sigma$ an ${\cal N}=1$ chiral multiplet survives the
projection and if it is anti-invariant we get an ${\cal N}=1$
vector multiplet. Note, that the surviving chiral multiplets
still contain the complexified K\"ahler moduli.

On the other hand the four scalars of the
$h^{(2,1)}$ hypermultiplets contain 2 scalars from the
ten-dimensional gravity field (the complex structure moduli)
equipped with 2 scalars arising from the dimensional reduction of
the RR 3-form along the two associated 3-cycles referring to
$H^{2,1}({\cal M}^3)$ and $H^{1,2}({\cal M}^3)$. Under the
$\Omega\o\sigma$ projection one of the two components of the complex
structure is divided out and moreover one linear combination of
the RR scalars survives, so that the former quaternionic
complex structure moduli
space gets reduced  to a complex moduli space of dimension $h^{(2,1)}$.

\subsec{Orbifold models}

Something  more can be said in the cases where ${\cal M}^{2d}$ is
a toroidal orbifold of K3 or a Calabi-Yau 3-fold. First of all,
Fix$(\o\sigma)$ can be determined rather generally. Let us restrict
to orbifold groups $\ZZ_N = \{ \Theta, \Theta^2, \, ... \, , 1
\}$, that act crystallographically on $T^{2d}$ which we assume to
be a direct product of two-dimensional tori  $T^2_I,\ I=1,\, ...
\, , d$. In chapter 7.2, we shall actually also discuss the orbifold
$T^6/\ZZ_2\times\ZZ_2$, but the generalizations required are
fairly obvious. When choosing local coordinates $z_I$, the
factorization implies a diagonal complex structure $\tau^{IJ}$,
\eqn\coord{
dz_I = dx_I + \sum_{J=1}^d \tau^{IJ} dy_J = dx_I + \tau^I dy_I , \quad
z_I \equiv z_I + 1, \quad z_I \equiv z_I + \tau^I .
}
These coordinates diagonalize the orbifold action by
\eqn\orb{
\Theta z_I = e^{2\pi iv_I } z_I, \quad \sum_{I=1}^d v_I = 0 ,
}
and $\o\sigma$ reflects $y_I$.
The orbifold action preserves the supersymmetry generators that
satisfy
\eqn\susyorb{
\Theta \epsilon = \pm \epsilon .
}
The entire orientifold group is now generated by $\Omega\o\sigma$ and $\Theta$.
By using
\eqn\commutator{
\Theta^{1/2} \o\sigma \Theta^{-1/2} = \Theta \o\sigma
}
it is evident that the fixed locus of $\o\sigma$ on the quotient
space can be understood as the orbit of its fixed locus ${\rm
Fix}(\o\sigma)\vert_{T^{2d}}$ on the tori, the product of the $d$
real circles, in addition to  the orbit of
Fix$(\Theta\o\sigma)\vert_{T^{2d}}$. Altogether we can write
\eqn\fixorb{
{\rm Fix}(\o\sigma) =
\bigcup_{i=0}^{N-1}{
\Theta^i \left( {\rm Fix}(\o\sigma)\vert_{T^{2d}} \right) \cup
\Theta^i \left( {\rm Fix}(\Theta\o\sigma)\vert_{T^{2d}} \right) }.
}
Since $\Theta^{1/2}\in\ZZ_N$ if and only if $N \in 2\ZZ +1$, the
two orbits are identical in these cases.

The condition that a pair of two flat D$q$-branes preserves any
supersymmetry is deduced from the Killing spinor equations \rangles
\eqn\killing{
\Gamma_{0\cdots q} \epsilon = \pm \tilde\epsilon, \quad
\Xi\, \Gamma_{0\cdots q}\, \Xi^{-1} \epsilon = \pm \tilde\epsilon,
}
which imply
\eqn\killb{
\Xi^2 \epsilon = \pm \epsilon .
}
Here, $\Xi$ denotes the relative rotation of the two branes. It is
evident, that D$q$-branes in an orbifold trivially preserve
supersymmetry as long as they are related by rotations $\Xi^2 =
\Theta$, but \susyorb\ and \killb\ have more interesting common
solutions. For the relative angles $\varphi_I$ of any one of the
two branes with respect to the orientifold plane the condition
\killb\ requires
\eqn\angletheta{
\sum_{I=1}^d{ \delta_I \varphi_I } = \theta ,
}
with some fixed angle $\theta$ and arbitrary signs $\delta_I=\pm
1$. If we further demand that they preserve the same supersymmetry
as the orientifold O$q$-plane, we have to impose
\eqn\angle{
\sum_{I=1}^d{ \varphi_I } = 0 .
}
This distinction corresponds to branes wrapping general sLag
cycles calibrated by $\Re(e^{i\theta}\Omega_d)$, respectively
branes calibrated by $\Re(\Omega_d)$. The relative signs $\delta_I
=1$ of the $d$ angles in \angle\ are dictated by those of the
$v_I$ in
\orb\ via \susyorb.

Furthermore, the action of $[\o\sigma]$ on the cohomology classes of degree $d$
can be classified in orbifold models. For the case of K3 it is evident that
$\o\sigma$ leaves only the two classes represented by
\eqn\kthreeinv{
dz_1 \wedge dz_2 + d\bar{z}_1 \wedge d\bar{z}_2 , \quad
dz_1 \wedge d\bar{z}_2 + d\bar{z}_1 \wedge dz_2
}
invariant among those classes of 2-forms, which descend from the
torus. Only the first one, in fact $\Omega_2 + \o{\Omega}_2$, is
also invariant under $\Theta$  with the exception of the case
$N=2$, where both 2-forms are invariant. Cycles that are
calibrated with respect to this second $\o\sigma$-invariant form
come with a relative sign switch in the angle criterion \angle\
and preserve supersymmetries with the opposite chirality as the
orientifold planes. As mentioned already, for the case of a
3-fold, the Dolbeault groups $H^{1,2}$ and $H^{2,1}$ get exchanged
by $\o\sigma$ such that one half of the linear combinations are
even. All the invariant combinations of $(2,1)$- and $(1,2)$-forms
are of the form
\eqn\cyinv{
dz_I \wedge dz_J \wedge d\bar{z}_K +
d\bar{z}_I \wedge d\bar{z}_J \wedge dz_K ,
}
which are invariant under $\Theta$ precisely if $2v_K =0\ {\rm mod}\ \ZZ$.
By examining the table of Calabi-Yau-orbifolds that admit
CFT descriptions \refs{\DixonJW,\DixonJC},
one finds examples that satisfy this condition for any even $N$ and for
exactly one $v_K$ in each case. Again, using these additional invariant forms
as calibrations would introduce different signs into \angle.

Blowing up any of the orbifold singularities introduces additional
cycles.
For instance, blowing up an orbifold singularity in a
K3 orbifold refers to gluing in $N-1$ copies of a $\IC\IP^1$. It
may be parameterized by an inhomogeneous coordinate $z_k$ as
$[1,z_k] \cup [0,1]$, using the index $k$ to label the set of such
$\IC\IP^1$. Turning on the blow-up mode means adding a term
proportional to $dz_k \wedge d\bar{z}_k$ to the K\"ahler form $J$.
Now $[\o\sigma]$ acts as $(-1) \otimes p$ on these forms, where the
$(-1)$ is the intrinsic $\Omega$ reflection of $J$ and $p$ denotes
a permutation matrix of order 2. It acts on the labels $k$ and
swaps the fixed points of the orbifold generators as a whole.
Thus, any doublet of $p$ will give rise to one even and one odd
class under $[\o\sigma]$, while any $p$-invariant fixed point will
simply be $\o\sigma$-odd. The action of $p$ is of course the place,
where the complex structure $\tau^I$ starts to play a decisive
role, as the relation $z_I\equiv z_I + \tau^I$ determines, if a
given fixed point of $\Theta$ is invariant under $\o\sigma$ or not.
These data are reflected in the CFT construction of K3 orbifolds
by noting that $\Omega\o\sigma$ does not exchange the sectors
twisted by $\Theta^k$ and $\Theta^{N-k}$ \rbgka.
This is in contrast to the situation in standard type IIB
orientifolds with a pure $\Omega$ projection.

\newsec{Intersecting brane worlds in six dimensions}

In this section we consider intersecting brane worlds on K3 manifolds.
We will provide evidence that the massless spectrum of chiral
fermions in the effective six-dimensional theory is entirely
determined by the topology of the configuration of orientifold
planes and D-branes. This result is supported by the partially
known orbifold limits of K3 and a discussion of its algebraic
realization as a quartic in $\IC\IP^3$. In addition, we show it to be
consistent with the compactification of F-theory on Voisin-Borcea
Calabi-Yau 3-folds, which in certain limits are believed to be
dual descriptions of orientifold vacua.
We also include some considerations on the F-theory lift of the
most simple $T^4/\ZZ_2$ orbifold. The cancellation of the
irreducible gravitational anomaly provides a link between closed
and open string spectra and the topology of the orientifold plane.
This extra condition is shown to hold in all the models.

\subsec{Chiral spectrum, anomalies and tadpoles}

Let us summarize the set-up: We compactify the type IIB string on
a K3 manifold which admits an anti-holomorphic involution
$\o\sigma$. Then, we can perform the orientifold projection by
$\Omega\o\sigma$,
which leads to a model with ${\cal N}=(0,1)$ supersymmetry in six dimensions.
In general the spectrum of fermions is chiral and the massless
closed string spectrum alone does not satisfy the cancellation of
the gravitational anomaly.
The condition for the vanishing of the coefficient of the
irreducible $R^4$ term is
\eqn\anomgrav{
n_{\rm H}-n_{\rm V}+29\,  n_{\rm T}=273 ,
}
stated in terms of the numbers of hyper-, vector- and
tensormultiplets. It puts constraints on the open string spectrum.

The O7-planes are located at the fixed locus Fix$(\o\sigma)$ of
$\o\sigma$ and are charged under some of the RR-fields. In terms of
the definitions \nform\ they wrap on sLag 2-cycles of the K3
manifold. Their RR-charge must be canceled by stacks of D7-branes
wrapping 2-cycles of the K3, as well,
\eqn\tadpole{
\sum_a  N_a\, (\pi_a +\pi'_a)=
          8\,   \pi_{{\rm O}7}.
}
The homology group $H_2({\rm K}3,\ZZ)$ is of dimension $b_2=22$
and comes with the intersection form, so it is a Lorentzian
lattice of signature $(3,19)$. By choosing a proper basis, it can
be put into the standard form
\eqn\lattstand{
H_2({\rm K}3,\ZZ) \cong \Gamma_{1,1}\oplus\Gamma_{1,1}\oplus\Gamma_{1,1}\oplus
\Gamma_{{\rm E}_8}\oplus\Gamma_{{\rm E}_8} ,
}
with $\Gamma_{1,1}$ denoting $2\times 2$ matrices with entries 0 
on the diagonal and 1 off-diagonal, and $\Gamma_{{\rm E}_8}$
the Cartan matrix of E$_8$. We shall find it an important point to normalize
the classes $\pi_{{\rm O}7}$ and $\pi_a$ appropriately, when going
from $H_2({\rm K}3,\IR)$ to $H_2({\rm K}3,\ZZ)$, which physically refers to
the proper Dirac quantization of charges.

The chiral part of the massless spectrum is expected to be determined
by the topology of the configuration due to index theorems.
In the case of intersecting branes the chiral matter is supported
by open strings localized at the intersection points of the
various branes \rangles. There are several cases to be distinguished: A
self-intersection of any D7$_a$-brane wrapping the cycle $\pi_a$
is never invariant under $\o\sigma$ by our convention, as it is
mapped to a self-intersection of the image wrapped on $\pi_a'$.
The excitations at this point therefore do not experience any
projection and their Chan-Paton index is an unconstrained $N_a\times
N_a$ matrix. This gives rise to a chiral fermion in the adjoint
representation of the $U(N_a)$ gauge group. However, an
intersection of a brane with its image can be invariant under
$\o\sigma$, which happens precisely if the intersection
locus lies within Fix$(\o\sigma)$. The fields at this
intersection would then be subject to a projection of the
Chan-Paton labels by $\Omega\o\sigma$.
Summarizing, intersection points which are invariant under $\Omega
\o\sigma$ provide  chiral fermions in the anti-symmetric
representation of $U(N_a)$, whereas $\Omega \o\sigma$ pairs of
intersection points lead to  chiral fermions in the anti-symmetric
and symmetric representation of the gauge group. Finally, open
strings stretched between two different D7-branes always give rise
to chiral fermions in the bifundamental representation of the two
gauge factors. Together one gets the chiral spectrum of table
1 (the subscripts below the gauge representations denote the
transformation properties under the little group $SO(4)\simeq SU(2)\times
SU(2)$ in six dimensions).
\vskip 0.8cm
\vbox{
\centerline{\vbox{
\hbox{\vbox{\offinterlineskip
\def\tablespace{height2pt&\omit&&
 \omit&\cr}
\def\tablerule{\tablespace\noalign{\hrule}\tablespace}

\hrule\halign{&\vrule#&\strut\hskip0.2cm\hfill #\hfill\hskip0.2cm\cr
& Representation  && Multiplicity &\cr
\tablerule
& [{\bf Adj}]$_{(1,2)}$  && $\pi_a\circ \pi_a$   &\cr
\tablerule
& $[{\bf A_a+\o{A}_a}]_{(1,2)}$  && ${1\over 2}\left(\pi_a\circ
\pi'_a+\pi_a\circ \pi_{{\rm O}7}\right)$   &\cr
\tablerule
& $[{\bf S_a+\o{S}_a}]_{(1,2)}$
     && ${1\over 2}\left(\pi_a\circ \pi'_a-\pi_a\circ \pi_{{\rm O}7}\right)$   &\cr
\tablerule
& $[{\bf (N_a,N_b)+(\o N_a,\o N_b)}]_{(1,2)}$  && $\pi_a\circ \pi_{b}$   &\cr
\tablerule
& $[{\bf (N_a,\o N_b)+( \o N_a, N_b)}]_{(1,2)}$
&& $\pi_a\circ \pi'_{b}$   &\cr
}\hrule}}}}
\centerline{
\hbox{{\bf Table 1:}{\it ~~ Chiral spectrum in $d=6$}}}
}
\vskip 0.5cm
\noindent
Here we denote by $I(x,y)=x \circ y$ the intersection number of $x$ and $y$, 
i.e. the product in the homology ring. If the intersection number is
negative, it is understood that one has to take the opposite chirality.
This spectrum will be shown to be completely generic for all
orbifold models we are going to consider, as well as for a purely
toroidal background.
We will also find it consistent with the algebraic model of K3 as
a quartic hypersurface in $\IC\IP^3$ and with F-theory compactifications
on Voisin-Borcea 3-folds.
Further, it is formulated
entirely in terms of topological quantities invariant under
continuous deformations of the moduli. We therefore propose it to
be valid throughout the moduli space of the background K3, beyond
the singular limits one can actually probe explicitly.

As a consistency check we find that all contributions to the irreducible
$F_a^4$ anomaly coefficients cancel automatically by the tadpole cancellation
\tadhom. This is necessary as there are no contributions
to the gauge anomalies from the closed string sector.
For the open string contribution to the gravitational $R^4$ anomaly we get
\eqn\ano{
A_{\rm op}= 14\,  \pi_{{\rm O}7}\circ \pi_{{\rm O}7} ,
}
when counted in $N=(0,1)$ supermultiplets. Intriguingly, the
anomaly in the open string sector only depends on the intersection
number of the ${\rm O}7$-plane, which is a closed string quantity.
On the other hand, a relation of this kind had to be expected as
\ano\ must cancel the contribution $A_{\rm cl}=273-n_{\rm H}-29\,
n_{\rm T}=28(9-n_{\rm T})$ to the anomaly from the closed string
sector. This implies the surprising relation
\eqn\rela{
\pi_{{\rm O}7}\circ \pi_{{\rm O}7}=2(9-n_{\rm T}) =
{1\over 32} \sum_{a,b}{ N_a N_b( \pi_a \circ \pi_b + \pi_a \circ \pi_b')}
}
between the self-intersection number of the ${\rm O}7$-plane,
the number of tensor multiplets and the total intersection of
all the D7-branes. At the end of this section we will prove
the relation \rela\ from pure mathematical reasoning.

At first glance, the statement that the spectrum in table 1 should
hold for each K3 orientifold  in general seems to be quite
surprising. For special orbifold limits of K3 the chiral massless
spectra  were obtained after a quite lengthy computation involving
the determination of the action of the discrete symmetries on the
Chan-Paton factors and the computation of the tadpole cancellation
conditions for the various twisted and untwisted sectors, using
the explicit one-loop partition function. In the upcoming
paragraphs,  we will compute the chiral massless spectrum in a
number of examples by using table 1 and
show that the chiral spectra  completely agree with those
obtained in the orbifold limits.

A particular class of models consists of those preserving
supersymmetry. As was discussed already, supersymmetry implies all
the D7-branes to wrap cycles calibrated with respect to
$\Re(\Omega_2)$, as do the O7-planes. This already implies the
very special form \dbic\ for the potential
and its vanishing via the tadpole cancellation
condition \tadhom. The most simple way to preserve supersymmetry
is to place the D7-branes right on top of the orientifold plane,
i.e. set all $\pi_a=\pi_{{\rm O}7}$. These models with maximal
gauge symmetry have been studied in
\refs{\rgimpol,\rgimjo,\rbluma,\DabholkarKA,\rblumb,\rdabol,\rbgka,\refPradisi}.
At least at the
orbifold points one can still check for supersymmetry, if the
D7-branes are not parallel to the orientifold planes, as here the
metric is given by the metric on the original torus.
Independently, one can explicitly determine the intersection
angles of the D7-branes and investigate whether they satisfy
\angle.
Since there does not exist a superpotential in six-dimensional
field theories with
${\cal N}=1$ supersymmetry, supersymmetry breaking can only occur
via Fayet-Iliopoulos terms. As we have seen, the scalar potential
that arises from the tension of the D7-branes by the reduction of
the DBI effective action only depends on
the complex structure moduli. For instance, blowing-up
the orbifold fixed points should therefore correspond to flat directions
and preserve supersymmetry.

\subsec{Mathematical derivation}

Suppose you have a smooth K3 surface ${\cal M}$ with smooth
antiholomorphic involution $\o\sigma$. The $\o\sigma$--fixed set
$\Sigma={\rm Fix}(\o\sigma)$ is either empty or a smooth embedded
surface (possibly with multiple disconnected components). It is
smooth because $(T_p{\cal M})^\sigma\simeq T_p\Sigma$ varies smoothly for
all $p\in \Sigma$. In section
\sIBWintro{} we already saw that $\Sigma$ is special
Lagrangian. Therefore the complex structure $J:T{\cal M}\to T{\cal M}$ maps
$T\Sigma$ to $N\Sigma$ and vice versa, so we get an isomorphism
\eqn\TSigmaNSigmaIso{ J_p: T_p\Sigma \simeq N_p\Sigma . }
Now by exponentiation we can associate a small deformation of
$\Sigma\subset {\cal M}$ to a section of $N\Sigma$, and therefore to a
section of $T\Sigma$, i.e. vector fields.

The self-intersection of $\Sigma$ is the intersection with some
deformed copy of $\Sigma$. Choose a vector field $\vec{v}\in
\Gamma(T\Sigma)$ with simple zeros then
\eqn\slagintersectionIndex{
  \Sigma \circ \Sigma =
  - \Big\{  \# \hbox{zeros of }\vec{v} \Big\}
  =
  - \chi(\Sigma)
}
by the Poincar\'e Hopf index theorem. Here the zeros of the vector
field of course have to be counted with signs and multiplicities, i.e
by their index. The overall sign in \slagintersectionIndex{} is subtle
but can easily be fixed by a local model. Note that it is not an
independent choice of convention, but rather fixed by sign conventions
for the index in vector fields and the convention for
self-intersections.

Now $\o\sigma$ acts freely on ${\cal M}-\Sigma$ and therefore
\eqn\eulerdivision{
  \chi({\cal M}/\o\sigma) =
  {\chi({\cal M}) - \chi(\Sigma) \over 2} + \chi(\Sigma) =
  12 + {\chi(\Sigma) \over 2} .
}
So the homology of ${\cal M}/\o\sigma$ is (since $\o\sigma$ preserves the top
class):
\eqn\QuotHomology{
  {\rm dim}_\IR \Big( H_i({\cal M}/\o\sigma; \IR) \Big) = \left\{
  \matrix{
    1 & i=4 \cr
    0 & i=3 \cr
    10+{\chi(\Sigma) \over 2} & i=2 \cr
    0 & i=1 \cr
    1 & i=0
  }
  \right.
}
and the dimension of the invariant subspace of $H_2(X;\IR)$ has to be
\eqn\invsubspacedim{
  b_2^+ =
  10+{\chi(\Sigma) \over 2}
  \quad \Rightarrow \quad
  b_2^- =
  22-b_2^+ =
  12+{\chi(\Sigma) \over 2} .
}
Combining
\invsubspacedim{} and
\slagintersectionIndex{} we
find
\eqn\GeometryAnomalyCancel{
  \Sigma \circ \Sigma =
  2 (10 - b_2^+)
  \quad \Leftrightarrow \quad
  \pi_{{\rm O}7}\circ \pi_{{\rm O}7} = 2 (9-n_{\rm T}) .
}

\newsec{Examples on K3}

We now present a number of examples of six-dimensional
orientifolds on K3 with intersecting D7-branes and O7-planes. We
start with a number of orbifold models, then discuss the quartic
in $\IC\IP^3$, and finally add a number of observations on the
lift of supersymmetric models to F-theory on a Calabi-Yau 3-fold.

\subsec{Preliminaries on K3 orbifolds}

In the following we study certain orbifold realizations of K3 and
provide evidence that the picture developed above applies. We
consider the $\{\ZZ_2,\ZZ_3,\ZZ_4,\ZZ_6\}$ orbifold limits of K3,
where the action of $\ZZ_N$ on the $z_1,z_2$ of
\coord\ is defined by $v_I=(1/N,-1/N)$ as in \orb.
Recall, that the spectrum in table 1 is meant to be computed using the
intersection numbers on the resolved orbifold and not on the
initial torus.
There are some 2-cycles $\pi_a$ on the K3 which are inherited
from the torus. In the KK reduction on the orbifold they
correspond to massless modes in the untwisted sector of the theory.
In general 2-cycles $\o\pi_a$
on the torus are arranged in orbits of length $N$
under the $\ZZ_N$ orbifold group, i.e.
\eqn\orbit{
\pi_a = \sum_{i=0}^{N-1} \Theta^i \o\pi_a .
}
Such an orbit can then
be considered as a 2-cycle of the orbifold, where the intersection
form is given by
\eqn\inta{
\pi_a\circ\pi_b={1\over N} \left(\sum_{i=0}^{N-1} \Theta^i \o\pi_a
  \right) \circ \left(\sum_{j=0}^{N-1} \Theta^j \o\pi_b \right) .
}
Beside these 2-cycles the fixed points of the orbifold action give
rise to exceptional 2-cycles, which correspond to massless fields
in the twisted sectors of the orbifold. Blowing up the $i$-th
$\ZZ_N$ singularity yields $(N-1)$ 2-cycles, $e_i^{(j)}$,
$j\in\{1,\ldots,N-1\}$, whose intersection matrix is given by the
Cartan matrix of the Lie algebra ${A}_{N-1}$, which is encoded in
its Dynkin diagram.
\fig{}{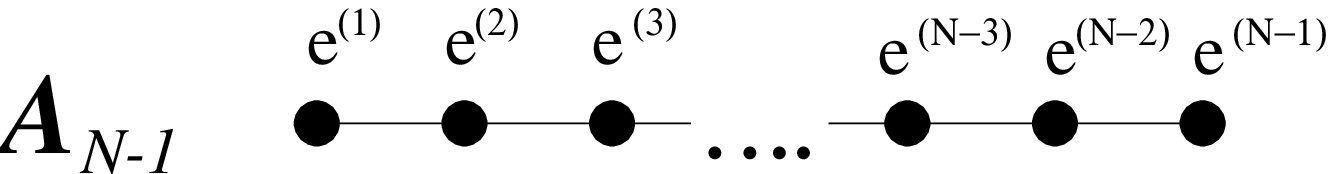}{8truecm}
\noindent
Note, that the cycles $\pi_a$ and  $e_i^{(j)}$ in
general do not yield an integral basis of $H_2({\rm K3},\ZZ)$, but
instead only generate a sub-lattice, while still of dimension 22, of
course.
Nevertheless, it provides the most convenient basis to work with when studying
orientifolds, as the homology class of the orientifold plane
and the action of the orientifold projection on these 2-cycles
can easily be determined.

As was anticipated in \rbgka, $\Omega\o\sigma$ leaves all twisted
sectors invariant and only leads to tadpoles from untwisted
sectors. Thus, after blowing up the orbifold singularities, the
${\rm O}7$-planes only wrap 2-cycles $\pi_a$ inherited from the
torus and no exceptional divisors. To find $\pi_{{\rm O}7}$ one
then only needs to apply \fixorb. By the relation \rela, knowing
only the orientifold plane is already sufficient to determine the
contribution of the chiral spectrum of open strings to the
gravitational anomaly. As an independent check, the action of
$\Omega\o\sigma$ on $H_2({\rm K3},\ZZ)$ can then be used to compute
$n_{\rm T}$ from \kthreespec.

The action $[\o\sigma]$ of $\o\sigma$ on the cohomology of K3 was
discussed in section 2.2 and can straightforwardly be determined.
At the orbifold point all twisted sectors organize into singlets
and doublets under $\Omega\o\sigma$. Taking the intrinsic reflection
into account as before, we write $[\o\sigma]=(-1)\otimes p$,
denoting the permutation of twisted sectors by $p$. In the smooth
case, we then expect $[\o\sigma]$ to act in an unchanged
manner,
\eqn\actio{
                 e_i^{(j)} \mapsto -e_{p(i)}^{(j)}\quad {\rm for}\quad
                     j\in\{1,\ldots,N-1\} .
}
Before moving forward to study explicit blown-up $\ZZ_N$ orbifolds
in more detail let us define a few useful conventions for the set
of 2-cycles $\o\pi_a$ on the torus. Within the framework of the
four $\ZZ_N$ orbifolds of K3 which we are going to consider it
turns out that each model allows two inequivalent complex
conjugations $\o\sigma$, depending on the complex structure of the
background torus. For the case of $\{\ZZ_2,\ZZ_4\}$ we shall
employ, up to rescaling, the root lattice of $SU(2)^4$, and for
$\{\ZZ_3,\ZZ_6\}$ that of $SU(3)^2$. The complex structure
$\tau^I$ of any single $T^2_I$ is defined by selecting two lattice
vectors ${\bf e}^I_1 = 1,\ {\bf e}^I_2 =
\tau^I$ as a basis, where $\o\sigma$ reflects along ${\bf e}^I_1$.
As in \rbgkb\  we define the choice {\bf A} for the two lattices by taking
\eqn\tauA{
\tau_{\bf A} = i,\quad \tau_{\bf A} = e^{2\pi i/6}
}
for $SU(2)^2$ and $SU(3)$, respectively. The elementary cells of
this type {\bf A} are displayed in figure 2.
\fig{}{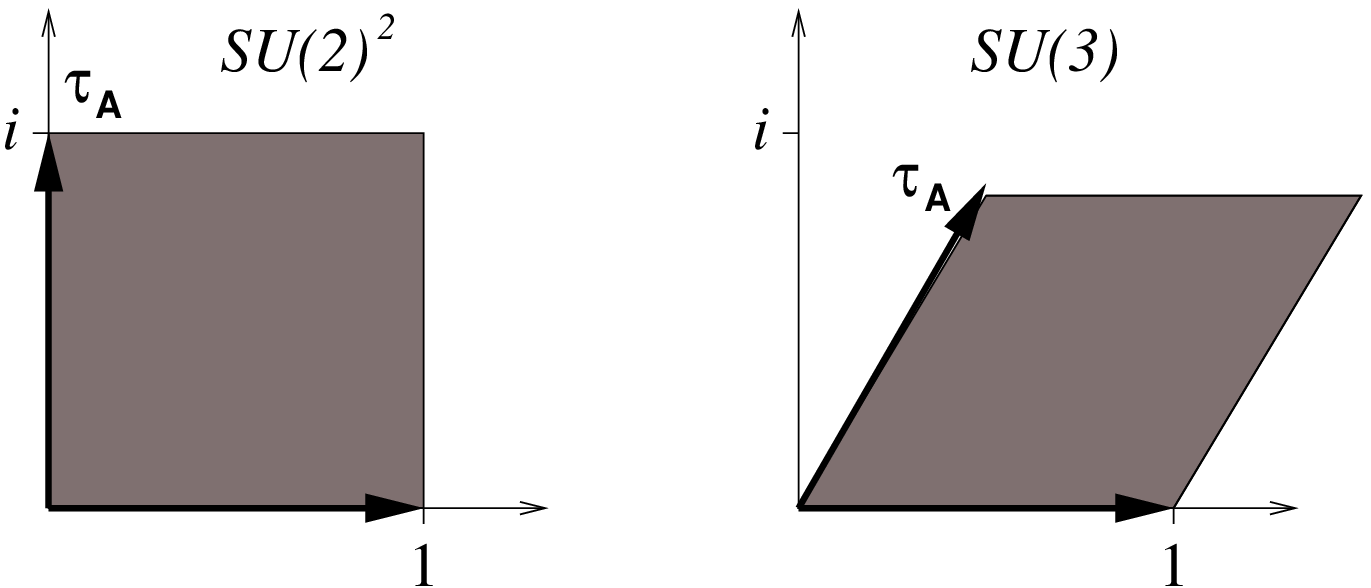}{10truecm}
The case B {\bf B} now refers to using the reflection symmetry of the lattices
along the diagonal of these cells {\bf A},
defined by ${\bf e}_1 + {\bf e}_2$. Therefore, the
{\bf B} type elementary cell is obtained by applying a rotation and
a rescaling to the {\bf A} type. The respective complex structures are
given by
\eqn\tauB{
\tau_{\bf B} = {1+i \over 2} ,\quad \tau_{\bf B}= {1 \over 2} +
{i \over 2\sqrt{3}}
}
for $SU(2)^2$ and $SU(3)$. The
{\bf B} type elementary cells are depicted in figure 3.
\fig{}{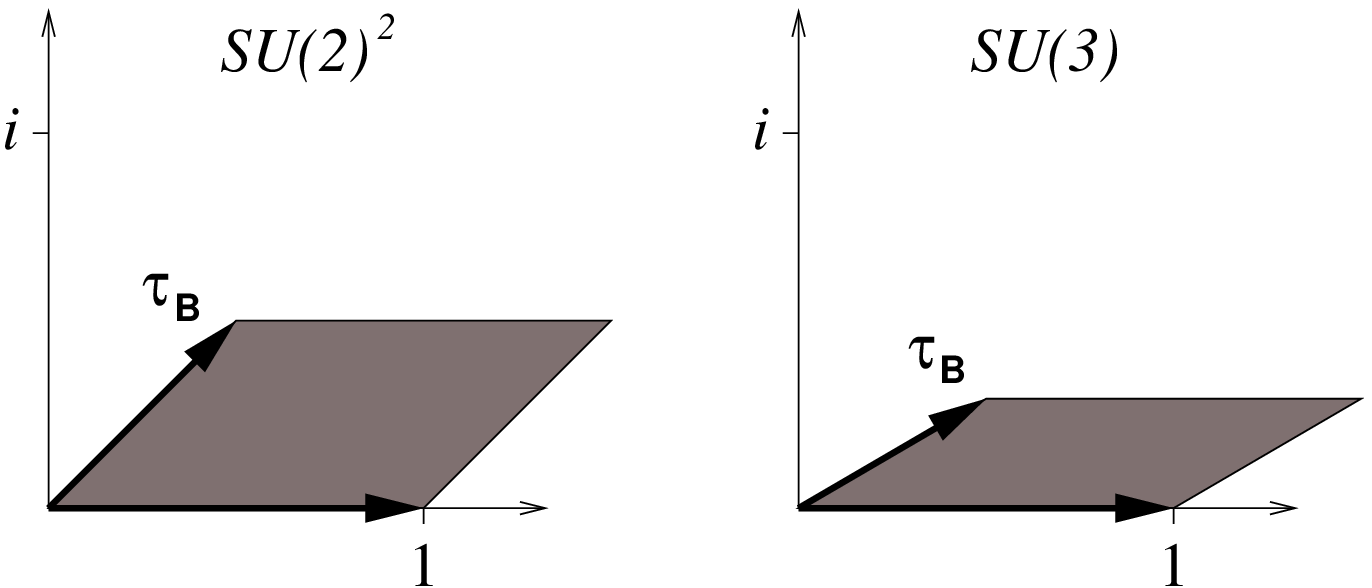}{10truecm}
Note, that in type I theory the real parts of the K\"ahler moduli
are frozen and may take values $0,1/2$ only. In the present dual
model, where $\Omega\o\sigma$ is projected out, this requirement
translates into demanding a crystallographic action of $\o\sigma$
leading to $\Re(\tau^I) =0,1/2$.

For convenience, we now pick a fixed basis for
$H_2(T^4;\ZZ)$ and use it in both cases, which implies slightly more
complicated formulas for the {\bf B} type models.
Given the {\bf A} type lattice ${\bf e}_1^I=1$ and
${\bf e}_2^I=\tau^I_{\bf A}$, we define
a basis $\{ \gamma_i \},\ i=1,\, ...\, ,4,$ of $H_1(T^4;\ZZ)$ by the four
circles spanned by the basis vectors,
$\gamma_{i+2(I-1)} = [\langle {\bf e}^I_i \rangle ]$.
Then
\eqn\torusbasis{
\o\pi_{ij} = \gamma_i \otimes \gamma_j
}
is a basis for $H_2(T^4;\ZZ)$. With this choice, the action of
$[\o\sigma]$ is very simple for the {\bf A} type cells, $[\o\sigma]$
leaves just the 1-cycles along ${\bf e}^I_1$ invariant, whereas
for the {\bf B} type, those along ${\bf e}^I_1 + {\bf e}^I_2$ are
invariant. By tensoring these together, one finds three
inequivalent combinations $\{ {\bf AA, AB, BB}\}$ for any orbifold
group. The three models will also differ in the transformation of
the fixed points of the orbifold group generator under $\o\sigma$,
which also depends on $\tau^I$.

\subsec{The orbifold limit $T^4/\ZZ_2$}

The simplest case to consider is the orbifold defined by $\Theta$
which reflects all four coordinates $\Theta : x_i\mapsto -x_i,\
i\in\{1,2,3,4\}$ of the $T^4$.
The sixteen fixed points under the $\ZZ_2$ are given by $\{
(0,1/2,\tau^1/2,(1+\tau^1)/2)\times
(0,1/2,\tau^2/2,(1+\tau^2)/2)\}$. We denote them as $P_{ij}$ with
$i,j\in\{1,2,3,4\}$. As defined by
\torusbasis, there are six elements $\o\pi_{ij}$ in
$H_2(T^4,\ZZ)$, which we denote as
\eqn\basist{ \{\o\pi_{13},\o\pi_{24},\o\pi_{14},\o\pi_{23},
               \o\pi_{12},\o\pi_{34} \} .}
Their intersection form reads
\eqn\inti{   I_{T^4}=
\left(\matrix{ 0 & 1 \cr 1 & 0 } \right)
\otimes \, {\rm diag}( 1,-1,-1) .
}
Each of these six 2-cycles $\o\pi_{ij}$ on $T^4$ gives rise to a
2-cycle on the orbifold, $\pi_{ij}$, by taking its orbit under
$\Theta$.
As elements of $H_2(T^4,\IR)$ they are simply invariant and thus
also generators of $H_2(T^4/\ZZ_2,\IR)$. But defining the image
classes in $H_2(T^4/\ZZ_2,\ZZ)$ requires the proper normalization.
It is evident, that a D7-brane wrapping along any of the
elementary cycles is in general mapped to a different position on
the torus by $\Theta$. The only exception occurs, if it is located
within Fix$(\o\sigma)|_{T^4}\cup {\rm Fix}(\Theta\o\sigma)|_{T^4}$. In
this case, it could only be moved outside of this locus, if there
were a second copy available, as its image under $\Theta$.
Otherwise it is stuck to the fixed locus of $\o\sigma$ and
fractional in the physical sense of carrying only $1/2$ of the
untwisted RR charge of a bulk brane. Thus, we conclude that the
cycles descending from the cycles of the $T^4$ all appear as
elements $\pi_{ij} = 2\o\pi_{ij}$ in $H_2(T^4/\ZZ_2,\ZZ)$,
corresponding to bulk D7-branes. Applying \inta\ yields their
intersection matrix
\eqn\inti{
I^{\rm Torus}_{T^4/\ZZ_2}= 2\, I_{T^4}
}
on the orbifold. The
remaining 16 2-cycles of the K3 are the blown-up $\IC\IP^1$ at the 16 fixed points
$P_{ij}$. Let us denote these sixteen exceptional divisors as $e_{ij}$ analogously.
They have vanishing intersection with the six 2-cycles inherited from the
$T^4$ and among themselves the intersections read
\eqn\intb{
e_{ij}\circ e_{kl}=-2\,\delta_{ik}\delta_{jl}   .
}
This is the Cartan matrix of $A_1^{16}$. Together this set
of 2-cycles $\{\pi_{ij}, e_{ij}\}$ generates the so-called Kummer
lattice, an index two sublattice of $H_2({\rm K3},\ZZ) \cong
\Gamma_{3,19}$.
In order to generate the entire lattice, we would also have to add
fractional branes wrapping on cycles of the kind ${1\over
2}\pi_{13} +{1\over 2}\left(e_{11}+e_{12}+e_{21}+e_{22} \right)$,
which is not part of the Kummer lattice. One always needs $N$
fractional branes to form a bulk D7-brane in a $\ZZ_N$ orbifold.

Now, we come to perform the orientifold by $\Omega\o\sigma$ of the smooth
K3 manifold.
As was explained above, we can define anti-holomorphic involutions $\o\sigma$
which act crystallographically on the $T^4$ with three different choices
of the complex structure of $T_I^2$, labeled by $\{ {\bf AA,AB,BB}\}$.

\vfill\eject
\bigno
{\it 4.2.1. {\bf AA}-orientifold}
\bigno

Let us first choose $\tau^I_{\bf A} =i$ on both $T_I^2$, the {\bf
AA} model. The homology class of the ${\rm O}7$-plane is
\eqn\oplane{
\pi_{{\rm O}7}=2( \pi_{13} + \pi_{24} ) .
}
With this choice of complex structure all the fixed points
$P_{ij}$ are geometrically invariant. Through the intrinsic
reflection of the blow-up mode, $[\o\sigma]$ then reflects all
$e_{ij}$. On $T^4$ only the cycles $\o\pi_{13}$ and $\o\pi_{24}$
are invariant. The action of $[\o\sigma]$ on the 22 two-cycles is
then given by
\eqn\inti{   [\o\sigma]_{\bf AA}=
{\rm diag}\left( {\bf 1}_2 , {\bf -1}_{20} \right)
, }
with ${\bf 1}_n$ denoting unit matrices of rank $n$.
This also implies, that the image under $\Omega \o\sigma$
of any brane with non-vanishing twisted charge always carries the opposite twisted
charge.
The number of tensor multiplets is now found to be $n_{\rm T}=1$.
Computing the self-intersection number of the orientifold plane \oplane\
we find $\pi_{{\rm O}7}\circ \pi_{{\rm O}7} =16 $ so that indeed the $R^4$ anomaly
is canceled. Now, we have all the ingredients required to compute the chiral
spectrum of a general intersecting brane world on the K3 given by a
blow-up of the orbifold $T^4/\ZZ_2$.
It is easy to see that as long as we only introduce bulk branes
the spectrum  in table 1 agrees with the spectrum found in \rbkl\ for this case.

In fact, for the simple case of a $\ZZ_2$ orbifold group, the
projection by $\Omega\o\sigma$ is equivalent to the standard
projection by $\Omega$ upon performing T-dualities along the two
circles parameterized by $\Im(z_i)$. Let us therefore show as a
check of our methods, that we can recover a T-dual version of the
original $\ZZ_2$ orientifold first discovered by Bianchi and
Sagnotti in
\sagn\ and described in terms of D-branes by Gimon and Polchinski
\rgimpol. The tadpoles can be canceled by introducing just two
stacks of fractional D7-branes with multiplicities $N_1=N_2=16$,
which support a gauge group $U(16)^2$. The cycles and their $\Omega\o\sigma$
images they are wrapped around are
\eqn\exama{\eqalign{
\pi_1&={1\over 2}\left( \pi_{13} \right)
                  +{1\over 2}\left(e_{11}+e_{12}+e_{21}+e_{22} \right), \cr
                   \pi_1'&={1\over 2}\left( \pi_{13} \right)
                  -{1\over 2}\left(e_{11}+e_{12}+e_{21}+e_{22} \right), \cr
                    \pi_2&={1\over 2}\left( \pi_{24} \right)
                  +{1\over 2}\left(e_{11}+e_{13}+e_{31}+e_{33} \right), \cr
                    \pi'_2&={1\over 2}\left( \pi_{24} \right)
                  -{1\over 2}\left(e_{11}+e_{13}+e_{31}+e_{33} \right). }
}
Their intersection numbers
\eqn\intnuma{\eqalign{  &\pi_1\circ\pi_1=-2, \quad\quad  \pi_2\circ\pi_2=-2,\quad\quad
                       \pi_1\circ\pi'_1=2, \quad\quad  \pi_2\circ\pi'_2=2, \cr
                       &\pi_1\circ\pi_{{\rm O}7}=2, \quad\quad  \pi_2\circ\pi_{{\rm O}7}=2,
         \quad\quad   \pi_1\circ\pi_2=0, \quad\quad  \pi_1\circ\pi'_2=1 \cr}}
yield the chiral massless spectrum shown in table 2
\vskip 0.8cm
\vbox{
\centerline{\vbox{
\hbox{\vbox{\offinterlineskip
\def\tablespace{height1pt&\omit&&\omit&\cr}
\def\tablerule{\tablespace\noalign{\hrule}\tablespace}

\hrule\halign{&\vrule#&\strut\hskip0.2cm\hfill #\hfill\hskip0.2cm\cr
& Representation && Multiplicity &\cr
\tablerule
& $[({\bf Adj,1})+({\bf 1, Adj})]_{(2,1)}$ && 2 &\cr
& $[({\bf A,1})+({\bf 1, A})+c.c.]_{(1,2)}$ && 2 &\cr
& $[({\bf 16},{\bf 16})+c.c.]_{(1,2)}$ && 1 &\cr
}\hrule}}}}}
\centerline{
\hbox{{\bf Table 2:}{\it ~~ GP model }}}
\vskip 0.5cm
\noindent
which agrees precisely with the massless spectrum determined in
\refs{\sagn,\rgimpol}.
Note, that in the orbifold limit the gauginos in the adjoint
representation of the gauge group appeared in the sector of open
strings with both ends on the same D7-brane. Chirality was induced
by a non-trivial projection on the Chan-Paton indices, but the
naive self-intersection of the brane was vanishing. However, in
the smooth case the gauginos are localized at the
self-intersections of the D7-branes. This self-intersection point
was somehow hidden in the orbifold singularity and encoded in the
action of the orbifold generator on the Chan-Paton indices. To
make this whole picture consistent, it was absolutely crucial that
the self-intersection number of an exceptional divisor is
$e_{ij}\circ e_{ij}=-2$. At the orbifold point the model is
supersymmetric, as here the D7-branes simply lie on top of the
orientifold plane. Changing the K\"ahler structure by blowing-up
the sixteen $\ZZ_2$ singularities does not break supersymmetry, as
long as each individual  D7-brane wraps a cycle calibrated by
$\Re(\Omega_2)$.

Since the D-branes in \exama\ do wrap exceptional divisors,
the tadpoles of the orientifold planes are not canceled
locally in this model. To cancel them locally, one must introduce
four stacks of $N=4$ D7-branes  wrapped on the cycles
\eqn\localta{   \widetilde\pi_1=\pi_1+\pi_1'=\pi_{13}, \quad\quad
                 \widetilde\pi_2=\pi_2+\pi_2'=\pi_{24}. }
Since these cycles are invariant under $\Omega\o\sigma$
one gets an overall gauge group $SO(4)^4\times SO(4)^4=SU(2)^8\times
SU(2)^8$. Note, that this should be precisely the model
for which Ashoke Sen presented an F-theory dual description
in \refs{\rsena,\rsenb}. We will discuss the implications of this
dual F-theory picture in section 4.7.1.

Let us consider another example with fractional D-branes that had
appeared in \rbkl. The
tadpoles can also be canceled by just a single stack of $N=16$
branes, together with its image under $\Omega\o\sigma$. It is
wrapped on
\eqn\exama{\eqalign{
\pi&={1\over 2}\left( \pi_{13}+\pi_{24}+\pi_{14}
                   +\pi_{23} \right)
                  +{1\over 2}\left(e_{11}+e_{44}+e_{14}+e_{41} \right) , \cr
                   \pi'&={1\over 2}\left( \pi_{13}+\pi_{24}-\pi_{14}
                   -\pi_{23} \right)
                  -{1\over 2}\left(e_{11}+e_{44}+e_{14}+e_{41} \right). }
}
Thus, we get a gauge group $U(16)$.
For the relevant intersection numbers we obtain
\eqn\intnum{ \pi\circ\pi=-2, \quad\quad  \pi\circ\pi'=4, \quad\quad
             \pi\circ\pi_{{\rm O}7}=4 ,
}
giving rise to the chiral massless spectrum
\vskip 0.8cm
\vbox{
\centerline{\vbox{
\hbox{\vbox{\offinterlineskip
\def\tablespace{height1pt&\omit&&\omit&\cr}
\def\tablerule{\tablespace\noalign{\hrule}\tablespace}

\hrule\halign{&\vrule#&\strut\hskip0.2cm\hfill #\hfill\hskip0.2cm\cr
& Representation && Multiplicity &\cr
\tablerule
& $[{\bf Adj}]_{(2,1)}$ && 2 &\cr
& $[{\bf A+\o{A}}]_{(1,2)}$ && 4 &\cr
}\hrule}}}}}
\centerline{
\hbox{{\bf Table 3:}{\it ~~ Chiral spectrum }}}
\vskip 0.5cm
\noindent
This is precisely the spectrum obtained in the orbifold limit
\rbkl. Computing the disc level scalar potential
\susy\ at the orbifold point one finds
\eqn\tadp{
{\cal V} = T_7 e^{-\phi_6}\left[
\prod_{I=1}^2\sqrt{\Im(\tau^I)+{1\over \Im(\tau^I)}}-
\left(\sqrt{\Im(\tau^1)\, \Im(\tau^2)}+{1\over
              \sqrt{\Im(\tau^1)\, \Im(\tau^2)}}\right) \right] }
which vanishes precisely if the two complex structure moduli
$\Im(\tau^1)$ and $\Im(\tau^2)$ for the two $T_I^2$ are equal.
This is the point, where the angle criterion
\angle\ for the D7-branes is satisfied and the configuration turns
supersymmetric.
Moving away from the supersymmetric $\Im(\tau^1)=\Im(\tau^2)$
locus, the intersection angles do not any longer satisfy \angle,
supersymmetry is broken and an open string tachyon appears.  In
the effective six-dimensional gauge theory this effect is
described by a Fayet-Iliopoulos term depending on
$\xi \sim \Im(\tau^1)-\Im(\tau^2)$ \refs{\rkachmca}.
Note, that in six dimensions the D-term potential for a $U(1)$ gauge field
has the general form
\eqn\dtermb{
{\cal V}_{\rm D-term} \sim
\biggl(\sum_i q^i |\phi_i|^2- \sum_i q^i |\widetilde\phi_i|^2-\xi\biggr)^2\, ,
}
where $\phi_i$ and $\widetilde\phi_i$ denote the two complex scalars
insider a hypermultiplet $H_i$ with charge $q_i$. Note  that independent of
the sign of $\xi$, if $\xi\ne 0$ one always gets a tachyonic mode,
which is in accord with the string theory picture.

Let us conclude this example with stating that here
supersymmetry and tadpole cancellation
conditions still
allow non-trivial D7-brane intersections with chiral matter.

\bigno
{\it 4.2.2. {\bf AB}-orientifold}
\bigno
In the case of an {\bf AB} complex structure, the O7-plane wraps
around the homological cycle
\eqn\oplane{
\pi_{{\rm O}7}= \pi_{13} + \pi_{24} +\pi_{14}-\pi_{23} .
}
On the exceptional divisors
$\{e_{11},e_{21},e_{31},e_{41},e_{14},e_{24},e_{34},e_{44}\}$
$\o\sigma$ acts with a minus sign, and the four cycles
$\{e_{12},e_{22},e_{32},e_{42}\}$ get exchanged with
$\{e_{13},e_{23},e_{33},e_{43}\}$. Its action on $H_2({\rm
K3},\ZZ)$ is summarized by
\eqn\inti{
[\o\sigma]_{\bf AB}=
 \left( \left(\matrix{ 0 & 1 \cr 1 & 0 } \right)
\otimes \, {\rm diag}( 1,-1) \right) \oplus \, (-{\bf 1}_2)
\oplus \bigoplus_{k=1}^4 \left(\matrix{ 0 & -1 \cr -1 & 0 } \right)_k
\oplus (-{\bf 1}_8) .
}
The first two terms stand for the cycles inherited from the torus,
the last two for the action on the exceptional cycles. Hence, the
number of tensor-multiplets is $n_{\rm T}=5$, one from the torus
and four from the blow-up modes, which is in accord with the
self-intersection number of the orientifold plane, $\pi_{{\rm
O}7}\circ \pi_{{\rm O}7} =8$.

\bigno
{\it 4.2.3. {\bf BB}-orientifold}
\bigno
If one chooses the {\bf B} type complex structure on both $T_I^2$
the orientifold plane is wrapping
\eqn\oplane{
\pi_{{\rm O}7}=\pi_{13} + \pi_{24}
}
with self-intersection number $\pi_{{\rm O}7}\circ \pi_{{\rm O}7} =4$.
On the 2-cycles $\o\sigma$ acts like
\eqn\inti{
[\o\sigma]_{\bf BB}=
\left(\matrix{ 0 & 1 \cr 1 & 0 } \right) \oplus
\left(\matrix{ 0 & -1 \cr -1 & 0 } \right) \oplus \, (-{\bf 1}_2)
\oplus \bigoplus_{k=1}^6 \left(\matrix{ 0 & -1 \cr -1 & 0 } \right)_k
\oplus (-{\bf 1}_4) .
}
This time, $\o\sigma$ acts on the four exceptional divisors
$\{e_{11},e_{14},e_{41},e_{44}\}$ with $-1$, and the six
exceptional 2-cycles $\{e_{12},e_{42},e_{21},e_{24},e_{22},e_{23}
\}$ get exchanged with
$\{e_{13},e_{43},e_{31},e_{34},e_{33},e_{32}\}$. We get $n_{\rm
T}=7$ tensor multiplets, again one from the torus plus six from
the blow-up modes. As an example for a solution to the tadpole
constraints, we introduce two stacks of fractional D7-branes with
$N_1=N_2=8$, which support a gauge group $U(8)^2$. They wrap the
following 2-cycles
\eqn\examb{\eqalign{
\pi_1&={1\over 2}\left( \pi_{13} \right)
                  +{1\over 2}\left(e_{11}+e_{12}+e_{21}+e_{22} \right), \cr
                   \pi_1'&={1\over 2}\left( \pi_{24} \right)
                  +{1\over 2}\left(-e_{11}+e_{13}+e_{31}+e_{33} \right), \cr
                   \pi_2&={1\over 2}\left( \pi_{13} \right)
                  -{1\over 2}\left(e_{11}+e_{12}+e_{21}+e_{22} \right), \cr
                   \pi_2'&={1\over 2}\left( \pi_{24} \right)
                  -{1\over 2}\left(-e_{11}+e_{13}+e_{31}+e_{33} \right). \cr  }
}
The relevant intersection numbers
\eqn\intnum{\eqalign{  &\pi_1\circ\pi_1=-2, \quad\quad  \pi_2\circ\pi_2=-2,\quad\quad
                       \pi_1\circ\pi'_1=1, \quad\quad  \pi_2\circ\pi'_2=1, \cr
                       &\pi_1\circ\pi_{{\rm O}7}=1, \quad\quad  \pi_2\circ\pi_{{\rm O}7}=1,
         \quad\quad   \pi_1\circ\pi_2=2, \quad\quad  \pi_1\circ\pi'_2=0 \cr}}
yield the chiral massless spectrum
\vskip 0.8cm
\vbox{
\centerline{\vbox{
\hbox{\vbox{\offinterlineskip
\def\tablespace{height1pt&\omit&&\omit&\cr}
\def\tablerule{\tablespace\noalign{\hrule}\tablespace}

\hrule\halign{&\vrule#&\strut\hskip0.2cm\hfill #\hfill\hskip0.2cm\cr
& Representation && Multiplicity &\cr
\tablerule
& $[({\bf Adj,1})+({\bf 1, Adj})]_{(2,1)}$ && 2 &\cr
& $[({\bf A,1})+({\bf 1,\bf A})+c.c.]_{(1,2)}$ && 1 &\cr
& $[{\bf(8,8)}+c.c.]_{(1,2)}$ && 2 &\cr
}\hrule}}}}
\centerline{
\hbox{{\bf Table 4:}{\it ~~ Chiral spectrum }}}}
\vskip 0.5cm
\noindent
This agrees completely with the spectrum derived in the orbifold
limit \rbkl.

\subsec{The orbifold limit $T^4/\ZZ_3$}

\noindent
For the orbifold of $T^4/\ZZ_3$ we in principle again have to
distinguish three inequivalent choices of the complex structure of
the $T^4$, which is now defined by the root lattice of $SU(3)^2$.
In any case only four 2-cycles are inherited from the torus, which
can be described as orbits under the action of the $\ZZ_3$ by
\eqn\cycles{\eqalign{
&\pi_1=  \o\pi_{13}+
    \o\pi_{14}+\o\pi_{23}-2\,\o\pi_{24}, \cr
     &\pi_2= 2\,\o\pi_{14}+
    2\,\o\pi_{23}-\o\pi_{13}-\o\pi_{24} , \cr
     &\pi_3= 3\, \o\pi_{12} , \cr
     &\pi_4= 3\, \o\pi_{34} .}
}
The intersection matrix for these four 2-cycles can be computed
using \inta\ to be
\eqn\inti{
I^{\rm Torus}_{T^4/\ZZ_3}=
\left(\matrix{ 2 & 1 \cr 1 & 2 }\right) \oplus
\left(\matrix{ 0 & 3 \cr 3 & 0 }\right).
}
The remaining 18 2-cycles result from blowing-up the 9 $A_2$
singularities. For each fixed point we get two 2-cycles
$e^{(1)}_{ij},\, e^{(2)}_{ij}$ with their intersection matrix
given by the Cartan matrix of $A_2$. The total intersection matrix
is given by
\eqn\torint{
I_{T^4/\ZZ_3}=\bigoplus_{k=1}^9 \left(\matrix{ -2& 1 \cr
                                                          1 & -2 \cr}\right)_k
                             \oplus I^{\rm Torus}_{T^4/\ZZ_3} .
}
In the following, we shall be very brief on the model defined by the
involution $\Omega\o\sigma$ applied to type IIB on $T^4/\ZZ_3$,
but add an excursion on a related class of models.

\bigno
{\it 4.3.1. Anti-holomorphic involution $\o\sigma$}
\bigno
For the orbifold
$T^4/\ZZ_3$ there again exist two possible choices for the complex structure
of each $T^2_I$
leading to three different models. The computation of the
homological cycle of the ${\rm O}7$-plane and the action of
$\Omega\o\sigma$ on the basis of $H_2({\rm K3},\ZZ)$ can be
determined straightforwardly. We summarize the main data in table
5.
\vskip 0.8cm
\vbox{
\centerline{\vbox{
\hbox{\vbox{\offinterlineskip
\def\tablespace{height2pt&\omit&&\omit&&\omit&&
 \omit&\cr}
\def\tablerule{\tablespace\noalign{\hrule}\tablespace}

\hrule\halign{&\vrule#&\strut\hskip0.2cm\hfill #\hfill\hskip0.2cm\cr
& Complex structure  && $\pi_{{\rm O}7}$
&& $\pi_{{\rm O}7}\circ \pi_{{\rm O}7}$ &&   $n_{\rm T}$ &\cr
\tablerule
& {\bf AA}   && $ \pi_1$  && $2$  &&  $8$ &\cr
\tablerule
& {\bf AB}   && $ \pi_1+\pi_2$  && $6$  &&  $6$ &\cr
\tablerule
& {\bf BB}   && $ 3\pi_2$  && $18$  &&  $0$ &\cr
}\hrule}}}}
\centerline{
\hbox{{\bf Table 5:}{\it ~~ $T^4/\ZZ_3$ orientifolds }}}
}
\vskip 0.5cm
\noindent
Using these data one can check that placing the D7-branes on top
of the orientifold planes leads precisely to the chiral massless
spectra obtained  in \refs{\rbgka,\refPradisi} in the orbifold limit.
More general models can easily be constructed in great numbers.

\bigno
{\it 4.3.2. Holomorphic involution $\sigma$}
\bigno

On a K3 surface the notion of a special Lagrangian cycle and a
complex curve, which refer to a particular choice of the K\"ahler
and the holomorphic 2-form, can get exchanged. This is done by a
rotation of the complex structure, which is discussed in some more detail
in section 4.6. This rotation maps the complex conjugation $\o\sigma$ to
\eqn\sigmabar{
\sigma : (z_1,z_2) \mapsto (z_1,-z_2) ,
}
which now reflects the holomorphic 2-form, not just its imaginary part.
In contrast to the earlier case,
we now have that
\eqn\holcycle{
\Omega_2 |_{{\rm Fix}(\sigma)} =0 ,\quad
J|_{{\rm Fix}(\sigma)} = d{\rm vol}|_{{\rm Fix}(\sigma)} ,
}
i.e. Fix$(\sigma)$ is a holomorphic cycle calibrated with respect to $J$.
If we were just considering toroidal compactifications, the two
descriptions would
be strictly equivalent. But the definition of the orbifolds via
\orb\ itself also implies a choice of a particular complex structure.
In order to compensate for the action on the zero modes of
the RR ground state, it is required to include an additional
sign $(-1)^{F_L}$, $F_L$ the left-moving world
sheet fermion number operator, with the world sheet parity $\Omega$.
Then, combining projections by $\Omega\sigma (-1)^{F_L}$ and $\Theta$
leads to inequivalent results, as compared to $\Omega\o\sigma$ and $\Theta$.
In fact, the group generated by
$\{ \Omega\sigma (-1)^{F_L},\Theta \}$ is the precise T-dual
of $\{\Omega,\Theta\}$, after T-dualities along the two circles
parameterized by $z_2$.
On the contrary, the T-dualities along the imaginary parts of $z_1,z_2$
take $\{ \Omega\o\sigma,\Theta \}$ to $\{ \Omega,\hat\Theta \}$, where
$\hat\Theta$ is an asymmetric operation that treats left- and right-moving
world-sheet fields with opposite phase factors \rbgkl.
Due to these relations, we
can expect, that the models defined by $\sigma$ reproduce the known
orientifold vacua of type I string theory \rgimjo. To demonstrate this fact, we
include the case of the $T^4/\ZZ_3$ orientifold with $\Omega\sigma$
projection here. For the even simpler case of $T^4/\ZZ_2$, the two types of
models are still completely equivalent and the definition of $\Theta$ does
not distinguish $\sigma$ and $\o\sigma$.

An important ingredient in the CFT construction of orbifold vacua
of type I is the action of
$\Omega$ on twisted sectors. A field twisted by $\Theta^k$
is mapped to one twisted by $\Theta^{N-k}$ via $\Omega$. In contrast to the
$\Omega\o\sigma$ models, one then also
gets twisted sector tadpoles. This means, that, after blowing up the orbifold
singularities, the orientifold planes also wrap exceptional divisors.
We propose that the action on the exceptional 2-cycles is then given by
\eqn\actior{
[\sigma] : e_i^{(j)} \mapsto -e_i^{(N-j)}\quad {\rm for}\
j\in\{1,\ldots,N-1\} .
}
Thus, $\Omega\sigma$ reflects the Dynkin diagram, as depicted in
figure 4.
\fig{}{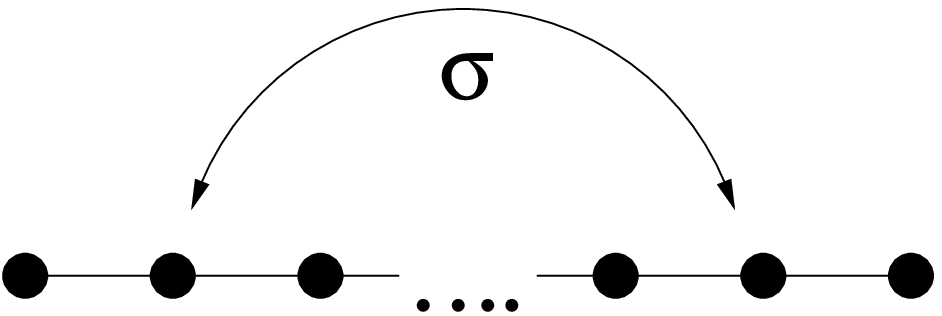}{8truecm}
\noindent

In other words, $\Omega\sigma$ exchanges the two blow-up modes
$e_{ij}^{(1)},\, e_{ij}^{(2)}$ of the two $\IC\IP^1$ which are
glued into any of the nine $A_2$ singularities at the fixed points
of the $\ZZ_3$. Taking also the action of $\Omega\sigma$ on the
four toroidal 2-cycles into account, which leaves one tensor
multiplet, we deduce that we get $n_{\rm T}=10$ tensor multiplets in the
closed string sector. At the four fixed points of $\sigma$ we
get an ${\rm O}7$-plane, which under the action of $\ZZ_3$ gives
rise to one bulk ${\rm O}7$-plane and one fractional ${\rm
O}7$-plane. The total homological cycle is given by
\eqn\oseven{
\pi_{{\rm O}7}=\pi_{12}+\left( {1\over 3}\pi_{12} +{1\over \sqrt{3}}
      \left[ e_{11}^{(1)} +e_{11}^{(2)} +e_{21}^{(1)} +e_{21}^{(2)} +
            e_{31}^{(1)} + e_{31}^{(2)} \right] \right)  .
}
Note, that the fractional ${\rm O}7$-plane is indeed invariant
under $\Omega\sigma$. The self-intersection number of this ${\rm
O}7$-plane can be computed to be $\pi_{{\rm O}7}\circ \pi_{{\rm
O}7}=-2$, which is consistent with having $n_{\rm T}=10$. One can cancel
the RR-tadpoles by introducing 8 bulk D7-branes and 8 fractional
D7-branes. Splitting the bulk branes into 3 fractional branes we
get three kinds of D7-branes
\eqn\branesyu{\eqalign{
N_1=16:\quad\quad
           \pi_1&={1\over 3}\pi_{12} +{1\over \sqrt{3}}
      \left[ e_{11}^{(1)} +e_{11}^{(2)} +e_{21}^{(1)} +e_{21}^{(2)} +
            e_{31}^{(1)} + e_{31}^{(2)} \right], \cr
     N_2=8:\quad\quad \pi_2&={1\over 3}\pi_{12} -{1\over \sqrt{3}}
      \left[ e_{11}^{(1)} +e_{21}^{(1)}  +
            e_{31}^{(1)}  \right], \cr
      N_2=8:\quad\quad \pi_2'&={1\over 3}\pi_{12} -{1\over \sqrt{3}}
      \left[ e_{11}^{(2)} +e_{21}^{(2)}  +
            e_{31}^{(2)}  \right]. \cr }}
Here, $\pi_1$ is $\Omega\sigma$-invariant, whereas $\pi_2$ and
$\pi'_2$ are mapped upon each other. The gauge group on an invariant
fractional brane is given by $SO(N)$, while a fractional brane that
belongs to an orbit of length 3 still carries a $U(N)$ gauge group. Together,
the gauge group is $SO(16)\times U(8)$. From the intersection numbers
\eqn\intnumberv{
\pi_i\circ \pi_i=-2,\quad \pi_2\circ \pi_{{\rm O}7}=1,\quad
                  \pi_2\circ \pi'_{2}=1,\quad  \pi_1\circ \pi_2=1
}
we deduce the chiral massless spectrum by using table 1,
\vskip 0.8cm
\vbox{
\centerline{\vbox{
\hbox{\vbox{\offinterlineskip
\def\tablespace{height1pt&\omit&&\omit&\cr}
\def\tablerule{\tablespace\noalign{\hrule}\tablespace}

\hrule\halign{&\vrule#&\strut\hskip0.2cm\hfill #\hfill\hskip0.2cm\cr
& Representation && Multiplicity &\cr
\tablerule
& $[({\bf Adj,1})+({\bf 1, Adj})]_{(2,1)}$ && 2 &\cr
& $[({\bf 1,A})+c.c.]_{(1,2)}$ && 1 &\cr
& $[{\bf(16,8)}+c.c.]_{(1,2)}$ && 1 &\cr
}\hrule}}}}
\centerline{
\hbox{{\bf Table 6:}{\it ~~ $\ZZ_3$ orientifolds }}}}
\vskip 0.5cm
\noindent
which agrees completely with the massless spectrum derived in \rgimjo\ for the
$\ZZ_3$ orientifold of type I string theory.
We consider this as a very neat confirmation that the formula for the chiral
massless spectrum is absolutely general and is expected to cover all
six-dimensional orientifolds computed in the literature.

\subsec{The orbifold limit $T^4/\ZZ_4$}

For the orientifold $T^4/\ZZ_4$ we can again use the lattice defined by $SU(2)^4$.
The resulting K3 manifold inherits the following four 2-cycles from the torus
\eqn\cyclesb{\eqalign{
&\pi_1= 2\,\o\pi_{13}-
    2\,\o\pi_{24} , \cr
     &\pi_2= 2\,\o\pi_{14}+2\, \o\pi_{23} , \cr
     &\pi_3= 4\, \o\pi_{12} , \cr
     &\pi_4= 4\, \o\pi_{34} , }
}
which have the intersection matrix
\eqn\inti{
I^{\rm Torus}_{T^4/\ZZ_4}=
\left(\matrix{ 2 & 0 \cr 0 & 2 }\right)\oplus \left(\matrix{ 0 & 4 \cr 4 & 0 }\right).
}
Moreover, one has four $A_3$ singularities $\{e_{11},e_{14},e_{41},e_{44}\}$
and six $A_1$ singularities, so that the complete intersection  matrix
is
\eqn\torinta{
I_{T^4/\ZZ_4}=\bigoplus_{k=1}^4 \left(\matrix{ -2& 1 & 0\cr
                                               1 & -2 & 1 \cr
                                               0 & 1 & -2 \cr }\right)_k
                          \oplus   \bigoplus_{l=1}^6 \left(\matrix{ -2 }\right)_l
        \oplus I^{\rm Torus}_{T^4/\ZZ_4} .}
In principle, we would have to distinguish the three choices of
complex structure for the $T^4$ as before. But one can observe
that the two models
{\bf AA} and {\bf BB} are equivalent, as they give rise to the same
orientifold plane. We summarize the main data for the two
inequivalent orientifolds in table 7.
\vskip 0.8cm
\vbox{
\centerline{\vbox{
\hbox{\vbox{\offinterlineskip
\def\tablespace{height2pt&\omit&&\omit&&\omit&&
 \omit&\cr}
\def\tablerule{\tablespace\noalign{\hrule}\tablespace}

\hrule\halign{&\vrule#&\strut\hskip0.2cm\hfill #\hfill\hskip0.2cm\cr
& Complex structure && $\pi_{{\rm O}7}$ && $\pi_{{\rm O}7}\circ \pi_{{\rm O}7}$ &&   $n_{\rm T}$ &\cr
\tablerule
& {\bf AA} or {\bf BB}  && $3\, \pi_1$  && $18$  &&  $0$ &\cr
\tablerule
& {\bf AB}   && $2(\pi_1+\pi_2)$  && $16$  &&  $1$ &\cr
}\hrule}}}}
\centerline{
\hbox{{\bf Table 7:}{\it ~~ $T^4/\ZZ_4$ orientifolds }}}
}
\vskip 0.5cm
\noindent
The result for the {\bf AB} model agrees with the result obtained
by an explicit CFT computation in \rbgka. However, for the
{\bf AA} and {\bf BB} model it was argued in \rbgka\ that
the corresponding cross-cap state was not well
defined\footnote{$^2$}{That argument stated that the relative
normalization of the cross-cap states $|\Omega\o\sigma
\Theta^k\ra$ must be fixed in such a way that in the  overall
tree-channel Klein-bottle amplitude only states invariant under
$\Theta$ propagate. In view of the results in this section, we
think this  condition was a too strong requirement. Note, that the
closed string exchange between two individual orientifold planes
is simply given by the overlap $\la \Omega\o\sigma \Theta^k |e^{-l
H_{\rm cl}} |\Omega\o\sigma \Theta^l\ra$, so that imposing
conditions on the overall tree-channel Klein-bottle amplitude has
no physical meaning.}.

\subsec{The orbifold limit $T^4/\ZZ_6$}

The torus defined by the lattice of $SU(3)^2$ does allow the action
of a $\ZZ_6$ symmetry group. The orbifold
inherits the following four 2-cycles from the $T^4$
\eqn\cyclesc{\eqalign{
&\pi_1= 2\,\o\pi_{13}+2\,
    \o\pi_{23}+2\,\o\pi_{14}-4\, \o\pi_{24} , \cr
&\pi_2= 4\,\o\pi_{14}+
    4\,\o\pi_{23}-2\,\o\pi_{13}-2\,\o\pi_{24} , \cr
      &\pi_3= 6\, \o\pi_{12} , \cr
     &\pi_4= 6\, \o\pi_{34} .}
}
They have an intersection matrix
\eqn\inti{
I^{\rm Torus}_{T^4/\ZZ_6}=
\left(\matrix{ 4 & 2 \cr 2 & 4 }\right) \oplus
\left(\matrix{ 0 & 6 \cr 6 & 0 }\right).
}
Further, $\Theta$ has just one fixed point, an $A_5$ singularity $\{e_{11}\}$,
$\Theta^2$ has four fixed points, $A_2$ singularities, and
$\Theta^3$ five $A_1$ singularities. Together, the intersection  matrix
reads
\eqn\torinta{
I_{T^4/\ZZ_6}=\left(
        \matrix{ -2& 1 & 0 & 0 & 0\cr
                  1 & -2 & 1 & 0 & 0\cr
                  0 & 1 & -2 & 1  & 0\cr
                  0 & 0 & 1 & -2 & 1 \cr
                  0 & 0 & 0 & 1 & -2 \cr }\right)  \oplus
              \bigoplus_{k=1}^4 \left(\matrix{ -2& 1 \cr
                                                          1 & -2 \cr}\right)_k
                          \oplus   \bigoplus_{l=1}^5 \left(\matrix{ -2 }\right)_l
        \oplus I^{\rm Torus}_{T^4/\ZZ_6} .
}
Again, the two models {\bf AA} and {\bf BB} are equivalent. We
then summarize the data for the two inequivalent orientifolds in
table 8.
\vskip 0.8cm
\vbox{
\centerline{\vbox{
\hbox{\vbox{\offinterlineskip
\def\tablespace{height2pt&\omit&&\omit&&\omit&&
 \omit&\cr}
\def\tablerule{\tablespace\noalign{\hrule}\tablespace}

\hrule\halign{&\vrule#&\strut\hskip0.2cm\hfill #\hfill\hskip0.2cm\cr
& Complex structure && $\pi_{{\rm O}7}$
             && $\pi_{{\rm O}7}\circ \pi_{{\rm O}7}$ &&   $n_{\rm T}$ &\cr
\tablerule
& {\bf AA} or {\bf BB}  && $2\, \pi_1$  && $16$  &&  $1$ &\cr
\tablerule
& {\bf AB}   && $ \pi_1+\pi_2$  && $12$  &&  $3$ &\cr
}\hrule}}}}
\centerline{
\hbox{{\bf Table 8:}{\it ~~ $T^4/\ZZ_6$ orientifolds }}}
}
\vskip 0.5cm
\noindent

\subsec{The quartic}

In this section we examine the most simple example of a K3 surface
defined as a hypersurface in a projective space, the quartic
polynomial in $\IC\IP^3$. While we only discuss three special cases,
these still turn out to give us rather exotic models very
different from the orbifolds above.

The quartic is defined by an arbitrary polynomial in the four
homogeneous coordinates $[z_0:z_1:z_2:z_3]$ that parameterize
$\IC\IP^3$, i.e.
\eqn\defquart{
\sum_{i,j,k,l=0}^3{ a_{ijkl}\, z_i z_j z_k z_l } =0 .
}
The $35=4\cdot 5\cdot 6\cdot 7/(2\cdot 3\cdot 4)$ parameters
$a_{ijkl}$ modulo the $16=4^2$ parameters of $GL(4,\IC)$ parameterize a
nineteen-dimensional family of algebraic K3 surfaces.  As long as all
coefficients $a_{ijkl}$ are chosen real, the conjugation on the four
$z_i$ is a symmetry, i.e. projecting by $\Omega\o\sigma$ removes one
half the moduli. The most simple case arises when there are no
real solutions to \defquart. This is precisely the case when only
quartic or quadratic terms have non-vanishing and positive
coefficients
\eqn\fermquart{
\sum_{i=0}^3{ \left( a_i\, z_i^4 + z_i^2 \sum_{j<i} b_{ij}\, z_j^2 \right) } =0 , \quad
a_i ,\, b_{ij} \ge 0,
}
which, for $b_{ij}=0$, includes the Fermat quartic. Here
Fix$(\o\sigma)$ is empty and thus $\pi_{{\rm O}7} =0$, so that
$n_{\rm T}=9$ tensor multiplets are required by the anomaly
cancellation. Therefore among the 22 two-cycles we expect 
10 cycles to be invariant under $\o\sigma$ and 12 cycles
to be anti-invariant.  Such a model
was inaccessible in orbifold vacua, where the orbit of the real
section on the torus always descends to an invariant locus under
$\o\sigma$. In fact there is a natural candidate for an orbifold
model of type I strings which can produce models without
orientifold planes.  In order to achieve this, one combines the
operation $\Omega$ with a freely acting shift. Its insertion in
the Klein bottle does not produce massless tadpoles which would
signal background charges of orientifold planes. In
\refs{\rbluma,\DabholkarKA} such an
orientifold model was constructed, inspired by the existence of an
F-theory compactification \MorrisonPP\ believed to be dual to a type I
vacuum without orientifold planes.

A slight modification of the above case leads to another model
given by flipping the sign of any one of the $a_i$ coefficients,
say $a_0=-1$. For simplicity, we also set $b_{ij}=0$ and scale
$a_i=1,\, i>0$ in \fermquart. By going to complex $a_0$ in
between, this deformation is continuous, but passing through a
region in the moduli space where $\o\sigma$ is no symmetry anymore.
Defining $x_i = \Re(z_i)$, the hypersurface intersects the real
section in a sphere, parameterized by
\eqn\spherehom{
- x_0^4 + x_1^4 + x^4_2 + x^4_3 = 0,
\quad [x_0:x_1:x_2:x_3]\in \IR\IP^3
.
}
For $x_0=0$ there is no solution, so we can rescale the homogeneous
coordinates to $x_0 =1$:
\eqn\spherehom{
x_1^4 + x^4_2 + x^4_3 = 1,
\quad (x_1,x_2,x_3)\in \IR^3
.
}
Topologically Fix$(\o\sigma)$ is thus given by a single calibrated
2-sphere. By the hyperk\"ahler property of K3, this is a
holomorphic $\IC\IP^1$ in another complex structure, which is
determined to have self-intersection $\pi_{{\rm O}7} \circ
\pi_{{\rm O}7}=-2$. This would require $n_{\rm T} =10$ in
accord with $\chi(\IC\IP^1)=2$. Thus in this model the {\rm
O}7-plane is wrapping a single exceptional 2-cycle, which looks
rather strange from the point of view of the orbifold models
studied earlier. The 22 two-cycles must arrange themselves into 11
$\o\sigma$ invariant ones and the same number of anti-invariant ones.

Finally, consider the following quartic
\eqn\eqQuarticTwoMinus{
-z_1^4 -z_2^4 +z_3^4 +z_4^4 = 0.
}
Here the real subspace ${\rm Fix}(\o\sigma)$ is a special Lagrangian
two-torus, as was remarked in \BryantTori. As such it has vanishing
self-intersection. So there are $n_{\rm T}=9$ tensor multiplets as
in the first case with empty ${\rm Fix}(\o\sigma)$. But of course now
there are orientifold planes and D-branes, making the physics much
more interesting.

\subsec{Lift to F-theory on Calabi-Yau 3-folds}

In this section we will analyze the perspectives to lift the type
IIB compactifications on K3 modded out by $\Omega\o\sigma$ to
F-theory \refs{\VafaXN,\MorrisonNA,\MorrisonPP}.
In particular we want to make contact to earlier works
on the analogous lift of the $T^4/\ZZ_2$ orientifold to an
F-theory compactification on an elliptically fibered Calabi-Yau
3-fold. In certain prominent cases it can be of the Voisin-Borcea
type. In general, the lift implies a geometric interpretation of
the ten-dimensional complexified string coupling $\lambda$ as the
complex structure of a complex torus fibered over the
compactification space of type IIB. The degeneration locus of the
fiber is then given by the locations of D7-branes and {\rm
O}7-planes. This implies that they wrap on complex cycles of the
K3, cycles calibrated with respect to the K\"ahler structure $J$.
As long as the RR charges carried by these objects are confined to
very small regions on the internal space and at least
approximately cancel out locally, the string coupling may be
assumed to be constant and small outside this region. Therefore,
the perturbative analysis in the CFT orbifold may be compared to
the geometric picture. This local charge cancellation can no
longer be achieved in general when the D7-branes and {\rm
O}7-planes wrap on different cycles of the K3.


\bigno
{\it 4.7.1. The Weierstra\ss\ model on $\IC\IP^1 \times \IC\IP^1$}
\bigno

In \rsenb\ Ashoke Sen gave a description of a Weierstra\ss\
model of the elliptic fibration on the base space $\IC\IP^1 \times
\IC\IP^1$, which makes up a Calabi-Yau 3-fold with Hodge numbers
$(h^{(1,1)},h^{(2,1)}) = (3,243)$. The model is believed to be
dual to the standard orientifold compactification of type IIB on
$T^4/\ZZ_2$ with local charge cancellation. It was first
considered in \sagn\ and subsequently discussed in \rgimpol, and was subject
of section 4.1. In the original formulation the group generator
$\Theta$ reflects all four coordinates of the K3, $\Theta : \hat
x_i \mapsto -\hat x_i,\, i=1,\, ...
\, ,4,$ and the orientifold group is generated by
$\Omega$ and $\Theta$. The tadpole cancellation conditions imply the
presence of 32 D9- and 32 D5-branes to cancel the background charges
of the O9-planes and the O5-planes localized at the 16 fixed points of
$\Theta$. The maximal gauge group is given by $U(16)\times U(16)$.
This formulation
is not quite adapted to an interpretation within the framework
of F-theory, and one first needs to apply T-dualities along two of the
circles of the $T^4$, say along $\hat x_1$ and $\hat x_3$. This maps
\eqn\tdual{
\Omega \mapsto \Omega \Theta_{13} (-1)^{F_L},\quad
\Theta \mapsto \Theta,
}
where $\Theta_{ij}$ now only reflects $\hat x_i$ and $\hat x_j$ and $F_L$ is
the left-moving world sheet fermion number operator. The resulting group
is generated by $\Omega\Theta_{13}(-1)^{F_L}$ and
$\Omega\Theta_{24}(-1)^{F_L}$. The former D9- and D5-branes, as well as the
O-planes, map to intersecting D7-branes and {\rm O}7-planes located in the
fixed loci of $\Theta_{13}$ and $\Theta_{24}$.

It is quite evident how this is related to type IIB
compactifications on K3 modded out by $\Omega\o\sigma$.
The key ingredient is the hyperk\"ahler nature of K3 which allows to
rotate its complex structure such that holomorphic and special
Lagrangian 2-cycles get exchanged.
For simplicity, let us choose a purely
imaginary complex structure $\tau^I=i$ for the original
coordinates $z_i$ of \coord, on which $\o\sigma$ acts by complex conjugation.
If we then rotate into
\eqn\newcoord{
z_1' = x_1 - i x_2,\quad z_2' = y_1 + i y_2,
}
and define the new holomorphic 2-form $\Omega_2'$ and the new
K\"ahler form $J'$ as in \nform, the fixed locus of $\o\sigma$, the
location of the {\rm O}7-planes, becomes a holomorphic cycle $\{
z_2'=0,1/2,i/2,1/2+i/2\}$, as well as the fixed locus $\{
z_1'=0,1/2,i/2,1/2+i/2\}$ of $\Theta\o\sigma$. Therefore, if we were
only considering D7-branes parallel to the {\rm O}7-planes at
$z_1'$ or $z_2'$ given above, we would have just recovered the
T-dual version of the standard $\ZZ_2$ orientifold by identifying
$z_1'=\hat x_1+i \hat x_3$ and $z_2'= \hat x_2 +i\hat x_4$. The
deformation we are going to introduce consists in allowing the
D7-branes wrap more general cycles in the K3. The former sLag
cycles, defined by \angle, i.e. $\varphi_1 = \varphi_2 = \varphi$,
become holomorphic cycles given by $z_1' = \tan (\varphi) z_2' +
c$ in the new complex structure. Any single such brane always
comes with its image under $\o\sigma$ at the opposite relative angle
$-\varphi$. As we will see, wrapping brane on invariant cycles,
symbolically $\pi_{\varphi}+\pi_{-\varphi}$, does have a nice lift to
F-theory.

We cannot redo the entire analysis of \rsenb\  but need
to collect a number of definitions and notations and the important results.
It was found that the orbifold was most directly identified with
the appropriate F-theory model after breaking the gauge symmetry down
to an $SU(2)^8\times SU(2)^8$ subgroup via turning on Higgs scalars.
We have shown in section 4.1 that this precisely matches the situation
where all the RR charges can be canceled locally.
Geometrically speaking this amounts to placing D7-branes pairwise on top
of each other, which are further identified under one of the $\Theta_{ij}$
with another pair. The holomorphic string coupling constant
$\lambda(u,v)=a+ie^{-\phi}$ enters
the model as the argument of the modular invariant $j$-function
\eqn\jfu{
j(\lambda(u,v)) = {4 (24 f)^3 \over 4f^3 + 27 g^2}
}
which defines the Weierstra\ss\ model
\eqn\weier{
y^2 = x^3 + f(u,v) x + g(u,v)
}
of the elliptic fiber. The functions $f(u,v)$ and $g(u,v)$, are
polynomials of degree $(8,8)$ and $(12,12)$. The vanishing locus
of the denominator $\Delta = 4f^3 + 27g^2$ defines the region on
the base, where the D7-branes and O7-planes are located. It is
required that  $j\rightarrow \infty$ nearly everywhere, in order
to have small string coupling. The coordinates $u,v$ parameterize
the base space $\IC\IP^1\times\IC\IP^1$. Further, $\tilde{u}_m,
\tilde{v}_m$ were used to denote the location of the {\rm
O}7-planes and $u_i,v_i$ that of the D7-branes of the original
model of \rsenb. There the D7-brane were only wrapping either one of
the two cycles $\pi_9 = [{\rm Fix}(\o\sigma)]$ and $\pi_5 = [{\rm
Fix}(\Theta\o\sigma)]$, where the indices are referring to the D5-
and D9-brane charges of the original type I model. Then, the
expected form of the discriminant for a configuration with the
given gauge group is
\eqn\deltaa{
\Delta = {P}_{(8,8)} \prod_{i=1}^8 (u-u_i)^2 (v-v_i)^2
}
with eight distinct $A_1$ singularities and an undetermined polynomial
of bi-degree $(8,8)$. A solution to this problem in terms of $f(u,v)$ and $g(u,v)$
could actually be found. It is given by
\eqn\poly{
{P}_{(8,8)} = 4C^3 \prod_{i=1}^8 (u-u_i)(v-v_i)  - 9C^2 h(u,v)^2 ,
}
with $h$ again defined by
\eqn\hpoly{
h(u,v) = \prod_{m=1}^4 (u-\tilde{u}_m) (v-\tilde{v}_m)
}
and $C$ a constant. This leads to
\eqn\jfub{
j(\lambda(u,v)) \sim \Delta^{-1} \left(
C \prod_{i=1}^8 (u-u_i) (v-v_i)  - 3h(u,v)^2 \right)^3 .
}
By scaling $C\rightarrow 0$ one can reach the small coupling
region of type IIB string theory. The degrees of the constituents
in $\Delta$ actually have a very intuitive interpretation: The
polynomial $P_{(8,8)}$ encodes the location of the $4+4$
O7-planes, each split off into two $A_1$ singularities around its
classical locus $h(u,v)=0$. This splitting of O-planes into
multiple singularities is due to the work of \refs{\SenVD,\BanksNJ},
an application of
the Seiberg-Witten curve. On the other hand, the remaining
polynomial of bi-degree $(16,16)$ encodes the location of $16+16$
D7-branes. The relative factor of 4 between the multiplicities is
made up for by the relative normalization of the charges of
O7-planes, $Q_7=-8$, and pairs of D7-branes. In this sense, the
degree of the polynomials directly relates to the charges.

Let us see, how the model should be modified in order to
incorporate D7-branes wrapping more general holomorphic cycles. In
analogy to the above, we put two branes on top of each other in each
stack and wrap them over cycles of the form
$\pi_a= r_a \pi_9 + s_a \pi_5$. Since these cycles lie on top
of the orientifold planes, supersymmetry is generically preserved and the
total space will be Calabi-Yau. The location
of the O7-planes is unaffected anyway. Thus, not only $h(u,v)$
should remain unchanged but the structure of $P_{(8,8)}$ as well.
For a general D7$_a$-brane wrapped on a cycle $\pi_a= r_a
\pi_9 + s_a \pi_5$ we now expect a factor $q_{(r_a,s_a)}$ in
$\Delta$ of degree $(r_a,s_a)$. The tadpole cancellation
condition \tadhom, translated into
\eqn\polytad{
\sum_a (2 r_a,2 s_a) = (16,16) ,
}
then insures, that $\Delta$ will be of the proper degree $(24,24)$
again,
\eqn\deltac{
\Delta = P_{(8,8)}' \prod_a q_{(r_a,s_a)}^2 .
}
The solution for $j(\lambda(u,v))$ in terms of appropriate $f$ and
$g$ is now be given completely analogously by just replacing the
factors for the D7-branes in
\jfub. Thus, we can obtain a small coupling region by scaling
$C\rightarrow 0$ as before, the vanishing locus of $\Delta$
referring to potentially smaller number of branes which intersect
in some pattern.

As a check, we can perform a similar analysis as was done in \rsenb.
Two stacks of D7-branes labeled by $a,b$ will intersect $\pi_a
\circ \pi_b = r_a s_b + r_b s_a$ times on the torus. Upon turning
on a scalar Higgs fields in the $({\bf 2}_a, {\bf 2}_b)$
representation at all of these intersections the gauge group is
broken from $SU(2)^2$ to the diagonal $SU(2)$. At each
intersection one neutral hypermultiplet survives as a modulus from
the ${\bf 2 \otimes 2 = 1 \oplus 3}$ decomposition. In the
F-theory interpretation this would be described by replacing
$q_{(r_a,s_a)}q_{(r_b,s_b)} \rightarrow q_{(r_a+r_b,s_a+s_b)}$.
Counting of the free coefficients in the three polynomials, up to
an overall rescaling, before and after the symmetry breaking gives
\eqn\countcoeff{
\left( ( r_a + r_b +1
)( s_a + s_b +1 ) -1 \right) - \left( ( r_a +1 )( s_a +1 ) -1 +(
r_b +1 )( s_b +1 ) -1  \right) = \pi_a \circ \pi_b ,
}
the expected result. In order to get a geometric intuition it is
instructive to look at the most simple example $r_a=s_a=r_b=1,
s_b=0$. In this case we have
\eqn\ployexa{
q_{(1,1)} q_{(1,0)}
\sim (uv + a_1 v + a_2 u + a_3)(u + b_1) = (v+ a(u))
(u+a_1)(u+b_1) = 0 .
}
This describes two stacks of D7-branes, the intersection point
given by $(v=-a(-b_1),u=-b_1)$. This intersection locus is being
resolved by adding the new parameter,
\eqn\polyexb{
q_{(2,1)} \sim (v+ a(u)) (u+a_1)(u+b_1) + c_1 =0 .
}
In a very similar fashion the location of the orientifold planes
gets smoothed out by turning on parameters in $h(u,v)=0$, i.e.
deforming like
\eqn\oplanesmooth{
(u-\tilde u_m)(v-\tilde v_m)=0 \quad  \rightarrow \quad (u-\tilde
u_m)(v-\tilde v_m) + \tilde a_1=0 .
}
It is interesting to note that this geometric realization in
F-theory suggests that the D7-branes are qualitatively unaffected by the
resolution of the orientifold singularities, and vice versa the
O7-planes by the resolution of intersections. This provides
confidence to trust the perturbative CFT analysis at the orbifold
point also after slightly deforming away from it. 

\bigno
{\it 4.7.2. F-theory on Voisin-Borcea Calabi-Yau 3-folds}
\bigno

The class of Voisin-Borcea Calabi-Yau 3-folds  \refs{\rv,\rb}\ provides
F-theory backgrounds, for which the spectrum and gauge group of the
low energy theory have been computed explicitly \MorrisonPP.
Furthermore, these
models have partly been interpreted as type I vacua
\refs{\rbluma,\DabholkarKA}.
In principle, they can be generalized orientifold vacua, where the coupling
constant may vary on the internal space according to the elliptic
fibration. In this section we point out that all the Voisin-Borcea
3-folds satisfy the topological tadpole relation \rela.

The most essential informations we are going to need have been given
in \MorrisonPP, which we refer to for the proper definitions and more details.
In general, Voisin-Borcea 3-folds are defined by taking a quotient
${\cal M}^6 = ({\rm K3}\times T^2)/\sigma$. Here, $\sigma$ reflects the
holomorphic 2-form of K3 and acts as a reflection on $T^2$,
\eqn\VBsigma{
\sigma\, :\, (\Omega_2,dz) \mapsto (-\Omega_2,-dz) ,
}
$z$ a coordinate on $T^2$. If this action has fixed points, they need to
be resolved in order to arrive at a smooth manifold with $SU(3)$
holonomy. This manifold is then elliptically fibered over
${\cal B} = {\rm K3}/\sigma$. All these spaces are classified in terms
of involutions of K3 surfaces \rnik, and are usually denoted by three
integers $(r,a,\delta)$. After the hyperk\"ahler rotation of the
complex structure $\sigma$ becomes a complex conjugation on the K3, which
makes this set of models accessible to our methods.
In order to check \rela\ we then need the number
of tensor multiplets in the effective six-dimensional theory and
the self-intersection of the fixed locus of $\sigma$.
According to \MorrisonPP, we first have
$n_{\rm T} = h^{(1,1)}({\cal B})-1$.

Let us for the moment restrict to all cases with
$(r,a,\delta) \not= (10,10,0)$ or
$(10,8,0)$. Excluding the two pathological choices, the fixed locus
Fix$(\sigma)\subset {\cal B}$ consists of a curve of genus $g=(22-r-a)/2$
and $k=(r-a)/2$ rational curves. Using \slagintersectionIndex, we can now
easily compute the self-intersection of
$\pi_{\rm O7} = [{\rm Fix}(\sigma)]$,
\eqn\VBselfint{
\pi_{\rm O7} \circ \pi_{\rm O7} = -\chi ( {\rm Fix}(\sigma) )
= 2g-2-2k = 2(10-r) .
}
This is related to $c_1^2({\cal B}) = 10-r$, which via
$c^2_1({\cal B}) + c_2({\cal B}) = 12$ allows to deduce
$h^{(1,1)}({\cal B}) = r$
and, thus, $r= n_{\rm T}+1$. Hence, all the cases considered
satisfy the anomaly cancellation condition derived from the chiral
fermion spectrum of table 1. Although it was explicitly checked that they are
free of irreducible gravitational anomalies, it is still
not automatic that table 1 applies. In principle, it is not even clear, if
all the F-theory compactifications on Voisin-Borcea 3-folds
have an interpretation as a six-dimensional orientifold.
And if so, the F-theory description may involve non-perturbative
degrees of freedom, which are beyond any CFT description.
In this sense the above analysis can be interpreted
as a confirmation that the chiral spectrum of table 1 is indeed valid
throughout the moduli space.

The remaining two cases $(r,a,\delta)=(10,10,0)$ and $(10,8,0)$ have been
argued to be related to certain orbifolds $T^6/ \ZZ_2\times\ZZ_2$.
For the first choice of $(r,a,\delta)$, $\sigma$ is freely acting.
Hence, the two generators $\Theta_1$ and $\Theta_2$ need to be reflections,
defined by $v_I=(1/2,-1/2,0)$ and $v_I=(0,1/2,-1/2)$, combined with
shifts of order 2 along $z_3$ and $z_1$ respectively. Then,
${\rm Fix}(\sigma)$ is empty and $n_{\rm T}=21-n_{\rm H}=9$.
This model has the same spectrum as the Fermat quartic equipped with
the involution $\o\sigma$, which was shown to be free as well in section 4.6.

For the second case, only one of the two generators is combined with a shift,
the other one given by the reflection. It has 16 fixed $T^2$, which
are identified pairwise by the free generator. Due to $b_1(T^2)=2$,
each contributes two scalars which form a hypermultiplet together
with their axionic partners. The singularities corresponding to the fixed
tori are of type $A_1$, leading to a gauge group $U(1)^8$.
Finally, all the tori are parallel, such that
$\pi_{\rm O7} \circ \pi_{\rm O7} =0$, consistent with $n_{\rm T}=9,\,
n_{\rm H}=12+8=20$. Still, this model requires D7-branes to cancel the
RR charge of the 16 O7-planes. To summarize, all the F-theory vacua on
Voisin-Borcea Calabi-Yau 3-fold can consistently be described as generalized
six-dimensional orientifolds with intersecting D7-branes, according to the
proposed chiral spectrum of table 1.

\newsec{Intersecting Brane Worlds in Four Dimensions}

In this section we generalize the results from the previous
section  to intersecting brane worlds on orientifolds of
Calabi-Yau 3-folds.

\subsec{Supersymmetric type II compactifications with D-branes}

The starting point is the type IIA superstring compactified on a
Calabi-Yau space ${\cal M}^6$. The non-trivial Hodge numbers of
${\cal M}^6$ are denoted by $h^{(1,1)}$ and $h^{(2,1)}$. As it is
well known, the closed string sector in four dimensions is built
by ${\cal N}=2$ supergravity coupled to $h^{(1,1)}$ vector
multiplets with gauge group $U(1)^{h^{(1,1)}+1}$. The bosonic
degrees of freedom of the vector multiplets are given by the
$h^{(1,1)}$ complex K\"ahler fields $T^k$, containing the K\"ahler
deformations of the internal metric $g_{ij}$ plus the internal
antisymmetric NSNS fields $b_{ij}$, as well as by $h^{(1,1)}$ RR
vector fields from the 3-form potential $C_{\mu ij}$. Second,
there arise $h^{(2,1)}+1$ neutral hypermultiplets. More
specifically, $h^{(2,1)}$ hypermultiplets are built from
$h^{(2,1)}$ complex structure moduli fields $U^i$ of $g_{ij}$ plus
$h^{(2,1)}$ further complex scalar fields from the RR 3-form
potential $C_{ijk}$. The universal hypermultiplet consists out of
the dilaton $\phi$, the $b_{\mu\nu}$ 2-form plus two RR scalars
from $C_{ijk}$.

One important issue  about  matter coupled to (ungauged) ${\cal N}=2$
supergravity is that the moduli space is a product space of the form
\eqn\modulspace{
{\cal M}_{\rm Moduli}={\cal M}_{\rm V}\lbrack h^{(1,1)}\rbrack
\otimes {\cal M}_{\rm H}\lbrack h^{(2,1)}+1\rbrack \, .
}
Due to the special K\"ahler property, the K\"ahler
potential of ${\cal M}_{\rm V}$ can be expressed as
\eqn\kpt{
{\cal K}(T,\bar T)
=-\log\Biggl( i\bar Y^K(\bar T)G_K(T)-iY^K(T)\bar G_K(\bar T)\biggr)\, ,}
where the $Y^K$ ($K=0,\dots ,h^{(1,1)}$)
are homogeneous coordinates of ${\cal M}_{\rm V}$ and, at least in
a certain symplectic basis, the $G_K$ are the first derivatives of a
holomorphic prepotential $G(Y)$, i.e. $G_K=\partial_KG(Y)$.
In the so-called special gauge the K\"ahler moduli
of ${\cal M}^6$ can be simply defined as $T^k=Y^k/Y^0$ ($k=1,\dots ,h^{(1,1)}$).
The holomorphic prepotential is determined by the classical
geometry of the Calabi-Yau space ${\cal M}^6$, namely by the cubic intersection
numbers $C_{KLM}$,  in addition to all world-sheet, rational (closed string)
instantons. However, there are no perturbative string
corrections to ${\cal M}_{\rm V}$, since the dilaton belongs to a hypermultiplet.

The couplings and self-interactions of the $h^{(2,1)}+1$ hypermultiplets
are determined by the quaternionic space
${\cal M}_{\rm H}$  of (real) dimension
$4h^{(2,1)}+4$. In general its structure is very hard to determine, as
${\cal M}_{\rm H}$ receives string loop corrections (but there are
no contributions due to world sheet instantons).
However, at least classically, ${\cal M}_{\rm H}$ contains a special K\"ahler
manifold $\widetilde{\cal  M}_{\rm H}$ of complex dimension $h^{(2,1)}$ as
a subspace, as it can be seen from the so-called c-map
(time-like T-duality to IIB on the same Calabi-Yau space ${\cal M}^6$).
This space $\widetilde{\cal M}_{\rm H}$ is parameterized by the complex structure
moduli fields $U^i$ ($i=1,\dots ,h^{(2,1)})$ of ${\cal M}^6$. Being
special K\"ahler, the same formalism as above applies for
$\widetilde{\cal M}_{\rm H}$, i.e. there exist a holomorphic prepotential
$F(X)$ with corresponding K\"ahler potential
\eqn\kpu{
{\cal K}(U,\bar U)
=-\log\Biggl( i\bar X^I(\bar U)F_I(U)-iX^I(U)\bar F_I(\bar U)\biggr)\, ,
}
with homogeneous fields $X^I$ ($I=0,\dots , h^{(2,1)}$) and first
derivatives $F_I=\partial_IF(X)$.
As before, in the special K\"ahler gauge the complex structure moduli
are given as $U^i=X^i/X^0$.
It is useful to
define an integral basis
for $H_3({\cal M}^6,{\ZZ})$, given by $(\alpha^I,\beta_J)$
($I,J=0,\dots ,h^{(2,1)}$) with the property
$\alpha^I\circ\alpha^J=\beta_I\circ\beta_J=0$ and
$\alpha^I\circ\beta_J=\delta^I_J$.
Sometimes the $\alpha^I$ are called electric cycles, and the $\beta_I$ are
called dual magnetic cycles, however this is clearly a basis dependent
statement.
Using the holomorphic 3-form $\Omega_{3}$,
one can define the period integrals
\eqn\periods{
X^I=\int_{\alpha^I}\Omega_{3},\quad
  F_I=\int_{\beta_I}\Omega_{3}\, 
}
and similarly the normalized periods
\eqn\periodsn{
\widehat X^I=\int_{\alpha^I}\widehat\Omega_{3},\quad
  \widehat F_I=\int_{\beta_I}\widehat\Omega_{3}.
}
In fact, the periods \periods\ just correspond to the homogeneous
coordinates and the
derivatives of the prepotential
of the special K\"ahler geometry
introduced before.
One can expand every homology 3-cycle $\pi_a$ in terms of the
basis cycles $\alpha^I$ and $\beta_J$,
\eqn\expand{
\pi_a=e^a_I\alpha^I+m_a^J\beta_J\, ,
}
where the $e^a_I$, $m_a^J$ are integer expansion coefficients.

If we would just consider intersecting D6-branes on the Calabi-Yau
manifold, this information would be enough to compute for instance
the massless spectrum of chiral fermions. Such models do have
${\cal N}=2$ supersymmetry in the bulk and the open string sector
does always break supersymmetry completely, which can be seen from
the fact that  D6-branes always have positive tension which cannot
be canceled. If we are interested in models which can preserve an
${\cal N}=1$ supersymmetry both in the closed and the open string
sector, we have to take an orientifold of the Calabi-Yau where now
the world-sheet parity transformation is combined with an
anti-holomorphic involution $\o\sigma$. Note, that combining
$\Omega$ with a  holomorphic involution  is not a symmetry of the
type IIA string. In fact, $\Omega\o\sigma$ breaks the bulk
supersymmetry in the closed string sector from ${\cal N}=2$ to
${\cal N}=1$. This implies that all bulk ${\cal N}=2$ superfields
are truncated to ${\cal N}=1$ superfields by the $\Omega\o\sigma$
projection. As discussed in section 2.2. there will be $h^{(2,1)}$
complex moduli
$U_i$, each consisting out of a real complex structure modulus and
a linear combination of RR scalars, plus $h^{(1,1)}-a$ complexified K\"ahler
moduli $T_k$, where $a$ is the number of the K\"ahler fields not invariant
under $\Omega\o\sigma$. In addition there will be in general $s_l$
($l=1,\dots ,b$) open string moduli describing the positions of the D-branes.
Due to the presence of these fields we do not expect that the total,
($h^{(1,1)}+h^{(2,1)}+b-a$)-dimensional moduli space has a direct
product structure.

The O6-planes are located at the fixed point set of the anti-holomorphic
involution, a sLag 3-cycle in the Calabi-Yau manifold.
The tadpoles induced by the O6-plane can be canceled by the D6-branes
and the positive and negative tensions can potentially compensate each other.
In the case that all D6-branes wrap sLag cycles calibrated with
respect to the same 3-form as the orientifold plane is, i.e.
$\Re(\Omega_3)$, the background preserves ${\cal N}=1$
supersymmetry. It has been established that for such branes on
sLag cycles, so-called A-type branes, the F- and D-terms in the
effective action receive the following contributions \refs{\rbdlr,\rkklmb}:
The F-terms, i.e. the superpotential, only depends
    on the K\"ahler moduli, getting contributions only from disc
     and $\IR\IP^2$ instantons.
The D-terms, including the Fayet-Iliopoulos term, only depends
                  on the complex structure moduli.
In section 2 we have determined the disc level scalar potential
and have seen explicitly that it only depends on the complex
structure moduli. Therefore, it is clear, that in the effective
theory this term appears as a Fayet-Iliopoulos term.
D-flatness then imposes conditions on
some of the complex structure moduli, freezing them dynamically.
Under small changes of the K\"ahler moduli the D-term clearly
still vanishes, but the F-term may not, as it gets contributions
from disc and $\IR\IP^2$ instantons. We will come back to this
point later.

If such type IIA vacua with configurations of D6-branes and
O6-planes do preserve ${\cal N}=1$
supersymmetry, one expects that there exists a lift
of the type IIA model to M-theory on a singular seven-dimensional
compact $G_2$ manifold \refs{\Jbook,\rkachmcb,\rcvetica,\rcveticb}.
In fact not very much is known about the explicit construction of such
manifolds, but one can consider the construction of supersymmetric
intersecting brane world models of D6-branes and O6-planes
as an implicit way of determining the
expected properties of such singular manifolds.

\subsec{Chiral fermion spectrum}

In order to compute intersection numbers involving the O6-planes
it is necessary that we choose generators (not necessarily a
basis) for $H_3({\cal M}^6,\ZZ)$, which allows us to determine
both  the homology class $\pi_{{\rm O}6}$ of the orientifold plane
and the action of $\Omega\o\sigma$ on this basis. As in the K3 case,
blown-up orbifold models do naturally give rise to such a basis by
combining the inherited toroidal and the exceptional cycles. The
intersection matrix for the 3-cycles is anti-symmetric, so that
all self-intersection numbers vanish. Since $Q_6=-4$, the
RR-tadpole cancellation condition
\tadhom\ reads
\eqn\tadpolec{
\sum_a  N_a\, ( \pi_a + \pi'_a) =
          4\,   \pi_{{\rm O}6} .
}
The gauge group in the generic case, when no D6-brane is invariant
under $\Omega\o\sigma$, is again given by \gauge. The spectrum of
left-handed massless chiral fermions is shown in table 9. It is
again motivated by computations at the orbifold limit and at large
volume, as well as by the general consideration on the topological
nature of the zero-modes of the Dirac operator.
\vskip 0.8cm
\vbox{
\centerline{\vbox{
\hbox{\vbox{\offinterlineskip
\def\tablespace{height2pt&\omit&&
 \omit&\cr}
\def\tablerule{\tablespace\noalign{\hrule}\tablespace}

\hrule\halign{&\vrule#&\strut\hskip0.2cm\hfill #\hfill\hskip0.2cm\cr
& Representation  && Multiplicity &\cr
\tablerule
& $[{\bf A_a}]_{L}$  && ${1\over 2}\left(\pi'_a\circ \pi_a+\pi_{{\rm O}6} \circ \pi_a\right)$   &\cr
\tablerule
& $[{\bf S_a}]_{L}$
     && ${1\over 2}\left(\pi'_a\circ \pi_a-\pi_{{\rm O}6} \circ \pi_a\right)$   &\cr
\tablerule
& $[{\bf (\o N_a,N_b)}]_{L}$  && $\pi_a\circ \pi_{b}$   &\cr
\tablerule
& $[{\bf (N_a, N_b)}]_{L}$
&& $\pi'_a\circ \pi_{b}$   &\cr
}\hrule}}}}
\centerline{
\hbox{{\bf Table 9:}{\it ~~ Chiral spectrum in $d=4$}}}
}
\vskip 0.5cm
\noindent
We therefore claim, that table 9 holds for every smooth
six-dimensional background space including Calabi-Yau manifolds
and the six-dimensional torus \rbgkl. Note  that in contrast to
the six-dimensional case, there are no chiral fields in the
adjoint representation of $U(N_a)$, as the self-intersection
numbers vanish. Computing the non-abelian gauge anomaly
$SU(N_a)^3$ one finds
\eqn\anapp{
A_{\rm non-abelian} \sim \pi_a\circ \pi_a =0.
}
Thus, there is no independent check of the construction by
testing the resulting spectra for anomalies.

In order to avoid confusion we would like to add the following comment.
As for instance pointed out in \rbdlr, due to the generation of an
${\cal N}=1$ superpotential,
it is not clear whether D-branes can be followed smoothly
from the large radius regime to regions deep in the interior of the
K\"ahler moduli space. In view of this, we are surely not stating
that the spectrum is invariant all over the moduli space.
If we go to a point in moduli space where an exact conformal field
theory description is available, one can compute the overlap of
the corresponding boundary states respectively cross-cap states
\refs{\rrecknagel,\rbdlr} and use them to define a stringy version
of the intersection form. On the other hand, in the large volume
regime the methods of classical geometry are available. When
passing from one regime to another, the formerly stable
configurations may decay into new ones and the spectrum and gauge
group changes. But the important point is, whenever the setting is
describable by a selection of D-branes on cycles of middle
dimension and with flat gauge bundle, table 9 may apply.

\subsec{Background 3-form flux}

To close this section we comment on the possibility to turn on
background fluxes\footnote{$^3$}{Type II models with different
sources of RR flux in four dimensions have been discussed e.g. in
\refs{\deWitXG\PolchinskiSM\MichelsonPN\TaylorII\MayrHH\CurioSC\HaackJZ\rgkp\CurioAE\AgataZH\LouisUY\rkst\ruranga\LouisNY\TatarFlux-\BeckerNN}.}, as well.
Due to the couplings
\eqn\couplij{
\int_{\IR^{1,3}\times {\cal M}^6} H_{3}^{\rm NS}
\wedge  G_{6-2n} \wedge C_{2n+1}
}
in the ten-dimensional type IIA string theory, there can exist one
more source for the RR 7-form gauge field,
\eqn\source{
\int_{\IR^{1,3}\times {\cal M}^6} G_0\,  H_{3}^{\rm NS}    \wedge C_{7}.
}
Here, $H_{3}^{\rm NS}$ is the NSNS 3-form and $G_0$ the RR 0-form
field strength, which can be considered as a cosmological constant.
This coupling induces the term
\eqn\rrterm{
G_0\, [H_{3}^{\rm NS}]
}
on the left hand side of the RR-tadpole cancellation condition
\tadpolec, where $[H_{3}^{\rm NS}]$ denotes the Poincar\'e dual 3-cycle
of the cohomology class of $H_{3}^{\rm NS}$. Thus, by turning on a
non-vanishing NSNS 3-form flux, more general configurations of
intersecting D6-branes can satisfy the RR-tadpole cancellation
conditions. Anomaly cancellation in such settings has been
discussed in \ruranga, and the possibility that such
couplings allow a new solution of the strong $CP$ problem
has been pointed out in \AldazabalPY.
An important result of the analysis of the
anomaly inflow into the brane world volume, which is generated by
\couplij, concerns the chiral matter. In fact, the anomaly
cancellation in the presence of such background fluxes appears to
work out without adding extra contribution from chiral fermions on
the world volume of the D6-branes. Therefore, the spectrum of
table 9 may be assumed to cover these cases just as well.

However turning on $0\not=H^{NS}_3 \in H^3(X,\IR)$ will change
the theory drastically if one goes away from the large volume limit
where the supergravity description is valid. The NSNS $B$-field
should then be thought of as a connection on a non-trivial gerbe,
and this will modify the K-theory groups that classify the charges.
Especially if the characteristic
class of this gerbe is not torsion then there is no Chern isomorphism
with ordinary (co)homology any more. This means that the homology
class of a brane configuration
is no longer an invariant, as was first discovered
in the context of WZW models in \SchomerusWZW. 
This can be understood as D-brane instantons \MooreKinst, 
where the instanton background violates the usual
conservation of the D-brane homology class.

\newsec{The Green-Schwarz mechanism}

We have seen that the non-abelian gauge anomalies of all $SU(N_a)$
factors in the gauge group vanish, given the chiral spectrum of
table 9. On the other hand, the mixed anomalies, denoted
$U(1)_a-SU(N_b)^2$, naively do not and a generalized Green-Schwarz
mechanism has to be invoked. Computing the $U(1)_a-SU(N_b)^2$
anomaly in the effective four-dimensional gauge theory for $a\ne
b$ one finds
\eqn\mixeda{
A^{(1)}_{ab}={N_a \over 2}\left( -\pi_a+\pi'_a\right)\circ \pi_b .
}
For $a=b$ one gets
\eqn\mixedb{
A^{(1)}_{aa}=\left({N_a-2\over 2}\right)\, 2\, \pi_{{\rm O}6}\circ \pi_a +
              \left({N_a\over 2}\right)\, 2\, \left( \pi'_a\circ \pi_a
           -\pi_{{\rm O}6}\circ \pi_a \right) -{1\over 2} \sum_{b\ne a} N_a\,
            \pi_a\circ \left( \pi_b+\pi'_b\right) ,
}
where we have used the fact that the $U(1)$ charge of the
symmetric and anti-symmetric representations of $U(N)$ are $\pm 2$.
Using the tadpole cancellation condition the above expression can be simplified
to precisely the form \mixeda\ for $b=a$.

The Green-Schwarz mechanism works quite analogously as in
\rafiru. On each
D6$_a$-brane there exist Chern-Simons couplings of the form
\eqn\cscoupl{
\int_{\IR^{1,3}\times\pi_a}  C_3\wedge {\rm Tr}\left(F_a\wedge F_a\right) , \quad\quad
\int_{\IR^{1,3}\times\pi_a}  C_5\wedge {\rm Tr}\left(F_a\right)
}
where $F_a$ denotes the gauge field on the D$6_a$-brane.
Now we expand every 3-cycle
$\pi_a$ and $\pi'_a$  in the basis $(\alpha^I,\beta_J)$
\eqn\expam{\eqalign{
&\pi_a=e^a_I\, \alpha^I+m^J_a\, \beta_J , \cr
                       &\pi'_a=(e^a_I)'\, \alpha^I+(m^J_a)'\, \beta_J .\cr }}
Moreover, we define the four-dimensional axions $\Phi_I$ and
2-form $B^I$, $I=0,\, ...\, , h^{(2,1)}$, as
\eqn\formsl{\eqalign{ \Phi_I=\int_{\alpha^I} C_3, \quad\quad
                      \Phi^{I+h^{(2,1)}+1}=\int_{\beta_I} C_3, \cr
                      B^I=\int_{\beta_I} C_5, \quad\quad
                      B_{I+h^{(2,1)}+1}=\int_{\alpha^I} C_5 .}}
In four dimensions $(d\Phi_I, dB^I)$ and
$(d\Phi^{I+h^{(2,1)}+1},dB_{I+h^{(2,1)}+1})$ are Hodge dual to each other.
The general couplings \cscoupl\ can now be expressed as
\eqn\dddd{\eqalign{
\int_{\IR^{1,3}\times\pi_a}  C_3\wedge
       {\rm Tr}\left(F_a\wedge F_a\right)=&
\sum_I \left( e_a^I+(e_a^I)'\right) \int_{\IR^{1,3}} \Phi_I \wedge {\rm Tr}\left(F_a\wedge
                 F_a\right) \cr
&+\sum_I \left( m^a_I+(m^a_I)'\right) \int_{\IR^{1,3}}
\Phi^{I+h^{(2,1)}+1}
    \wedge {\rm Tr}\left(F_a\wedge F_a\right), \cr
\int_{\IR^{1,3}\times\pi_a}  C_5\wedge {\rm Tr}\left(F_a\right)
     =&N_a \sum_I \left( m^a_I-(m^a_I)'\right) \int_{\IR^{1,3}} B^I \wedge F_a \cr
&+N_a \sum_I \left( e_a^I-(e_a^I)'\right) \int_{\IR^{1,3}}
B_{I+h^{(2,1)}+1} \wedge F_a,}
 }
where we have used that the gauge field on the $\Omega\o\sigma$ mirror brane is
$F'_a=-F_a$.
The tree-level contribution to the mixed gauge anomaly described by these couplings
takes the form depicted in figure 5.
\fig{}{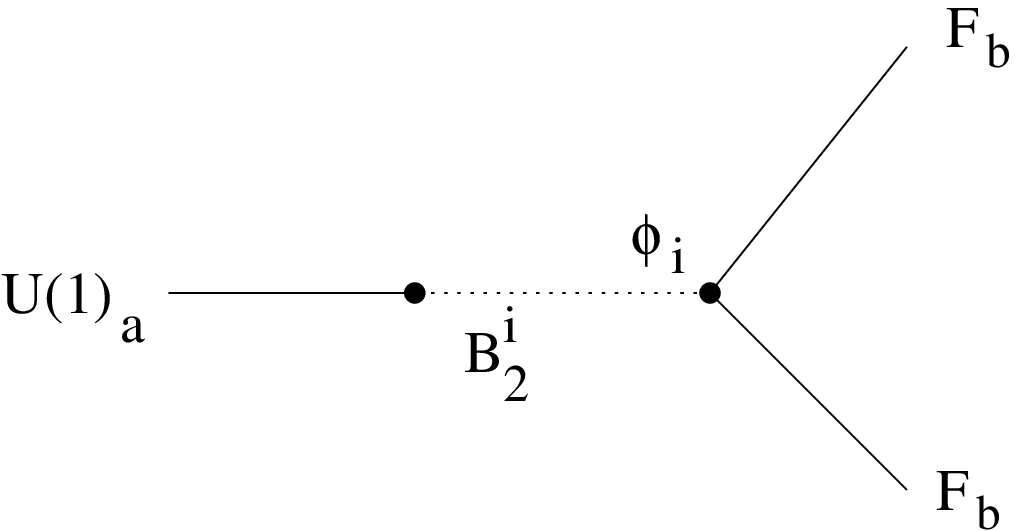}{8truecm}
\noindent
Adding up all terms for the  $U(1)_a-SU(N_b)^2$
anomaly
we get
\eqn\aadito{\eqalign{ A^{(2)}_{ab}&\sim N_a\, \sum_I \left(e_a^I+(e_a^I)'\right)
               \left(m^b_I-(m^b_I)'\right)+
                N_a \sum_I \left(m^a_I+(m^a_I)'\right)\left(e_b^I-(e_b^I)'\right)\cr
             &\sim 2\,N_a\, \left( \pi_a - \pi'_a\right)\circ\pi_b,} }
which has just the right form to cancel the anomalous contribution \mixeda\
of the chiral fermions.

\newsec{Examples on Calabi-Yau 3-folds}

Here, we collect examples of intersecting brane worlds in four
dimensions. They are given by orbifolds and the algebraic
realization of the quintic in $\IC\IP^4$.

\subsec{Preliminaries on Calabi-Yau orbifolds}

There is a certain number of orbifolds of $T^6$ preserving
supersymmetry in four dimensions. These have been classified in
\refs{\DixonJW,\DixonJC}.
The construction of orbifold examples is very similar to the 6D case.
Therefore, we restrict ourselves to the
study of only three examples, namely the $\ZZ_2\times \ZZ_2$, the
$\ZZ_3$ and the $\ZZ_4$ orbifolds. 
For the first two examples all 3-cycles are inherited from the $T^6$,
where for the $\ZZ_2\times \ZZ_2$ orbifold
one gets $h^{(2,1)}=3$ and for the $\ZZ_3$ orbifold $h^{(2,1)}=0$.

Such orbifolds  are defined by $v_I =
(1/N,1/N,-2/N)$ for the two $\ZZ_N$ and by
$v^{(1)}_I=(1/2,-1/2,0)$ and $v^{(2)}_I=(0,1/2,-1/2)$ for the two
generators of $\ZZ_2^2$. Note, that the ordinary $\Omega$ orientifolds
of type I string theory, e.g. discussed in
\refs{\BerkoozDW,\rstanev,\AldazabalMR,\KakushadzeEG,\AngelantonjCT},
are T-dual to $\Omega\o\sigma$ orientifolds
of asymmetric orbifolds and are not considered in this paper.

We shall employ completely analogous conventions as in the case of
six-dimensional models to denote the orbifold backgrounds
\refs{\rbgkc,\rfhs,\rfhstwo}. The
background tori $T^6$ for the groups $\ZZ^2_2$ and $\ZZ_4$ will be
defined by a factorized $SU(2)^6$ lattice, while the $\ZZ_3$ torus
is given by the $SU(3)^3$ lattice. On any single $T^2_I$ there can
be a distinction of two in general inequivalent models according
to the complex structures $\tau_{\bf A}$ and $\tau_{\bf B}$. This
results in a total of six models, $\{
{\bf AAA, ABA, BBA, AAB, ABB, BBB}\}$, for the $\ZZ_N$ and the
four $\{ {\bf AAA, AAB, ABB, BBB}\}$ for $\ZZ_2^2$. Actually, we
are not going to elaborate all the examples in detail, but simply
look for special cases to illustrate the construction. We again
pick a basis of 3-cycles on the torus by tensoring the 1-cycles
$\gamma_i$, which are defined by the basis vectors ${\bf e}_i^I$
of the {\bf A} type elementary cells. As in \torusbasis\ we define
\eqn\basisb{
\o\pi_{ijk} = \gamma_i \otimes \gamma_j \otimes \gamma_k.
}
The 3-cycles on the orbifold are then given by taking the
independent orbits of the $\o\pi_{ijk}$ under the orbifold
generators.

At the orbifold point, where the complete CFT description is available,
not only the D-term but also the F-term vanishes
in the case of supersymmetric configurations. However, blowing up
the orbifold singularities might not be a flat direction of the exact
potential, as new
non-trivial discs and $\IR\IP^2$ appear in the background
which might lead to non-vanishing contributions to the superpotential via
open string world sheet instantons. We shall discuss the corrections to the
superpotential and the D-terms later in some more detail.

\subsec{The orbifold $T^6/\ZZ_2\times \ZZ_2$}

In this case the number of 3-cycles is given by $b_3=2+2 h^{(2,1)}=8$.
They can be represented by
$\{\pi_{135},\pi_{246},\pi_{245},\pi_{136},\pi_{236},\pi_{145},\pi_{146},\pi_{235}\}$,
which have the intersection form
\eqn\inty{
I^{\rm Torus}_{T^4/\ZZ_2\times \ZZ_2}=\bigoplus_{k=1}^4 \left(\matrix{ 0 & 4 \cr
                                            -4 & 0 \cr}\right)_k .
}
For the four different orientifolds the homology classes of the
O6-planes are summarized in table 10.
\vskip 0.8cm
\vbox{
\centerline{\vbox{
\hbox{\vbox{\offinterlineskip
\def\tablespace{height2pt&\omit&&
 \omit&\cr}
\def\tablerule{\tablespace\noalign{\hrule}\tablespace}

\hrule\halign{&\vrule#&\strut\hskip0.2cm\hfill #\hfill\hskip0.2cm\cr
& Complex structure  && $\pi_{{\rm O}6}$ &\cr
\tablerule
& {\bf AAA} && $2\left( \pi_{135}+\pi_{245}+\pi_{146}+\pi_{236} \right)$ &\cr
\tablerule
& {\bf AAB} && $ \pi_{135}+\pi_{246}+\pi_{136}+\pi_{245}+\pi_{235}-\pi_{146}+
             \pi_{145}-\pi_{236}$  &\cr
\tablerule
& {\bf ABB} && $\pi_{135}-\pi_{246}+\pi_{235}+\pi_{146}$ &\cr
\tablerule
& {\bf BBB} && $\pi_{135}+\pi_{246}$ &\cr
}\hrule}}}}
\centerline{
\hbox{{\bf Table 10:}{\it ~~ $T^6/\ZZ_2\times \ZZ_2$ orientifolds }}}
}
\vskip 0.5cm
\noindent
The action of $\o\sigma$ on the basis for the homology lattice is
evident, regarding earlier results for the six-dimensional
$T^4/\ZZ_2$ orbifold of section 4.1. Using these data one can
construct a variety of intersecting brane world models. A couple
of explicit examples including a quasi-realistic three-generation
supersymmetric Standard Model can be found in
\refs{\rcvetica,\rcveticb,\rcls}. There, the computation was performed in the orbifold
limit and the spectra were computed by some effort in solving
constraints for the Chan-Paton matrices. Actually, the results for
the massless spectra precisely match with table 9.

\subsec{The orbifold $T^6/\ZZ_3$}

The orbifold $T^6/\ZZ_3$ has Hodge numbers $(h^{(1,1)}, h^{(2,1)})=(36,0)$,
so that the number of 3-cycle is $b_3=2$.
A basis is given by
\eqn\threecycl{\eqalign{
&\pi_1=  \o\pi_{136}+\o\pi_{145}+\o\pi_{235}-
                                    \o\pi_{146}+\o\pi_{236}+\o\pi_{245} , \cr
                          &\pi_2=  \o\pi_{135}+\o\pi_{246}-\o\pi_{136}-
                                    \o\pi_{145}+\o\pi_{235} . \cr}}
It leads to the intersection matrix
\eqn\inti{
I^{\rm Torus}_{T^6/\ZZ_3}=\left(\matrix{ 0 & 1    \cr
                               -1 & 0   \cr }\right).
}
For the choice ${\bf AAA}$ of the complex structure the homology class
of the orientifold plane is
\eqn\oripl{
\pi_{{\rm O}6}=\pi_1 ,
}
and the action of $[\o\sigma]_{\bf AAA}$ on $H_3(T^6/\ZZ_3,\ZZ)$ reads
\eqn\inti{
[\o\sigma]_{\bf AAA}=\left(\matrix{  1& 0   \cr
                                    1 & -1   \cr  } \right) .
}
We skip the computation for the other choices of complex
structure. Instead, we show that one can indeed recover the 3
generation $SU(5)$ GUT model found in \rott\ as a $T^6/\ZZ_3$
orbifold compactification. Choosing $N_1=5$ and $N_2=1$ D6-branes
wrapped on the cycles
\eqn\zetdrei{  \eqalign{
N_1=5:\quad\quad
           \eta_1&= -\pi_1+3\pi_2 , \cr
           N_1=5:\quad\quad
           \eta'_1&=  2\pi_1-3\pi_2 , \cr
          N_2=1:\quad\quad
           \eta_2&= -2\pi_1+3\pi_2 , \cr
          N_2=1:\quad\quad
           \eta'_2&= \pi_1-3\pi_2 . \cr}
}
These branes support a gauge group $U(5)\times U(1)$. Computing
the relevant intersection numbers we obtain the chiral spectrum
\vskip 0.8cm
\vbox{
\centerline{\vbox{
\hbox{\vbox{\offinterlineskip
\def\tablespace{height1pt&\omit&&\omit&\cr}
\def\tablerule{\tablespace\noalign{\hrule}\tablespace}

\hrule\halign{&\vrule#&\strut\hskip0.2cm\hfill #\hfill\hskip0.2cm\cr
& Representation && Multiplicity &\cr
\tablerule
& $[({\bf 10,0})+({\bf 1, -2})]_{L}$ && 3 &\cr
& $[({\bf \o 5,1})]_{L}$ && 3 &\cr
}\hrule}}}}
\centerline{
\hbox{{\bf Table 11:}{\it ~~ $\ZZ_3$ orientifolds }}}}
\vskip 0.5cm
\noindent
which agrees with the spectrum found in \rott. There, the gauge
symmetry breaking patterns, the perturbative stability of the
model and some phenomenological implications have been studied.

\subsec{The orbifold $T^6/\ZZ_4$}

In this case, there are four 3-cycles inherited from the
$T^6$ corresponding to one untwisted complex structure deformation in  
the orbifold model. A basis is given by
\eqn\cyclesb{\eqalign{
&\pi_1=  2\,\o\pi_{135}-
    2\,\o\pi_{245} , \cr
     &\pi_2=  2\,\o\pi_{136}-2\, \o\pi_{246} , \cr
    &\pi_3=  2\,\o\pi_{145}+
    2\,\o\pi_{235} , \cr
     &\pi_4=  2\,\o\pi_{146}+2\, \o\pi_{236} , \cr }
}
with the intersection matrix
\eqn\intya{
I^{\rm Torus}_{T^6/\ZZ_4}=\bigoplus_{k=1}^2 \left(\matrix{ 0 & -2 \cr
                                            2 & 0 \cr}\right)_k .
}
Let us just present one example for an ${\bf ABA}$ model, which
has a supersymmetric groundstate. In this case, the O6-plane is
wrapping the 3-cycle
\eqn\oyuj{
\pi_{{\rm O}6}=2\left( \pi_1+\pi_2+\pi_3-\pi_4 \right) .
}
Introducing the four stacks of D6-branes
\eqn\branesyu{\eqalign{  N_1=2:\quad\quad
           \eta_1&=3\left( \pi_1+\pi_2 \right) -\left(\pi_3+\pi_4 \right) , \cr
           N_1=2:\quad\quad
           \eta'_1&=3\left( \pi_3-\pi_4 \right) -\left(\pi_1-\pi_2 \right) , \cr
          N_2=2:\quad\quad
           \eta_2&=\pi_1+\pi_3 , \cr
          N_2=2:\quad\quad
           \eta'_2&=\pi_1+\pi_3 \cr}}
cancels the RR-tadpole. the gauge group given by $U(2)^2$.
Computing the intersection numbers leads to the spectrum
\vskip 0.8cm
\vbox{
\centerline{\vbox{
\hbox{\vbox{\offinterlineskip
\def\tablespace{height1pt&\omit&&\omit&\cr}
\def\tablerule{\tablespace\noalign{\hrule}\tablespace}

\hrule\halign{&\vrule#&\strut\hskip0.2cm\hfill #\hfill\hskip0.2cm\cr
& Representation && Multiplicity &\cr
\tablerule
& $[({\bf A,1})]_{L}$ && 16 &\cr
& $[({\bf S,1})]_{L}$ && 8 &\cr
& $[({\bf \o{2}},{\bf 2})]_{L}+[({\bf \o{2}},{\bf \o{2}})]_{L}$ && 4 &\cr
}\hrule}}}}
\centerline{
\hbox{{\bf Table 12:}{\it ~~ $\ZZ_4$ orientifolds }}}}
\vskip 0.5cm
\noindent
which is, of course, free of non-abelian gauge anomalies.
We now fix the complex structures of the first two $T^2_I$ and allow
$\Im(\tau^3)=U$ to vary. At the orbifold point, the scalar potential reads
\eqn\tadp{
{\cal V}= T_6 e^{-\phi_4}\left[ 16\, \sqrt{2\, U}+16\, \sqrt{10\left(U+{1\over
U}\right)}-
           32\sqrt{2}\left(\sqrt{U}+{1\over \sqrt{U}}\right) \right]
}
which can be shown to be positive, ${\cal V}\ge 0$. It vanishes
for $U=1/2$. Thus, we have found another ${\cal N}=1$
supersymmetric intersecting brane world model which should lift in
M-theory to a $G_2$ manifold.

\subsec{The Quintic}

As an example of a Calabi-Yau manifold that is not a torus orbifold
consider the Fermat quintic $Q$
\eqn\fermatquintic{
  \sum_{i=1}^{5} z_i^5 = 0
  \quad \subset \IC \IP^4
}
together with the obvious involution from the complex conjugation
of the coordinates $z_i\to \o{z}_i$. The fixed points are the real
quintic $\sum_{i=1}^{5} x_i^5 = 0 \subset \IR\IP^4$. Topologically
this is a sLag $\IR\IP^3$.

Now the Fermat quintic has $\ZZ_5^5$ acting on it via
\eqn\quinticaction{
  z_i \mapsto \omega^{k_i} z_i
  \qquad \omega=e^{2\pi i\over 5},~k_i\in \ZZ_5
}
The diagonal $\ZZ_5$ acts trivially, therefore we get $5^4=625$
different minimal $\IR\IP^3$s
\eqn\quinticRPs{
  \left|k_2,k_3,k_4,k_5\right>
  \mathrel{\buildrel{\rm def}\over=}
  \Big\{
  [x_1:\omega^{k_2}x_2:\omega^{k_3}x_3:\omega^{k_4}x_4:\omega^{k_5}x_5]
  \Big| x_i\in \IR,~\sum_{i=1}^5 x_i^5=0
  \Big\}
}
The intersection number of $\left|1,1,1,1\right>$ and
$\left|k_2,k_3,k_4,k_5\right>$ was determined in \rbdlr, it can be
written as the coefficient of the monomial
$g_2^{k_2}g_3^{k_3}g_4^{k_4}g_5^{k_5}$ in
\eqn\quinticIntersect{
  I_{\IR\IP^3} =
  \prod_{i=1}^5 \big( g_i + g_i^2 - g_i^3 - g_i^4 \big)
  ~\in
  \ZZ[g_1,g_2,g_3,g_4,g_5]
  \Big/
  \big< g_i^5=1, \prod g_i=1 \big>
}
All the other intersection numbers are then determined by the
$\ZZ_5^4$ symmetry. The ensuing intersection matrix
$M\in {\rm Mat}(625,\ZZ)$ has rank $204=b_3$, so the
$\left|k_2,k_3,k_4,k_5\right>$ generate
the full $H_3(Q;\IR)$.

Of course, not all those minimal $\IR\IP^3$ are $\Re(\Omega_3)$ calibrated,
but rather $\Re (\omega^k \Omega_3)$ calibrated. Indeed the
$\Re(\Omega_3)$ calibrated sLag cycles $L$ cannot span $H_3(Q)$ since
$\Im(\Omega_3)|_L = 0$ $\Rightarrow$
$L$ is orthogonal to ${\rm PD}(\Im (\Omega_3))$. To determine the
$\Re(\Omega_3)$ sLags we need to know how $\ZZ_5^4$ acts on the
holomorphic volume form. This is clear from the residue formula
\eqn\residueformula{
  2 \pi i \Omega_3 =
  \oint_\Gamma
  { {
    \epsilon^{i_1 \cdots i_5} z_{i_1}
    {\rm d}z_{i_2}\wedge {\rm d}z_{i_3}\wedge
    {\rm d}z_{i_4}\wedge {\rm d}z_{i_5}
  } \over {
    \sum_{i=1}^5 z_i^5
  } }
}
So $\Omega_3$ transforms as $\Omega_3 \mapsto (\prod_i\omega^{k_i}) \Omega_3$,
and the $\Re(\Omega_3)$ calibrated $\IR\IP^3$s are the
$\left|k_2,k_3,k_4,k_5\right>$ with $\sum_{i=2}^5 k_i = 0~{\rm mod}~5$.
There are ${1\over 5} 5^4=125$ such sLags, and using the intersection
matrix one can check that they generate a $101$-dimensional subspace of
$H_3(Q)$. Furthermore those $125$ cycles have vanishing intersection
numbers among themselves, so by wrapping branes on these sLag cycles
one cannot obtain chiral fermions.

Even without chiral fermions, is it at least possible to
find supersymmetric tadpole-free models besides the one with all branes
sitting on top of the orientifold plane? First consider the RR tadpole.
Since the $125$ sLag $\IR\IP^3$s span only a $101$-dimensional
subspace of $H_3(Q)$ there are $24$ relations between them. This yields a
$24$ dimensional family of brane configurations that formally cancel the
RR tadpole \tadhom. But even if the overall homology class adds up
correctly we have to make sure that we use only branes, and no
antibranes (whether some cycle is a brane or antibrane is determined
by its orientation relative to the orientation induced by the calibration).
So in the $24$ dimensional space of relations we may
only use the positive coefficients if we want to preserve supersymmetry.
It can easily be seen that there is no solution besides the
zero relation (for example note that the coefficients in the relations
add up to $0$).

Secondly let us consider the NSNS tadpole.
For this we need to know the volume of some linear combination of
the sLags. But because we got all the $\IR\IP^3$s
by acting with a symmetry they all have the same volume (say, $1$):
\eqn\quinticRPvol{
  {\rm Vol}\left(
    \sum_{a=1}^{125} n_a
    \big|k_2^{(a)},k_3^{(a)},k_4^{(a)},k_5^{(a)}\big>
  \right)
  = \sum_{a=1}^{125} | n_a |
}
So let $\sum \tilde{n}_a
\left|k_2^{(a)},k_3^{(a)},k_4^{(a)},k_5^{(a)}\right>$ be an
arbitrary (nonzero) relation, i.e. homologous to zero, and let
$\tilde{n}_1$ be the coefficient of $\left|1,1,1,1\right>$. Then
it is not difficult\footnote{$^4$}{For example, note that there is
an integral basis for the relations such that the coefficient of
$\big|1,1,1,1\big>$ is always $-1$, $0$, or $1$, and such that for
each of the $24$ generators there is some $\IR\IP^3$s with
coefficient $\pm 1$ that does not occur in the other relations.
} to see that
\eqn\quinticvolineq{
  |Q+\tilde{n}_1| + \sum_{a=2}^{125} | \tilde{n}_a |
  ~>~ |Q|
}
So the volume of the cycle always increases if we go away from the
configuration with all branes wrapping the orientifold plane. This means
that there is also no NSNS tadpole free configuration besides the trivial
one, at least not without introducing other sLag cycles besides the
$125$ $\IR\IP^3$s.

\subsec{The Quintic Standard Model}

We have seen that using the 3-cycles in \quinticRPs\ 
we cannot obtain interesting brane configurations
if we want to preserve supersymmetry. However one might hope that one could
give up supersymmetry and then at least realize the Standard Model
with intersecting branes. This is indeed possible, as we will show in this
section.

First note that all intersection numbers of the $625$ minimal
$\IR\IP^3$s are in the range $-2,\dots,2$, while we need $\pm 3$
for some cycles. So we cannot simply wrap branes on the
$\IR\IP^3$s, we must use linear combinations. Fortunately one can
represent every class in $H_3(Q;\ZZ)$ by a connected $Spin^{\IC}$
submanifold for dimensional reasons. In addition to the O6-plane
on the cycle $\pi_{{\rm O}6}=\left|0,0,0,0\right>$ we wrap
D6-branes on the following 3-cycles
%
\eqn\quinticcycles{\eqalign{
  \eqalign{
    \pi_a =&~    \left|0,0,3,1\right> \cr
    \pi_b =&~    \left|4,3,0,3\right> \cr
    \pi_c =&~    \left|3,0,1,1\right>- 2\left|4,3,0,3\right> \cr
    \pi_d =&~    \left|4,2,4,4\right>- 2\left|0,0,3,1\right> \cr
  } \qquad \Rightarrow \qquad
  \eqalign{
    \pi_a'  =&~    \left|0,0,2,4\right> \cr
    \pi_b'  =&~    \left|1,2,0,2\right> \cr
    \pi_c'  =&~    \left|2,0,4,4\right>- 2\left|1,2,0,2\right> \cr
    \pi_d'  =&~    \left|1,3,1,1\right>- 2\left|0,0,2,4\right> \cr
  }
}}
It is straightforward to check that the homology classes are primitive, 
i.e. not a multiple of some other class in $H_3(Q,\ZZ)$, by finding one 
$\IR\IP^3$ such that the intersection number is $\pm 1$.
The intersection numbers of the given $3$-cycles are
shown in table 13.
\vskip 0.8cm
\vbox{
\centerline{\vbox{
\hbox{\vbox{\offinterlineskip
\def\tablealign{&\omit&&\omit&&\omit&&\omit&&\omit&&
     \omit&&\omit&&\omit&&\omit&&\omit&\cr}
\def\tablespace{height2pt\tablealign}
\def\tablerule{\tablespace\noalign{\hrule}\tablespace}

\def\tableruleB{\tablespace\noalign{\hrule}height0.5mm\tablealign\noalign{\hrule}}
\hrule\halign{
&\vrule#&\strut\hskip0.2cm\hfill $#$\hfill\hskip0.2cm
&\vrule#\hskip0.5mm\vrule&\strut\hskip0.2cm\hfill $#$\hfill\hskip0.2cm
&\vrule#&\strut\hskip0.2cm\hfill $#$\hfill\hskip0.2cm
&\vrule#&\strut\hskip0.2cm\hfill $#$\hfill\hskip0.2cm
&\vrule#&\strut\hskip0.2cm\hfill $#$\hfill\hskip0.2cm
&\vrule#&\strut\hskip0.2cm\hfill $#$\hfill\hskip0.2cm
&\vrule#&\strut\hskip0.2cm\hfill $#$\hfill\hskip0.2cm
&\vrule#&\strut\hskip0.2cm\hfill $#$\hfill\hskip0.2cm
&\vrule#&\strut\hskip0.2cm\hfill $#$\hfill\hskip0.2cm
&\vrule#&\strut\hskip0.2cm\hfill $#$\hfill\hskip0.2cm
\cr
& \circ
&&  \pi_a &&  \pi_b &&  \pi_c &&  \pi_d
&& \pi_a' && \pi_b' && \pi_c' && \pi_d'
&& \pi_{{\rm O}6}
&\cr
\tableruleB
& \pi_a  &&  0 &&  -1 && 3 &&  0 &&  0 &&  -2 && 3 &&  0 &&  0 &\cr
\tablerule
& \pi_b  && 1 &&  0 &&  0 &&  0 &&  -2 &&  0 &&  0 && 3 &&  0 &\cr
\tablerule
& \pi_c  &&  -3 &&  0 &&  0 && 3 && 3 &&  0 &&  0 &&  -3 &&  0 &\cr
\tablerule
& \pi_d  &&  0 &&  0 &&  -3 &&  0 &&  0 && 3 && -3 &&  0 &&  0 &\cr
\tablerule
& \pi_a' &&  0 && 2 &&  -3 &&  0 &&  0 && 1 &&  -3 &&  0 &&  0 &\cr
\tablerule
& \pi_b' && 2 &&  0 &&  0 &&  -3 &&  -1 &&  0 &&  0 &&  0 &&  0 &\cr
\tablerule
& \pi_c' &&  -3 &&  0 &&  0 && 3 && 3 &&  0 &&  0 &&  -3 &&  0 &\cr
\tablerule
& \pi_d' &&  0 &&  -3 && 3 &&  0 &&  0 &&  0 && 3 &&  0 &&  0 &\cr
\tablerule
&\pi_{{\rm O}6}&&  0 &&  0 &&  0 &&  0 &&  0 &&  0 &&  0 &&  0 &&  0 &\cr
}\hrule}}}}
\centerline{
\hbox{{\bf Table 13:}{\it ~~ Intersection numbers}}}
}
\vskip 0.5cm
\noindent
Note the symmetries of the intersection matrix
\eqn\intersectsymmetries{\eqalign{
  \pi_i \circ \pi_j = -\pi_j \circ \pi_i = \pi_j' \circ \pi_i' = -\pi_i' \circ \pi_j' , \cr
  \pi_i \circ \pi_j' = \pi_j \circ \pi_i' = -\pi_i' \circ \pi_j = \pi_j \circ
  \pi_i'.
}}
The intersection pattern in table 13 is precisely the one required
for the Standard Model as proposed in \rimr: If one wraps 
four different stacks of D6-branes, namely a stack
of $3$ branes on $\pi_a$, a stack of $2$ branes on $\pi_b$, and a
single brane on $\pi_c$ and $\pi_d$ then the gauge group
is $U(3)\times U(2)\times U(1)^2$; in addition
one gets just the
necessary bifundamental chiral fermions. Since $\pi_i\circ
\pi_i'=\pi_i \circ \pi_{{\rm O}6}=0$ there are no chiral fermions
in the symmetric or antisymmetric representations of the gauge
groups. The chiral massless spectrum is shown in table 14.
\vskip 0.8cm
\vbox{
\centerline{\vbox{
\hbox{\vbox{\offinterlineskip
\def\tablespace{height2pt&\omit&&\omit&&\omit&&\omit&&
 \omit&\cr}
\def\tablerule{\tablespace\noalign{\hrule}\tablespace}

\hrule\halign{&\vrule#&\strut\hskip0.2cm\hfil#\hfill\hskip0.2cm\cr
\tablespace
& Sector && Field &&  $SU(3)\times SU(2)\times U(1)^4$ && $U(1)_Y$
&& Multiplicity &\cr
\tablerule
& $(ab)$ && $Q_L$ && $({\bf 3},{\bf 2})_{(1,-1,0,0)}$ &&  $1/3$
&& $1$ & \cr
\tablespace
& $(a'b)$ && $Q_L$ &&  $({\bf 3},{\bf 2})_{(1,1,0,0)}$ &&  $1/3$
&& $2$ & \cr
\tablespace
& $(ac)$ && $U_R$ && $(\o{\bf 3},{\bf 1})_{(-1,0,1,0)}$ &&
$-4/3$ && $3$ & \cr
\tablespace
& $(a'c)$ && $D_L$ && $(\o{\bf 3},{\bf 1})_{(-1,0,-1,0)}$ &&
$2/3$ && $3$ & \cr
\tablerule
& $(b'd)$ && $L_L$ && $({\bf 1},{\bf 2})_{(0,-1,0,-1)}$ &&  $-1$
&& $3$ & \cr
\tablespace
& $(cd)$ && $E_R$ && $({\bf 1},{\bf 1})_{(0,0,-1,1)}$ &&  $2$ &&
$3$ & \cr
\tablespace
& $(c'd)$ && $N_L$ && $({\bf 1},{\bf 1})_{(0,0, 1,1)}$ &&  $0$ &&
$3$ & \cr
\tablespace}\hrule}}}}
\centerline{
\hbox{{\bf Table 14:}{\it ~~ Chiral left-handed fermions for the 3 generation
model.}}}
}
\vskip 0.5cm
\noindent
The anomaly-free hypercharge is given by
\eqn\hyper{  U(1)_Y={1\over 3} U(1)_a - U(1)_c + U(1)_d .}
The phenomenological properties of this model are analogous to the
one described in detail in \rimr.

Of course, so far our model does have a nonvanishing RR tadpole:
The sum of the homology classes
\eqn\quintictadhom{\eqalign{
\pi ~{\buildrel {\rm def} \over =}&~
\sum_a  N_a\, (\pi_a + \pi'_a)
          +Q_q\,   \pi_{Oq}= \cr
=&~
3 (\pi_a+\pi_a') + 2(\pi_b+\pi_b') + (\pi_c+\pi_c') + (\pi_d+\pi_d')
-4 \pi_{{\rm O}6}
}}
does not vanish. For example, one can easily find other cycles that have
nonzero intersection with $\pi$. However the intersection number of
$\pi$ and the standard model branes does vanish. This is not surprising
since the intersection numbers were precisely chosen to yield the Standard
Model so that this set of intersecting branes yields an anomaly-free
spectrum by themselves.

Thus, one can introduce a hidden brane that
carries the right charge to cancel the tadpole but does not intersect the
Standard Model branes, so there is no bifundamental matter charged under both
the SM and the hidden gauge group. For this we simply add one brane
in the homology class $\pi_H{\buildrel {\rm def} \over =}-\pi$.
As already noted it does not intersect the SM branes, so this is
really a hidden brane.

We find it quite amusing that it is fairly easy to construct  a
non-supersymmetric three generation standard model on the quintic
Calabi-Yau. Using different 3-cycles on the quintic it remains to
be seen whether it is also possible to construct a realistic
supersymmetric intersecting brane model coming as close as
possible to the MSSM.

\newsec{Stability, scalar potential and gauge couplings}

Non-supersymmetric brane configurations are in general unstable.
On the one hand, depending on the intersection angles there can be tachyons
localized at the intersection points. Phenomenologically it was
suggested that these tachyons might be interpreted as Standard
Model Higgs fields \refs{\rbachas,\rbgklnon}, where
in particular in \rqsusyb\  it was demonstrated that the
gauge symmetry breaking is consistent with this point of view.
On the other hand, even if tachyons are absent one generally
faces uncanceled NSNS tadpoles, which might  destabilize the
configuration \refs{\berlin,\rott,\BlumenhagenUA,\belrab}.
In \berlin\ it was shown that it is at least
possible for appropriate
choices of the D-branes that the complex structure moduli
are stabilized by the induced  open string tree level potential.
The stabilization of the dilaton remains a major challenge
as in all non-supersymmetric string models.

For supersymmetric intersecting brane worlds we can
expect much better stability properties.
First tachyons are absent in these models due to the
Bose-Fermi degeneracy. However, since for orientifolds on Calabi-Yau spaces
the configuration only preserves ${\cal N}=1$ supersymmetry,
in general non-trivial F-term and D-term potentials
can be generated. In the case that an exact CFT description
is available, like in the orbifold examples,
it is clear that at that point in moduli space
the F- and D-terms vanish, but in general such terms
will be generated.
Luckily, the generation of D-term and F-term potentials
for D-branes wrapping sLag cycles of some underlying
Calabi-Yau manifold has been the subject of intense
study during the recent years. Therefore,
in the remainder of this section we will
collect partially known results for the
contributions to the F-term and D-term potential.

\subsec{F-term superpotential}

The effective ${\cal N}=1$ superpotential of a type II
compactification on a Calabi-Yau 3-fold with D6-branes and
O6-planes on sLag cycles receives contributions from different
sources. Due to supersymmetry there are no string loop corrections
to the superpotential, and the Peccei-Quinn symmetry forbids
perturbative $\alpha'$ corrections. The only corrections possible
arise from  world-sheet instantons. They are provided by oriented
closed strings with the topology of a sphere, wrapping
homologically non-trivial 2-cycles in the Calabi-Yau manifold, and
by unoriented closed strings with the topology of $\IR\IP^2$,
ending on the orientifold plane \AcharyaAG. Finally, the superpotential can
receive contributions from open strings discs glued  to the
various D6-branes
\refs{\OoguriBV,\AganagicGS,\AganagicNX,\LercheCW}.

It is known that closed string instantons on a type II Calabi-Yau
background do not generate a superpotential, so let us discuss the
disc instanton corrections a bit more detailed. Consider a
D6-brane which is wrapped on a supersymmetric Lagrangian 3-cycle
$\Sigma\subset {\cal M}^6$. In general, all holomorphic open
string disc instantons will contribute to the superpotential,
which will be given in terms of an instanton sum of the form \rakv
\eqn\discw{
w(s,T^k)=\sum_{k,n,\vec m}{1\over n^2}N_{k,\vec m}
\exp(n\lbrack ks-\vec m\cdot\vec T\rbrack )\, .
}
Here the open string modulus $s$ parameterizes the size of the
holomorphic disc, $N_{k,\vec m}$ are integers counting the number of domain
wall D4-branes ending on the D6-brane, which wrap a certain 2-cycle class
captured by $\vec m$, and $k$ finally denotes the
wrapping number around the boundary.
In the large volume
limit, $T^k\rightarrow\infty$, the classical superpotential is
vanishing.
Thus, very similar to heterotic string models with $(0,2)$ world-sheet
supersymmetry, also for supersymmetric intersecting brane models
one expects the background to be destabilized by world-sheet
instanton corrections. However, in the $(0,2)$ context is
is known that there exist a subclass of models, the so-called
Distler-Kachru models \rdka, which are not destabilized by
world-sheet instantons \refs{\rdkb,\rsw,\rbw,\rbsw}. It would be interesting
to determine a stable class of ${\cal N}=1$ supersymmetric
intersecting brane models, as well.

To compute this instanton expansion directly is a horrendous task,
as one does not understand  the background well  enough to
determine all non-trivial holomorphic discs. However, the
instanton expansion for A-type branes and orientifold planes,
D-branes and O-planes on sLag cycles, can be computed in simple
cases by the use of mirror symmetry. The contribution from the
disc and $\IR\IP^2$ world-sheets arise from domain walls in the
type IIB model \refs{\rgukova,\rgukovb}. Such domain-walls are given by D5-branes and
O5-planes wrapping supersymmetric 3-cycles in ${\cal W}^6$, the
mirror dual of ${\cal M}^6$. Therefore the superpotential can
only depend on the complex structure of ${\cal W}^6$ and there are
no world-sheet instanton corrections on the type IIB side. The
superpotential simply reads
\eqn\superpot{
w=\int_{{\cal W}^6} G\wedge \Omega_{3} ,
}
where $G=dC_2$ is the RR 3-form field strength. Note, that
\superpot\ is more general in the sense that even without explicit
D5-branes all sources for the RR flux $G$ also contribute to this
superpotential. For some non-compact Calabi-Yau 3-folds and
certain D6-brane configurations \superpot\ has indeed been
computed, leading to highly non-trivial results on the number of
holomorphic discs in the dual backgrounds ${\cal M}^6$.

For the particular case we are going to study in section 10, where
we consider the mirror situation with
D9-branes with gauge fluxes ${\cal F}$, the equation
of motion for $G$ can be obtained from the formulas given there.
It reads
\eqn\geqaa{\eqalign{    {1\over \kappa^2}\, d\,G&=-{\mu_9\over 8\pi^2} \biggl[
              \sum_{a} 2\, n_a\, {\rm Tr} \left({\cal F}_a\wedge {\cal F}_a \right)
                 - {\rm Tr} \left({\cal R}_a\wedge {\cal R}_a \right)\biggr]+
                      2\mu_5 \delta({\rm O}5) . \cr
                          }}
Here, $\delta({\rm O}5)$ denotes a form which is supported at the location
of the O5-planes.
Note, that \geqaa\  also includes the contribution from the O9-plane to the
curvature term.
It can be integrated and using the result from \raahv\ for the O5-plane
we get
\eqn\geqab{\eqalign{
w=-{\mu_9 \kappa^2\over 8\pi^2}&\biggl[ \sum_a
            \int_{{\cal W}^6} 2 n_a\, {\rm Tr}\left(A_a\wedge d\, A_a+{2\over 3}
              A_a\wedge A_a\wedge A_a\right) \wedge \Omega_{3} \cr
             &- \int_{{\cal W}^6} {\rm Tr}\left(\omega\wedge d\, \omega+{2\over 3}
              \omega\wedge \omega\wedge \omega\right) \wedge \Omega_{3} \biggr]
             +2\mu_5\kappa^2\, \int_C \Omega_3,\cr }  }
where $\omega$ denotes the spin-connection and $C$ a 3-cycle satisfying
$\partial C=\pi_{{\rm O}5}$ .
The superpotential \geqab\ looks very similar to the superpotential for the
type I or heterotic
string. From this discussion it is
clear that to really compute the superpotential
for the A-type models with intersecting D6-branes it is essential
to gain a better understanding  of the detailed mirror map for the
D6-branes.

\subsec{D-term potential at tree level}

The self-energy of the D-branes is due
to the tension of the wrapped D-branes. Since the open string tree-level
scalar potential only depends on the complex structure moduli it is clear that
in the effective ${\cal N}=1$ field theory this potential does not arise
from a holomorphic superpotential but from a D-term.
The D-term scalar potential has the following general form
\eqn\dterm{
{\cal V}_{\rm D-term} =
\sum_a {1\over 2g_a^2}\biggl(\sum_i q^i_a|\phi_i|^2+\xi_a\biggr)^2\, .
}
Here $g_a$ is the gauge coupling of some $U(1)_a$ gauge group, and
$\xi_a$ is the Fayet-Iliopoulos term associated to $U(1)_a$. The
scalar fields $\phi_i$ will obtain a positive or negative
quadratic mass term for non-vanishing $\xi_a$, depending on the
sign of their $U(1)_a$ charges $q^i_a$. In the intersecting brane
worlds the potential energy of the wrapped branes corresponds to
the term ${1\over g_a^2}\xi_a^2$ in \dterm, as we will discuss in
the following, and the scalars $\phi_i$ are open strings fields
which are massive, massless or tachyonic depending on the choice
of moduli parameters. ${\cal N}=1$ supersymmetry will be unbroken,
if the potential vanishes in the groundstate. In particular this
can be achieved for $\xi_a=0$, where the scalars $\phi_i$ are all
massless. If, however, $\xi_a > 0$, a supersymmetric groundstate is only
accessible after the condensation of a tachyonic scalar.

Recall that the disc level scalar potential for D6-branes wrapping sLag
cycles can be written as
\eqn\dbia{\eqalign{
{\cal V}&=
T_6\, e^{-\phi_4}\, \left[
\sum_a{N_a \left| \int_{\pi_a} \widehat\Omega_3 \right|} +
 \sum_a{N_a \left| \int_{\pi'_a} \widehat\Omega_3 \right|}-
4\, \int_{\pi_{{\rm O}6}} \Re(\widehat\Omega_3) \right] \cr
&=
2\,T_6\, e^{-\phi_4}\,
\sum_a N_a \left( \left| \int_{\pi_a} \widehat\Omega_3 \right|-
              \int_{\pi_a} \Re(\widehat\Omega_3) \right)   ,
}}
where we have used \dbic\ and \tadhom.
On the other hand, the $U(1)_a$ gauge coupling is given by
\eqn\gaugec{
{1\over g_{U(1)_a}^2}={N_a\over g_a^2}={N_a\, M_s^3\over (2\pi)^4}\,
                     e^{-\phi_{4}}
                \left| \int_{\pi_a} \widehat\Omega_3 \right|.
                }
Hence, by direct comparison with \dterm\ we see that the FI-term
can be identified as
\eqn\fiterm{
\xi^2_a={M_s^4 \over 2\pi^2}\left( { \left| \int_{\pi_a} \widehat\Omega_3 \right|-
               \int_{\pi_a} \Re(\widehat\Omega_3) \over
                \left| \int_{\pi_a} \widehat\Omega_3 \right|} \right) ,}
which apparently vanishes if the D-brane is calibrated with respect
to the same 3-form as the orientifold plane.

Since the FI-term is not a holomorphic quantity one expects
higher loop corrections to the classical result \fiterm.
Using the expansion \expand\ of the 3-cycle
$\pi_a$ in terms the basis 3-cycles $\alpha^I$ and $\beta_J$,
one can write
\eqn\vcya{\left| \int_{\pi_a} \widehat\Omega_3 \right| = |e^a_I\hat
  X^I(U)-m_a^I\hat F_I(U)|\, .
  }
This expression is invariant under modular symplectic
transformations in $Sp(2h^{(2,1)}+2,\ZZ)$. Note that this type of
potential was already discussed in the context of automorphic
functions on Calabi-Yau spaces in \FerraraUZ. Moreover a very
similar expression appears as the superpotential due to RR- and
NSNS-fluxes on Calabi-Yau spaces, e.g. discussed in
\refs{\TaylorII,\MayrHH,\CurioSC,\CurioAE}. The vacuum structure of this
type of potentials was investigated in some detail in
\CurioSC\ with result that at a generic point in the moduli space
supersymmetry is completely broken. However the minima of ${\cal
V}$ are generically such that the system is dynamically driven to
those degeneration points in the moduli space (large complex
structure limit, conifold points) where full ${\cal N}=2$
supersymmetry is restored. In addition, there might by isolated
singular points (field theory Seiberg-Witten points), where only
${\cal N}=1$ supersymmetry is unbroken. A general analysis of
those supersymmetric attractor points was also provided  in a
beautiful work of Moore \refs{\MooreZU,\MoorePN}\ and also in the
context of supersymmetric ${\cal N}=2$ black holes in
\refs{\BehrndtJN,\DenefSV}.

The tree level scalar potential ${\cal V}$ was computed in
\refs{\rott,\BlumenhagenUA} (see also \belrab) for the case of
intersecting D6-branes on $T^6$ and also for the case of the
$T^6/\ZZ_3$ orbifold. For the case of the toroidal background,
${\cal V}$ only depends on the three complex structure moduli
$U^I$ of the three two-tori. For the $\ZZ_3$ orbifold with
$h^{(2,1)}=0$, it is simply constant. Let us ``check" the general
Calabi-Yau formula \vcya\ for the case of the torus $T_1^2\times
T_2^2\times T_3^2$ with three complex structure moduli $U^I,\,
I=1,2,3$. This discussion also applies for every orbifold with
$h^{(2,1)}=3$ like the $\ZZ_2\times \ZZ_2$ orbifold. The moduli
space for these three fields is $\widetilde{\cal M}_{\rm
H}=({SU(1,1)/U(1)})^3$. The corresponding prepotential for the
complex structure moduli is given by the expression
\eqn\torusf{
F(X)={X^1X^2X^3\over X^0}=(X^0)^2U^1U^2U^3,\quad U^I={X^I\over
X^0}
\, .
}
Then, first without any further restriction on the expansion
coefficients $e_I$ and $m^I$, the scalar potential takes the following form
\eqn\pota{
{\cal V}= T_6 e^{-\phi_4}
{|e_0+e_1U^1+e_2U^2+e_3U^3+m^0U^1U^2U^3+
m^1U^2U^3+m^2U^1U^3+m^3U^1U^2|\over\prod_{I=1}^3\sqrt{{\Im(U^I)}}\,
} . }
However, we want to assume that the D6-brane wraps such homology
3-cycles which factorizes into three 1-cycles around each of the
three tori; hence this 3-cycle can be characterized by three pairs
of unrestricted wrapping numbers $(r_I,s_I)$. This means that the
$e_I$ and $m^I$ in \pota\ are not completely arbitrary but the
following relations have to hold:
\eqn\relationem{\eqalign{
& e_0=r_1r_2r_3,\quad
 e_1=s_1r_2r_3,\quad e_2=s_2r_1r_3,
\quad e_3=s_3r_1r_2,\cr
& m^0=s_1s_2s_3,\quad
m^1=r_1s_2s_3,\quad m^2=r_2s_1s_3,\quad m^3=r_3s_1s_2.}
}
Then, the potential \pota\ can be rewritten as
\eqn\potb{
{\cal V} =
T_6 e^{-\phi_4} \prod_I{|r_I+s_IU^I|\over \sqrt{{\Im(U^I)}}\, .
}}
This expression agrees with the potential obtained in
\refs{\rott,\BlumenhagenUA}\ and in \belrab.

\subsec{Gauge couplings}

Each gauge factor supported on a stack of D6-branes comes with its
own gauge coupling $g_a$, which can be deduced from the
Dirac-Born-Infeld action to be
\eqn\gaugec{
{1\over g_a^2}={M_s^3 \over (2\pi)^4}\,  e^{-\phi_4} \left\vert
                           \int_{\pi_a}   \widehat{\Omega}_3 \right\vert .
                           }
Thus, the gauge coupling also depends on the volume of the sLag
3-cycle the D6-brane is wrapping around. In a supersymmetric
theory the gauge coupling can be combined with an axionic
theta-angle into the holomorphic gauge kinetic function $f$. The
relevant axionic coupling is contained in the Chern-Simons part of
the D6-brane action and looks like
\eqn\axion{
{\cal S}_{\rm axion}=\mu_6 \int_{\IR^{3,1}\times\pi_a}
  2\pi^2(\alpha')^2\,  C_3\wedge {\rm Tr}({\cal F}_a\wedge {\cal F}_a) .
  }
Hence, the axion is given by
\eqn\axi{
\lambda_a={2\, M_s^3\over (2\pi)^4}\, \int_{\pi_a}  C_3  .
}
These two couplings combine into a complex coupling
\eqn\coml{
f={M_s^3\over (2\pi)^4} \left[  e^{-\phi_4}
                           \int_{\pi_a}   \Re(\widehat{\Omega}_3) + 2i\,
                           \int_{\pi_a}  C_3 \right] ,
                           }
where the real part depends on the $h^{(2,1)}$ real complex
structure moduli. This was already shown for the toroidal case in
\rqsusy. For a general Calabi-Yau, the complex gauge coupling depend
holomorphically on the $h^{(2,1)}+1$ chiral superfields that
contain the complex structure moduli as scalar components.
Moreover, it is known that for ${\cal N}=1$ supersymmetric
theories the gauge kinetic function receives quantum correction
only at the one loop level. Thus, as in \rmayrb\ one will get
corrections from world-sheet instantons with annulus, M\"obius
strip and Klein-bottle topology. Since the gauge couplings depend
on the volume of the 3-cycles, it is clear that the constraints
one gets for realistic intersecting brane world models to meet the
phenomenological gauge couplings at  the weak scale are rather
moderate and at least at the classical level easy to satisfy by
choosing appropriate values for the complex structure moduli.

Finally, we would like to comment on the possibility to obtain
scenarios with a low string scale and large extra dimensions
\refs{\radd,\raadd}
with intersecting D6-branes. Since the four-dimensional Planck
mass is given by $M_{pl}^2=M_s^8\, e^{-2\phi_4}$, the following
relation holds
\eqn\largeex{   {1\over g_a^2}\sim {M_{pl}\over M_s} \left\vert\int_{\pi_a}
  \widehat{\Omega}_3 \right\vert ,}
so that gauge couplings of order one can be obtained for small
values of the string scale, if we can choose
\eqn\beding{  \left\vert \int_{\pi_a} \widehat{\Omega}_3 \right\vert \ll 1 }
for all 3-cycles ${\pi_a}$. In the toroidal case with factorizable
D6-branes it was not possible to meet the requirement \beding\ for
all D6-branes while preserving chirality.
This was due to the fact, that a large transverse
volume requires at least some of the complex structure moduli to
become large. Then, \largeex\ was violated, because in order to preserve
chirality not all D6-branes on a torus could avoid the large direction. This
obstruction does no longer appear for general Calabi-Yau models,
where low scale string models hence seem to be  accessible.

\newsec{Mirror symmetry}

So far we have considered type IIA orientifolds with A-type
D6-branes wrapping sLag 3-cycles in the Calabi-Yau. We have
formulated the consistency conditions for  the absence of
RR-charges on the internal Calabi-Yau and have presented a general
purely topological formula for the chiral massless spectrum
localized at the intersection points of the D6-branes. In this
section we will give the mirror symmetric formulation of such
intersecting brane world models.

Here, one starts with the type IIB string on the mirror Calabi-Yau
space ${\cal W}^6$ with Hodge numbers $\widetilde
h^{(2,1)}=h^{(1,1)}$ and $\widetilde h^{(1,1)}=h^{(2,1)}$.
Let us briefly discuss the fate of the intersecting D6-branes and
the orientifold O6-plane under mirror symmetry. It is known that
A-type branes are mapped under mirror symmetry to B-type branes \rvafa,
which are wrapped on even-dimensional holomorphic cycles in the
$H_{2p}({\cal W}^6;\ZZ)$ homology of the mirror manifold and
carry a holomorphic stable gauge bundle with curvature 2-form
${\cal F}$ on their world volumes. The calibration condition
$\Re(e^{i\theta}\Omega_3) = d{\rm vol}|_{{\rm D}q}$ of the
A-branes, is now replaced by the
MMMS equation \MarinoAF
\eqn\mmms{
{1\over 3!} {\cal F}\wedge {\cal F}\wedge {\cal F} -
{1\over 2!} J\wedge J \wedge {\cal F} =
\tan(\theta) \left(
{1\over 2!} J\wedge {\cal F}\wedge {\cal F}  -
{1\over 3!} J\wedge J \wedge J \right),
}
together with the requirement, that the $(2,0)$ component of
${\cal F}$ vanishes. The parameter $\theta$ takes the role of
the free phase in the angle condition \angle\ of the sLag cycles.
For simplicity, we will take all branes to be D9-branes. The
precise map of intersecting D6-branes to D9-branes with constant
gauge field strength is known for the torus, but for a general
Calabi-Yau it is hard to find explicitly.
In the case of orbifold realizations
mirror symmetry is simply given by T-duality along three of the
six compact directions which maps a symmetric $\ZZ_N$ orbifold to
an asymmetric $\hat{\ZZ}_N$ orbifold.
Moreover, intersecting D6-branes are mapped to D9-branes
with constant magnetic flux as has been studied in
\refs{\rbgklnon,\rbgkl,\KorsKU}.

Even without knowing the precise mirror map for a generic
Calabi-Yau model we can independently formulate the models in the
mirror symmetric setting. Thus, we start with the type IIB string
theory compactified on a Calabi-Yau manifold ${\cal W}^6$. Now, we
divide out this model by the world-sheet parity transformation
$\Omega$. As a result we get an orientifold O9-plane wrapping the
entire Calabi-Yau manifold and O5-planes wrapping holomorphic
2-cycles $\pi_{{\rm O}5}$ of the Calabi-Yau. In principle, a mirror
configuration could also contain O7- and O3-planes, however, not
together with O9- and O5-planes, as this would break
supersymmetry.
Now we introduce stacks of D9$_a$-branes\footnote{$^5$}{As in the
toroidal case we expect that D-branes of smaller dimension can be
considered as special degeneration limits of bundles over
D9-branes. The appropriate framework to describe  such degenerate
bundles is given by coherent sheaves \rdouglas.} wrapping the
entire threefold $n_a$ times and each stack carrying a $U(N_a)$
gauge bundle defined by ${\cal F}_a$. Via the Chern-Simons
couplings
\bornina\ in the Born-Infeld action
these fluxes introduce charges under the lower degree RR-forms as well.
Under $\Omega$ a stack of D9-branes with a $U(N_a)$ gauge connection ${\cal
A}_a$ is mapped to another stack of D9-branes with a $U(N_a)$ gauge
connection
\eqn\gaugeb{ {\cal A}'_a=-\gamma_{a}\, {\cal A}_a^T\,
                  \gamma_{a}^{-1}, }
where the $\gamma_{a}$ denote unitary matrices.
Since the ${\cal A}_a$ are skew-Hermitian this is identical to
\eqn\gaugeba{
{\cal A}'_a=\gamma_{a}\, {\cal A}_a^*\,
                  \gamma_{a}^{-1} ,
                  }
implying that the action on the gauge field strength
${\cal F}_a=d {\cal A}_a+{\cal A}_a\wedge {\cal A}_a$ is also given by
\eqn\gaugeb{ {\cal F}'_a=\gamma_{a}\, {\cal F}_a^*\,
                  \gamma_{a}^{-1}. }
The RR-tadpole
cancellation conditions can be derived from the Chern-Simons couplings
\eqn\tadio{\eqalign{
{\cal S}_{\rm CS}&= \mu_9 \sum_p \Big(
 \int_{{\cal W}^6}  C_p \wedge \sum_a n_a\, \bigl[ {\rm ch}({\cal F}_a)+
               {\rm ch}({\cal F}^*_a) \bigr]    \wedge \sqrt{\hat {\cal A}({\cal R})} \cr
& \hskip4cm  + Q_9
                \int_{{\cal W}^6}    C_p\wedge 
\sqrt{\hat {\cal L}({\cal R} /4)} \Big) + \mu_5 Q_5 \, \int_{\pi_{{\rm O}5}} 
                C_6
}}
and can be written as
\eqn\tafg{   \sum_a n_a\, \bigl[ {\rm ch}({\cal F}_a)+
               {\rm ch}({\cal F}^*_a) \bigr]    \wedge 
\sqrt{\hat {\cal A}({\cal R})} -32 \, \sqrt{\hat {\cal L}({\cal R}/4)} - 2
                {\mu_5 \over \mu_9}  \delta(\pi_{{\rm O}5}) =0
                  }
where $\delta(\pi_{{\rm O}5})$ denotes the Poincar\'e dual 4-form
of $\pi_{{\rm O}5}$. Since on a Calabi-Yau manifold one has
$h^{(0,0)}=h^{(3,3)}=1$ and $h^{(1,1)}=h^{(2,2)}$, expanding the
tadpole condition
\tafg\ into a basis one gets $2h^{(1,1)}+2$ different RR-forms and their
respective  tadpole cancellation conditions. However, similar to the  type IIA
orientifolds only half of these conditions are non-trivial, as
one half of the RR-forms are projected out.

In the large volume regime
the chiral massless spectrum can be deduced from the Atiyah-Singer
index theorem. The four-dimensional chiral fermions transforming
in the bi-fundamental representation of two factors of the total
gauge group are sections in a spinor bundle associated to the
tensor product of the two individual gauge bundles. Due to the
property ${\rm ch}( {\cal F}_a \otimes {\cal F}_b) = {\rm ch}({\cal
F}_a)
\wedge {\rm ch}({\cal F}_b)$
of the Chern character, their multiplicity is given by
\eqn\intis{\eqalign{
I(a,b)&= {n_a\, n_b\over N_a\, N_b}\,
                \int_{{\cal W}^6} {\rm ch}_{(\o{N}_a,{N}_b)}
({\cal F}^*_{a}\otimes {\cal F}_b) \wedge
                          {\hat {\cal A}({\cal R})} \cr
            &={n_a\, n_b\over N_a\, N_b}\, \int_{{\cal W}^6}
{\rm ch}({\cal F}^*_{a})\wedge
                                {\rm ch}( {\cal F}_{b})\wedge
                          {\hat {\cal  A}({\cal R})} .}
                          }
For the case of flat D9-branes with constant gauge curvatures this
formula can also be produced from the open string quantization, of
course. Moreover, the number of $\Omega$ invariant
self-intersection points is given by the M\"obius amplitude, which
was derived in
\roemel,
\eqn\intisb{\eqalign{  I(a)&= 8\, {n_a\over N_a}\,
                \int_{{\cal W}^6} {\rm ch}({\cal F}_{a}) \wedge
                          \sqrt{\hat {\cal A}({\cal R})}\wedge
                          \sqrt{\hat {\cal L}({\cal R}/4)}+
                          {1\over 2}\, {n_a\over N_a}\, \int_{\pi_{{\rm O}5}} {\rm ch}({\cal
                            F}_{a}).}}
Thus, the spectrum of chiral fermions for a set of D9-branes with
gauge fluxes is now given by table 15.
\vskip 0.8cm
\vbox{
\centerline{\vbox{
\hbox{\vbox{\offinterlineskip
\def\tablespace{height2pt&\omit&&
 \omit&\cr}
\def\tablerule{\tablespace\noalign{\hrule}\tablespace}

\hrule\halign{&\vrule#&\strut\hskip0.2cm\hfill #\hfill\hskip0.2cm\cr
& Representation  && Multiplicity &\cr
\tablerule
& $[{\bf A_a}]_{L}$  && ${1\over 2}\left(I(a^*,a)+I(a)\right)$   &\cr
\tablerule
& $[{\bf S_a}]_{L}$
     && ${1\over 2}\left(I(a^*,a)-
                           I(a)\right)$   &\cr
\tablerule
& $[{\bf (\o N_a,N_b)}]_{L}$  && $I(a,b)$   &\cr
\tablerule
& $[{\bf (N_a, N_b)}]_{L}$
&& $I(a^*,b)$   &\cr
}\hrule}}}}
\centerline{
\hbox{{\bf Table 15:}{\it ~~ Chiral spectrum in $d=4$}}}
}
\vskip 0.5cm
\noindent
Using the RR-tadpole cancellation condition \tafg\ it is
straightforward to confirm that the generic massless spectrum in
table 15 is free of non-abelian gauge anomalies.

As a simple check one can verify that the Atiyah-Singer index
theorem \intis\ applied to $U(1)$ gauge bundles with constant
field strength on a flat six-torus $T^6=T_1^2\times T_2^2\times T_3^2$
indeed gives the result derived in
\rbgklnon. The following constant $U(1)$ gauge fluxes are T-dual
to factorizing intersecting D6-branes on $T^6$
\eqn\sfkd{  {\cal F}_a=-{2\pi i} \sum_{I=1}^3 {m_a^I\over n_a^I
                 {\rm Vol}(T^2_I)}
        \, {\bf 1}_{(N_a,N_a)}\, dx_I\wedge dy_I ,}
where $x_I,y_I$ are Cartesian coordinates on the torus $T^2_I$.
The evaluation of the integral in \intis\ for the number of chiral
fermions
\eqn\erwsa{   I(a,b)= \prod_{I=1}^3  \left( n_a^I\,
m_b^I-m_a^I\, n_b^I\right)
}
which agrees with the formula deduced in \rbgklnon\ by CFT
methods. One can push this list of analogies further, for instance
in evaluating the potential in the dual frame. In the end, the two
mirror sides are just equivalent descriptions of identical physics.

\newsec{Lift to M-theory and related issues}

In the previous sections we have described in quite general terms
intersecting D6-brane models on general orientifolds of Calabi-Yau manifolds.
These models provide a huge class of four-dimensional
string compactifications with quasi-realistic gauge groups
and chiral matter content, where the gauge degrees of freedom are localized
on the D6-branes with co-dimension three and the chiral matter sits
at the intersections of these D6-branes of  co-dimension
six. In general, such models break supersymmetry completely, but this is
not automatic. Demanding supersymmetry actually fixes some of the
complex structure moduli dynamically  
and deviating from the supersymmetric
locus introduces a D-term potential, such that the breaking can be
described spontaneously.
In order to have any chance to preserve supersymmetry in these models,
it is essential that we are dealing with orientifolds
containing O6-planes with negative tension, so that the overall tension
can potentially vanish.

An interesting issue in type IIA vacua with only D6-branes, and possibly
O6-planes, is the fact that the eleven-dimensional lift to M-theory
is purely geometrical in the sense that the eleven-dimensional 3-form is
trivial. This can be easily seen by looking at the fields which couple
to the 6-brane background. The D6-branes are the magnetically charged
monopoles of the KK vector, and thus only couple to components of the
eleven-dimensional metric, the RR 1-form $C_1$ and the dilaton of type IIA.
The geometric lift of isolated D6-branes in flat ten-dimensional space-time
is then a non-trivial $U(1)$ fibration over an $S^2\subset\IR^3$,
a Taub-Nut space. The base $S^2$ is given by the asymptotic infinity of
their transverse space $\IR^3$ in
ten-dimensions, and the $U(1)$ fiber defines the dilaton of type IIA.
Similarly, the O6-planes lift to an Atiyah-Hitchin space.
It is important that these spaces are equipped with metrics that allow a
consistent ten-dimensional interpretation of the configuration.
This requires in particular, that the fiber circle is regular everywhere and
approaches a finite radius at infinity. Then, the geometry corresponds to a type IIA
vacuum with a finite coupling everywhere, which upon rescaling may be assumed
to be small at infinity.

In order to get a chiral model it is not sufficient to consider
parallel D6-branes and O6-planes, as we have discussed at length.
In chiral backgrounds the charges cannot cancel locally and
the dilaton and the RR 1-form potential vary.
Supersymmetric settings of this type preserve exactly four supercharges,
i.e. ${\cal N}=1$ supersymmetry
in four dimensions. By the general classification of the number of
Killing spinors on a seven-dimensional
manifold equipped with the Levi-Civita connection,
they lift to an M-theory compactification on a manifold with holonomy $G_2$.
Due to the fact, that smooth $G_2$-manifolds only have abelian isometries,
to lift a chiral model with a non-abelian gauge group, we need to allow for
singularities \refs{\PapadopoulosDA,\AcharyaGB,\rcveticb,\rwitten}.
In practice, little is known about the lift of compact six-dimensional
backgrounds to eleven dimensions, except for orbifold models \KachruJE\
and CFT results
\refs{\ShatashviliZW,\BlumenhagenJB,\RoibanCP,\EguchiIP,\NoyvertMC},
which, however, do not directly relate to a geometric
interpretation of the manifold. In the following section we just
like to collect a number of observations and speculations starting
from the known cases of local models. In a sense, we would like to
take the attitude that supersymmetric orientifolds with
intersecting branes in ten dimensions provide an alternative
approach to the study of M-theory compactifications on certain
singular $G_2$ manifolds.\footnote{$^6$}{It would be also
interesting to get a better understanding about the relation
between M-theory on a compact $G_2$ manifold respectively
intersecting D6-branes in type IIA and orbifold compactifications
of heterotic M-theory to six and four dimensions with chiral
matter fields located at the intersections of the various orbifold
fixed planes
\refs{\FauxHM\KaplunovskyIA\FauxDV\FauxSP\DoranVE-\GorbatovPW}.}

\subsec{Local models of intersecting D6-branes}

It has been argued, that the proper lift of two intersecting D6-banes
in $\IR^6$ is given by a non-compact seven-dimensional cone \AtiyahQF.
Concretely, it was observed by M. Atiyah and E. Witten
that one can define a $U(1)$ action on
$\IC\IP^3$ such that $\IC\IP^3/U(1) = S^5$ and Fix$(U(1)) = S^2 \cup S^2$.
Taking the seven-dimensional cone on $\IC\IP^3$ leads to a
$U(1)$-manifold with a fiber that degenerates exactly along
two intersecting sLag 3-planes within $\IR^6$. Smoothing out this geometry
is believed to describe the geometric lift
of two intersecting D6-branes located at the
fixed locus of the $U(1)$ that defines the dilaton.
Actually, a metric on the cone on $\IC\IP^3$ with holonomy $G_2$
is known \refs{\rbs,\GibbonsER,\CveticZX}.
It describes a space topologically $\IR^3\times S^4$, where
the co-associative 4-cycle $S^4$ replaces the singular tip of the cone.
Unfortunately, this metric can not function as a candidate for the lift
of the type IIA vacuum without modifications, as the circle meant to define the
string coupling is not finite at infinity.

In \rwa\ (see also \rberbra)
a generalization of the construction towards multiple intersecting
D6-branes was proposed. Using $\IC\IP^3/U(1) = S^7/U(1)^2$, one can divide out
by any linear combination $U(1)'$ of the two $U(1)$ in the first step to define the
base of another seven-dimensional cone. The linear combination of
the two generators
is defined by two integers $(n,m)$, which simultaneously play the role of the
weights of the weighted projective space
$S^7/U(1)' = {\bf W}\IC\IP^3_{n,n,m,m}$ and of the multiplicities
of the two D6-branes. These are now modeled by the $A_{n-1}$ and $A_{m-1}$
singularities of the weighted projective space.
Not even knowing the proper metric to describe the lift of a single pair of
D6-branes, there appear to be little clues, how to determine a sufficiently regular
metric for this more demanding case (see e.g. \GubserMZ\ for some discussion).
Still, one may want to add more physical observations about what we expect these
spaces to behave like, even without knowing their precise definition.

\subsec{Geometric transitions}

In particular, in the type IIA background we encounter a number
of transitions which should lift to viable topology changing
transitions of the $G_2$ manifold.
At a supersymmetric intersection of two D6-branes the bosonic
superpartner of the chiral fermion is a marginal operator transforming
in the bi-fundamental representation of the two gauge factors.
This flat direction indicates a deformation of the two intersecting
D6-branes towards a single D6-brane with the same volume
while preserving the homology class \rurangab. This transition can
be thought of as the six-dimensional analogue of the deformation of sLag
2-cycles, that was studied in section 4.7.1. In the effective
field theory it can be considered as moving on a Higgs-branch, where the
gauge symmetry of the two individual stacks is broken to the diagonal.
It is clear that in the M-theory such a smooth transition must
lift to a topology change between the two different singular $G_2$ manifolds
described above. In a local model,
we then expect a smooth transition between seven-dimensional cones on
different three-dimensional weighted projective spaces.

An even more dramatic change is given by the conifold transition,
where a 3-cycle in the Calabi-Yau shrinks to zero size and the resulting singularity
is resolved by a 2-cycle. Following C. Vafa \VafaWI\ D6-branes
which are wrapped around the shrinking 3-cycle become some RR
2-form flux through the resolving 2-sphere. In fact, this type of
topology change is quite well understood for
the deformed conifold geometry $T^*S^3$  with $N$ wrapped D6-branes on the
sLag $S^3$, and
the resolved conifold ${\cal O}(-1)\oplus{\cal O}(-1)\rightarrow
\IC\IP^1$ with $N$ units of flux for the RR 2-form $G_2$.
In the corresponding M-theory lift, this
situation was geometrically interpreted by a flop transition between
two $S^3$ in the corresponding non-compact $G_2$ manifold \AtiyahZZ.
Here, the situation is slightly better in that an appropriate metric on one
side of the transition is known, while for the metric after the
flop numerical studies have been
performed \refs{\BrandhuberYI,\BrandhuberKQ}.

For compact Calabi-Yau manifolds the conifold transition affects
the whole Calabi-Yau manifold. Then,  in general we have to expect that
after the transition all the D6-branes and the O6-planes are
arranged in a completely new way. Whether this quite complicated
type IIA transition lifts to a simple flop transition on the $G_2$
manifold is clearly a difficult question. But one may try to derive at least
some consistency requirement from the tadpole conditions analogous
to \tadhom.
So let us assume that we are
left after the conifold
transition with some remaining 3-cycles $\pi_a$ ($a=1,\dots ,N'$) with
$N_a$ D6-branes wrapped around as well as with the O6 homology
3-cycle $\pi_{{\rm O}6}$. In addition $N''$ new 2-cycles $S^2_b$
($b=1,\dots ,N''$) have emerged after the conifold transition,
which each supports $N_b$ units of Ramond 2-form flux:
\eqn
\twoflux
{N_b=\int_{S^2_b}G_2,\quad G_2=dC_1\, .}

Before we try to generalize the type IIA tadpole condition including the
Ramond 2-form fluxes, let us
switch to the type IIB
mirror picture. Here the conifold transition replaces D5-branes
wrapped around a 2-sphere by some flux of the 3-form gauge field strength
$H_3^{\rm R}+\tau H_3^{\rm NS}$ through some appropriate dual 3-cycles
$C^3_b$ and $\tilde C^3_b$.
Then we will get the following tadpole condition
for the 3-form fluxes in the absence of space-time filling D3-branes
(see e.g. the discussion in \rkst):
\eqn
\threeflux
{\sum_{b=1}^{N''}\int_{C^3_b\times \tilde C^3_b}H_3^{\rm R}\wedge H_3^{\rm NS}
=0\,
.}

Coming back to the type IIA side, it was argued
in \VafaWI\ (see also the comments in
\CurioDZ) that the mirror description of the type IIB NS 3-form flux is
given in type IIA in terms of
a non-vanishing flux of a NS 4-form field
strength $H^{\rm NS}_4$. Actually the role of this NS 4-form field
strength is still not very well understood in the literature. It
is associated with the lack of integrability of the complex
structure of the Calabi-Yau space due to the presence of the
Ramond 2-form flux. Hence $H_4^{\rm NS}$ reflects the back-reaction of
turning on the Ramond 2-form flux on the six-dimensional
Calabi-Yau metric. In any case, if this interpretation is correct,
the following tadpole condition for the Ramond 2-form fluxes suggests itself:
\eqn\tadpoleca{
\sum_{b=1}^{N''} \int_{S^2_b\times C^4_b}G_2\wedge H^{\rm NS}_4
=
         0 .}
Here the sum is meant to be over the relevant homology 2-cycles $S^2_b$.
In addition we have also to satisfy the tadpole condition for the
still remaining wrapped D6-branes:
\eqn\tadpolecaa{
\sum_a^{N'}  N_a\, ( \pi_a + \pi'_a)
=
          4\,   \pi_{O6} .}
These two equations constitute two sets of independent conditions,
one for the wrapped D6-branes on the homology 3-cycles, and a second for the
fluxes through the homology 2-cycles. Therefore due to these two equations
we expect that the conifold transition with D6-branes on a compact
CY-space is only possible under strong restrictions, namely only if
the disappearing D6-branes, and hence  at the same the
newly emerging 2-form fluxes satisfy the Ramond tadpole conditions
on their own.

\newsec{Conclusions}

In this paper we have investigated aspects of intersecting
brane worlds on orientifolds of K3 and Calabi-Yau manifolds.
As our main result we have provided a general formula
for the chiral massless spectrum in terms of topological
invariants. In view of the Atiyah-Singer index theorem
it is required that the chiral fermions
are determined entirely by the topology of the configurations
when classical geometry is valid.
However, after having witnessed the construction of various
orientifold models over the years using explicit and elaborate
conformal field theory techniques, it appears to be
quite amusing that the most important part of the
massless spectra can be so easily determined from
the intersection numbers of cycles in the blown-up
geometries. As a new example,
which mainly just demonstrates the
power of the general topological formulas, we
derived  a ``realistic'' three generation Standard-like Model from
wrapping D-branes on 3-cycles of the quintic Calabi-Yau.

We have pointed out that these intersecting brane world
models are interesting for many reasons. From
the phenomenological point of view they
generically produce effective six- and four-dimensional models
with chiral fermions and do provide an
intrinsic mechanism to break supersymmetry.
In order for realistic non-supersymmetric models
to make sense it is viable that the string
scale is not much large than the TeV scale.
It would be interesting to construct concrete models for which
such a low string scale scenario is realized. But the central issue in
non-supersymmetric model building still remains to be the
problem of stabilizing the scalar fields, in the first place the
dilaton, of course.

For specific choices of the intersecting branes
it is  possible to get models with
${\cal N}=1$ supersymmetry in four dimensions.
For these models it would be a major step forward
if one could determine the instanton generated F-term superpotential
using mirror symmetry. In particular, it would be interesting to see in how far
there exist intersecting brane worlds which are
not destabilized by such potentials. At least we know
that supersymmetric intersecting brane worlds  on orbifold backgrounds
are exact minima of the scalar potential.
Of course, the issue of supersymmetry breaking is crucial to relate these
models to the low-energy world. Moreover,
every supersymmetric intersecting brane world is expected to
lift to an M-theory compactification
on a singular compact $G_2$ manifold.
Since not very much is known about such $G_2$ manifolds,
intersecting branes might provide an important approach
to detect at least part of their structure.

Another important problem in four-dimensional intersecting brane world
models is the question of gauge coupling unification, or respectively,
to obtain realistic values for the gauge couplings. This seems to be
quite non-trivial in non-supersymmetric models with a TeV string scale,
but probably easier in supersymmetric models with a high string scale and
low-energy supersymmetry breaking. In any case, to analyze this
problem one has to compute the higher loop threshold corrections
to the gauge couplings of the open strings on the Calabi-Yau manifolds.

Finally, let us mention the possibility of spontaneously
breaking the gauge group of the Standard
Model by a tachyonic scalar field which has
the quantum numbers of the Standard Model Higgs field. This mechanism
was discussed in \refs{\rbgklnon,\rafiruph,\rimr,\rott,\rqsusyb}
and it is described in terms of brane language by the recombination of the
$SU(2)_L$ and $U(1)_Y$ D-branes into a single $U(1)_Q$ D-brane.
Following the change of intersecting numbers triggered by the
D-brane recombination process,
information about the mass generation for quarks and leptons
can be in principle obtained. Therefore, a careful examination of the
tachyon potential and the rearrangement of topological
D-brane intersection numbers
due to tachyon condensation on Calabi-Yau spaces would be worth
to be investigated.
In summary, a lot of theoretical and phenomenological problems
related to D-branes on Calabi-Yau spaces are still to be solved.

\vskip2cm

\centerline{{\bf Acknowledgments}}\pano
We would like to thank A. Klemm for useful discussions.
The work is supported in part by the EC under the RTN project
HPRN-CT-2000-00131. R.B. would like to thank the Theoretical Physics groups
at the University Bonn and at the Universidad Autonoma de Madrid in 
addition to ISAS-SISSA Trieste for hospitality.
We thank L. G\"orlich and T. Ott for pointing out an error in an earlier
version of this paper. 

\noindent
R.B. would like to dedicate this paper to the memory of his colleague Sonia Stanciu.
He shared a creative  time with her as graduate students at the
University Bonn in the early nineties.

\vfill\eject

\listrefs

\bye
\end